# H I Observations of Giant Low Surface Brightness Galaxies

Philip Lah[1,2], Nikhil Arora[3,4], Ivan Yu. Katkov[1,5], Joseph D. Gelfand[1], Anna S. Saburova[5], Igor V. Chilingarian[5,6], Ivan Gerasimov[5,7], and Damir Gasymov[5,8]

[1] Center for Astrophysics and Space Science (CASS), New York University, Abu Dhabi PO Box 129188, Abu Dhabi, UAE; philiplah@gmail.com
[2] Sydney Institute for Astronomy, School of Physics A28, University of Sydney, NSW 2006, Australia
[3] Arthur B. McDonald Canadian Astroparticle Physics Research Institute, Queen's University, Kingston, ON K7L 3N6, Canada
[4] Department of Physics, Engineering Physics and Astronomy, Queen's University Kingston, ON K7L 3N6, Canada
[5] Sternberg Astronomical Institute, Lomonosov Moscow State University, Universitetskij pr., 13, Moscow, 119234, Russia
[6] Center for Astrophysics—Harvard and Smithsonian, 60 Garden Street MS09, Cambridge, MA 02138, USA
[7] Université Côte d'Azur, Observatoire de la Côte d'Azur, CNRS, Laboratoire Lagrange, 06000 Nice, France
[8] Astronomisches Rechen-Institut, Zentrum für Astronomie der Universität Heidelberg, Mönchhofstr. 12–14, 69120 Heidelberg, Germany

## Abstract

Giant low surface brightness (gLSB) galaxies are galaxies with extremely extended, faint, optical disks over 50 kpc in radius and have high total masses, which can reach $10^{12}$ $M_{\odot}$. The existence of such galaxies is problematic for current models of galaxy formation, since the major mergers responsible for the large total mass would likely have destroyed the extended optical disk. Examining the gas content of these galaxies is an important step in determining their formation mechanism, whether it be through slow gas accretion or the large disk (re) forming after a major merger. We present neutral atomic hydrogen (H I) observations of 19 gLSB galaxies identified with the Hyper Suprime-Cam Subaru Strategic Program survey. Although most have high H I masses, they are generally lower than expected based on their large optical sizes, and we do identify some gLSB galaxies with unusually low gas content. The H I spectra of these galaxies show evidence for a rotational disk, though these disks are more asymmetric than those of other galaxies with comparable mass. Four galaxies with surface brightness profiles similar to those of the gLSB galaxies have also been selected from the Numerical Investigation of a Hundred Astrophysical Objects (NIHAO) simulation for comparison. There is evidence for significant galaxy mergers in the past for three of these NIHAO galaxies, and these three galaxies show similar asymmetry in their H I spectra. Together, these results could indicate that the large optical disk of a gLSB galaxy is the result of a recent merger.

## 1. Introduction

Giant low surface brightness (gLSB) galaxies are systems that present extended, optically faint disks over 50 kpc in size, as well as having a high total mass that can reach $10^{12}$ $M_{\odot}$ (Saburova et al. 2021). The prototype of this class of galaxy is Malin 1 discovered by G. D. Bothun et al. (1987), which, besides its extended optical disk, also has a very high neutral atomic hydrogen (H I) gas mass of $M_{\rm H\,I} = 4.57 \times 10^{10}$ $M_{\odot}$ (L. D. Matthews et al. 2001). Usually, gLSB galaxies have a high central surface brightness with a low surface brightness disk component extending out to a large radius. In the standard hierarchical merging paradigm, galaxies as massive as gLSB galaxies are expected to have undergone several major mergers to build up their mass. However, such major mergers would have likely destroyed the extended disks (J. E. Barnes 1992; P. J. Quinn et al. 1993; R. Jesseit et al. 2007; V. Rodriguez-Gomez et al. 2015).

There are two major theories about the formation of gLSB galaxies. First, within the noncatastrophic scenario, extended disks are proposed to form through continuous accretion of gas and/or small satellites (J. Peñarrubia et al. 2006; A. V. Kasparova et al. 2014). Within the catastrophic scenario, the disk is (re-) formed after a major merger (M. Mapelli et al. 2008; A. S. Saburova 2018; Q. Zhu et al. 2018). Understanding which scenario, or possibly both scenarios, is responsible for such objects is a critical test of galaxy formation models. In both cases, measuring the gas properties of the galaxies is the first step to determining the formation scenario responsible. In particular, if the gas is asymmetrically distributed within the galaxy, this may favor the major merger scenario, as we will show in our work on simulated gLBS galaxies.

One of the defining elements of gLSB galaxies in the literature is their large H I mass, which underlies many H I studies of gLSB galaxies (D. Sprayberry et al. 1995; T. E. Pickering et al. 1997, 1999; L. D. Matthews et al. 2001; D. Monnier Ragaigne et al. 2003a, 2003b, 2003c; K. O'Neil et al. 2004, 2023; A. Mishra et al. 2017). One of the methods commonly used in these studies is to take a large sample of low surface brightness galaxies that could be gLSB galaxies, observe them in the radio, and then define the gLSB galaxies in the sample based on their H I mass. In this work, we start with a sample of known gLSB galaxies from their optical properties and then measure their H I properties. This should eliminate any bias in assuming that gLSB galaxies all have high H I mass. Thus, we can answer the question of whether all gLSB galaxies have massive H I disks.

Simulations have produced galaxies similar to gLSB galaxies. Q. Zhu et al. (2023) searched the IllustrisTNG100 cosmological simulation for gLSB galaxies, but they used a large H I radii as their defining characteristic for being a gLSB galaxy. They did not consider galaxies with large optical disks without corresponding large H I disks. A. S. Saburova et al. (2023) found 44 gLBS galaxies in the EAGLE reference simulation using their optical selection criteria (see below).

Our sample of gLSB galaxies comes from a survey carried out with Hyper Suprime-Cam (HSC) Subaru Telescope observations of 120 $\mathrm{deg}^2$ of sky (A. S. Saburova et al. 2023). The optical selection criteria for being a gLSB galaxy were that the galaxy has a $g$-band 27.7 mag $\mathrm{arcsec}^{-2}$ isophotal radius $\geqslant$50 kpc or has four disk scale lengths of 4, $h \geqslant 50$ kpc. The scale length is defined at the radius where the surface brightness is a factor of $e$ ($\sim$2.7) fainter than its center. The criterion also included the surface brightness cut for the deprojected $g$-band central surface brightness of the disk: $\mu_{0,g} > 22.7$ mag $\mathrm{arcsec}^{-2}$. This survey found 37 gLSB galaxies in the observations, suggesting a number density of $4.04 \times 10^{-5}$ $\mathrm{Mpc}^{-3}$. In this paper, we report the results of Green Bank Telescope (GBT) H I observations of 15 galaxies from this sample. These findings, in addition to H I measurements of four other galaxies in this sample in the literature, allow us to test whether the high H I mass of GBT optical gLSB galaxies is a defining characteristic. Additionally, we compare the gLSB galaxies of A. S. Saburova et al. (2023) with similar galaxies from the simulation Numerical Investigation of a Hundred Astrophysical Objects (NIHAO; L. Wang et al. 2015; M. Blank et al. 2019).

In Section 2, we discuss the details of the observations. In Section 3, two examples of observed gLSB galaxies are presented along with the results of all the observations displayed in tables. In Section 3.1, the optical luminosity H I mass relation is investigated. In Section 3.2, the defining characteristic of gLSB galaxies, their optical size, is examined. In Section 3.3, the properties of the shape of the gLSB's H I spectra are explored. In Section 4, we find galaxies similar to gLSB galaxies in the NIHAO simulations and then examine their H I properties, comparing them to what was found for the observed gLSB galaxies. In Section 5, we present our conclusions.

In this work, we used a Hubble constant of $H_0 = 69.6$ km $\mathrm{s}^{-1}$ $\mathrm{Mpc}^{-1}$, $\Omega_M = 0.286$, and $\Omega_\lambda = 0.714$ (C. L. Bennett et al. 2014).

## 2. Observations and Data Analysis

We selected 15 galaxies for our observations from the 37 gLSB galaxies from the optically selected sample by A. S. Saburova et al. (2023). These were chosen to be close enough to measure a H I 21 cm signal within a reasonable observing time with the GBT ($vr < 24{,}000$ km $\mathrm{s}^{-1}$; $z < 0.08$). Observations were performed in the $L$ band using the Versatile GBT Astronomical Spectrometer backend tuned to the known frequency of the H I 21 cm emission (1420.405 MHz) based on the galaxies' known optical redshifts. At the frequencies of the observation, the GBT beam has a full width at half-maximum (FWHM) of $\sim$9$'$.3 (the angular size of the observed galaxies is at most 30$''$). A bandwidth of 23.44 MHz with 4096 channels was used. Position switching was used with scans of 5 minutes on a source and then 5 minutes off the source, pointed at the empty sky. Data were collected every 10 s. A pilot program of four galaxies was observed in 2024 February. An additional 11 galaxies were observed in 2025 February. Observation lengths varied from 30 to 140 minutes depending on the redshift of the galaxy. See Section 3 for the length of each observation for a galaxy. Data were reduced with GBTIDL,[9] the custom data reduction package at the GBT that interfaces with the Interactive Data Language. The observations were Gaussian smoothed over five channels and then decimated. This gave the final spectra a velocity step of $\sim$7 km $\mathrm{s}^{-1}$. A polynomial was fitted to the spectra away from the H I signal to flatten out the bandpass. This was usually a fifth-order polynomial except for a few cases with more structure in the bandpass where a 10th-order polynomial was necessary. This polynomial fit was subtracted from the spectrum.

There are four galaxies in the A. S. Saburova et al. (2023) sample of gLSB galaxies that had preexisting H I observations (see Appendix A for references for each galaxy). These galaxies have been included in our analysis.

The emission from the H I gas was measured, and the H I mass of the galaxies was determined using the equation below:

$$M_{\mathrm{HI}} = \frac{2.36 \times 10^5}{1 + z}\left(\frac{F_{\mathrm{HI}}}{\mathrm{Jy\ km\ s^{-1}}}\right)\left(\frac{d_L^2}{\mathrm{Mpc}^2}\right), \quad (1)$$

where $z$ is the redshift, $d_L$ is the luminosity distance, and $F_{\mathrm{H\,I}}$ is the summed H I flux with $M_{\mathrm{H\,I}}$ measured in solar masses (M. H. Wieringa et al. 1992).

The rms value for the H I spectrum is determined in the region away from where the H I emission signal is present. The H I flux error is determined by

$$(F_{\mathrm{HI}}\mathrm{err})^2 = \Sigma(\mathrm{RMS}^2 \times \mathrm{velstep}^2), \quad (2)$$

where you are summing over the number of measurements within the region that contains the H I flux, and velstep is the velocity step between each of the measurements.

$W_{50}$ was measured from the H I spectra, where $W_{50}$ is defined as the velocity width measured at 50% of the peak flux of the spectrum.

To estimate the dynamical mass, we used the approach defined by N. Yu et al. (2020). The first step was to measure $v_{85}$, the velocity width that captures 85% of the total H I flux. This needed to be corrected for the inclination of the galaxy, which was determined from the ratio of the optical axes $(b/a)$ as measured by A. S. Saburova et al. (2023). The inclination was determined by the formula below:

$$i = \mathrm{acos}\sqrt{\frac{(b/a)^2 - q_0^2}{1 - q_0^2}}. \quad (3)$$

$q_0$, the axial ratio, was set as 0.13, which is a reasonable number for spiral galaxies (M. Hall et al. 2012). The velocity corrected for inclination follows the formula below:

$$v_{85c} = \frac{(v_{85} - v_{\mathrm{inst}})/(1 + \mathrm{z}) - v_{\mathrm{turb}}}{2\sin(\mathrm{i})}, \quad (4)$$

where $v_{\mathrm{inst}}$ is the instrumental resolution and $v_{\mathrm{turb}}$ is the broadening due to turbulence. Calculating these quantities using the method outlined by N. Yu et al. (2020), we determined that they were negligible for the $v_{85}$ of our sample.

[9] https://gbtidl.nrao.edu/

To convert to the rotational velocity, the formula from N. Yu et al. (2020) below was used:

$$v_{\rm rot} = (0.94 \pm 0.02) v_{85c} + (13.33 \pm 3.31). \quad (5)$$

To determine the dynamical mass, one needs the radius of the H I gas, $R_{\rm H\,I}$. The definition of $R_{\rm H\,I}$ is the radius of the H I disk defined at a surface density ($\Sigma_{\rm H\,I}$) of 1 $M_\odot$ pc$^{-2}$. There is a very tight relationship for isolated galaxies between $M_{\rm H\,I}$ and $R_{\rm H\,I}$ from J. Wang et al. (2016) of the form

$$\log\left(\frac{R_{\rm HI}}{\rm kpc}\right) = (0.51 \pm 0.00) \log\left(\frac{M_{\rm HI}}{\rm M_\odot}\right) - (3.59 \pm 0.01). \quad (6)$$

For our sample of gLSB galaxies, this relationship may not hold, but there are a few gLSB galaxies for which the $R_{\rm H\,I}$ has been measured directly. Malin 1 has an $R_{\rm H\,I}$~ 2 times bigger than what this relationship suggests (J. Wang et al. 2016). Therefore, the dynamical masses we measure will be a lower limit and could be as much as 2 times higher. With this caveat, we proceed to estimate the dynamical masses of the galaxies using the following equation from N. Yu et al. (2020):

$$M_{\rm dyn} = 2.3 \times 10^5 M_\odot \left(\frac{v_{\rm rot}}{\rm km\ s^{-1}}\right)^2 \left(\frac{R_{\rm HI}}{\rm kpc}\right). \quad (7)$$

From the H I spectra for each gLSB galaxy, we measure the flux asymmetry parameter as defined by M. P. Haynes et al. (1998). This is the ratio of the flux in the H I spectra on either side of the central velocity.

$$A_F = \begin{cases} F_b/F_r, & \text{if } F_b \geqslant F_r \\ F_r/F_b, & \text{if } F_b < F_r \end{cases}, \quad (8)$$

where $F_b$ and $F_r$ are the integrated fluxes of the blue and red sides of the profile, respectively. The error in $F_b$ and $F_r$ is determined using Equation (2), except that the region summed over is half the full H I emission region. To get the error in $A_F$, the following equation was used:

$$\left(\frac{A_F\ {\rm err}}{A_F}\right)^2 = \left(\frac{F_b\ {\rm err}}{F_b}\right)^2 + \left(\frac{F_b\ {\rm err}}{F_r}\right)^2. \quad (9)$$

Following N. Yu et al. (2020), we also measure the degree of concentration of the gLSB galaxy line profiles using the formula

$$C_v = V_{85}/V_{25}, \quad (10)$$

where $V_{25}$ and $V_{85}$ are the line widths enclosing 25% and 85% of the total flux, respectively. Double-horned profiles have $C_v < 3.1$, Gaussian profiles have $C_v > 3.7$, and boxcar profiles have $C_v$ between these two limits. Additionally, we estimate another nonparametric measure of profile shape K, as described by N. Yu et al. (2022). To measure this parameter, the curve of growth of the flux in the line width is normalized at $V_{85}$, and the integrated flux is normalized at 85% of the total flux. The profile shape K is defined as the integrated area between this normalized curve of growth and the diagonal line of unity. The value of K is positive if the area is above the line of unity, negative if the area is below the line of unity. A double-horned profile has a K < 0.02, a perfectly flat-topped profile has K ~ 0, and a single-peaked line has K > 0.02.

## 3. Results

The 15 galaxies observed, along with four other galaxies with spectra available in the literature, have their properties displayed in the tables in this section. Two example galaxies are presented here, and comments on the rest are presented in Appendix A along with their optical images and H I spectra.

For the example gLSB galaxies presented here, the H I spectrum for each galaxy is displayed next to the optical image. The ellipse around the galaxy in the optical image shows the extent of the four exponential disk scale lengths (4 hr). The central velocity of each H I spectrum is the optical redshift, which often is not exactly the same as the H I redshift. The error on the optical redshift is displayed as dotted lines on either side of the main optical redshift line. Green dotted lines show the H I redshift.

Figure 1 shows the properties of galaxy MCG-01-06-068, which is a fairly normal gLSB galaxy with a double-horned H I profile and high H I mass. Figure 2 shows LEDA 1208506, which is an abnormal gLSB galaxy with no evidence for H I emission.

In Table 1 we show the GBT observing time, optical redshift, and difference between the optical and H I redshift for the gLSB galaxies.

WISEA J020019.56-012210.9 is an unusual galaxy. It was observed for 60 minutes with the GBT, but there was contamination by companion galaxies in the H I spectrum. Very Large Array (VLA) observations of 11 hr were used to quantify the H I emission from this galaxy, and the results in the tables are from those observations. In Table 2, the various properties derived from the H I spectra are displayed, including the H I mass. Table 3 displays the H I shape parameters and the environment of the galaxy as taken from A. S. Saburova et al. (2023).

LEDA 1208506 had a nondetection of H I emission. A 2$\sigma$ upper limit for this galaxy was determined by assuming the signal was contained in a 200 km s$^{-1}$ velocity width (this is a reasonable default width for a galaxy as used by K. L. Masters et al. 2019). What signal there was within this window was summed, and then 2 times the noise value across this region was added to give a 2$\sigma$ upper limit of the H I flux. This was then converted to H I mass using the known distance to the galaxy from its optical redshift.

The H I mass distribution of the observed distribution is highly biased toward high mass galaxies. 13 of the 19 observed gLSB galaxies have H I masses above $10^{10}$ $M_\odot$. However, there are a fair number of lower mass galaxies; five of the gLSB galaxies have H I masses below $10^{10}$ $M_\odot$. There is also one galaxy already mentioned that has a nondetection of H I emission, indicating that it has a very low H I mass, a 2$\sigma$ upper limit of 0.075 × $10^{10}$ $M_\odot$. Thus, from our optically selected sample, it is clear that while most gLSB galaxies have high H I masses, not all do, and they can even have very little H I gas at all.

### 3.1. Optical Luminosity and H I Mass

There is a correlation between optical luminosity and H I mass as derived by A. Durbala et al. (2020) using Sloan Digital Sky Survey (SDSS) photometry and H I masses from

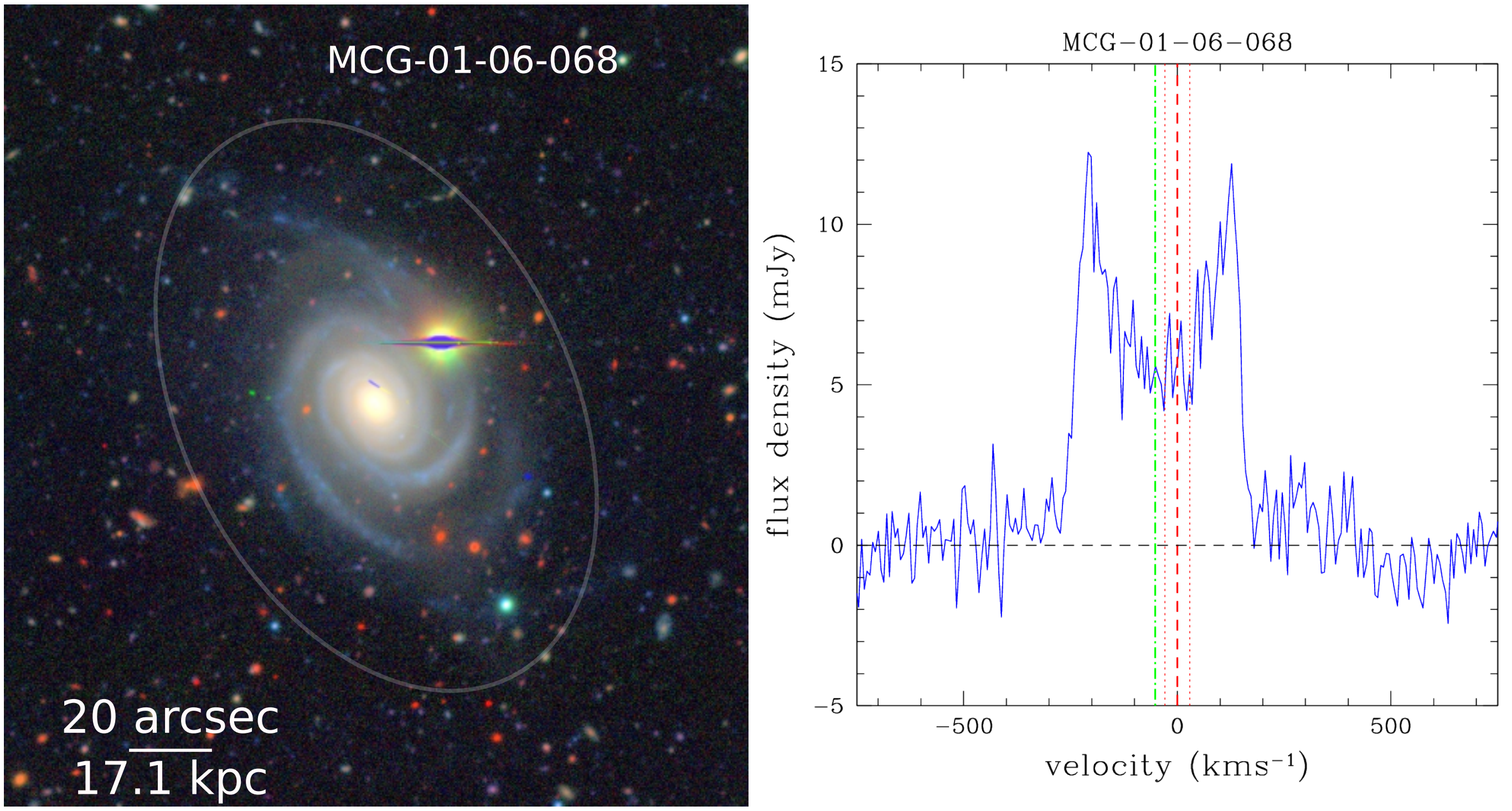

**Figure 1.** This figure displays the properties of galaxy MCG-01-06-068. On the left is a color composite of the HSC images. On the right is the GBT H I spectrum centered at the optical redshift. The galaxy has H I mass of $2.49 \times 10^{10}\, M_\odot$ and a $g$-band absolute magnitude of −21.67. MCG-01-06-068 is the classical double-horned profile, indicating a rotating disk of H I gas. The only thing to note here is that the H I mass of this gLSB galaxy is indeed high above $10^{10}\, M_\odot$.

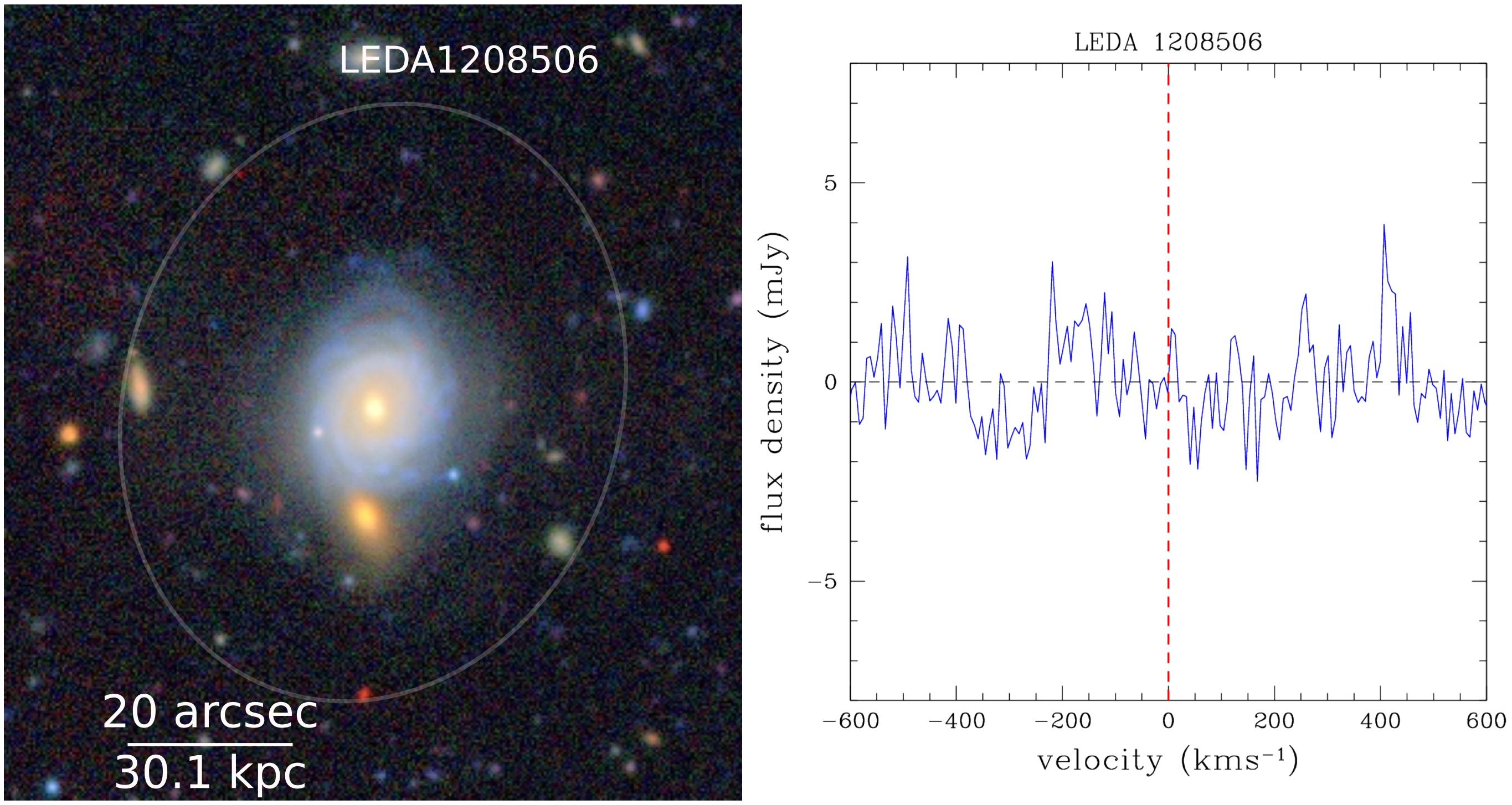

**Figure 2.** This figure displays the properties of galaxy LEDA 1208506. On the left is a color composite of the HSC images. On the right is the GBT H I spectrum centered at the optical redshift. The galaxy has a $g$-band absolute magnitude of −21.74. LEDA 1208506 is a very unusual gLSB galaxy as there does not appear to be any H I emission in its spectrum. The H I mass $2\sigma$ upper limit is $7.5 \times 10^{8}\, M_\odot$, assuming a velocity width of 200 km s$^{-1}$. Unlike WISEA J020019.56-012210.9, the other unusual gLSB galaxy, LEDA 1208506, is fairly isolated. There is another galaxy at a similar redshift, but it is ∼500 kpc away. The optical image shows blue starlight from the disk, indicating the presence of young, high-mass stars. This indicates that it likely had H I gas recently and evolved to this low gas state.

the Arecibo Legacy Fast Arecibo $L$-band Feed Array (ALFALFA) survey. This relationship with A. Durbala et al. (2020) is shown in Figure 3. It should be noted that this relationship only applies to galaxies detected in ALFALFA; there are early-type galaxies with large optical luminosities that have little to no H I gas. Most of the gLSB galaxies lie

**Table 1**
gLSB Aalaxies with Their Relevant Observation Details

| Galaxy | GBT On-off Observing Length | Optical Redshift (km $s^{-1}$) | H I Redshift Offset (km $s^{-1}$) |
|---|---|---|---|
| MCG-01-06-068 | 60 minutes | 12,968 ± 29[a] | −52 |
| 2MASXJ02274954-0526053 | 80 minutes | 13,839 ± 45[b] | −10 |
| UGC 2061 | 60 minutes | 14,764 ± 4[c] | −25 |
| WISEA J020019.56-012210.9 | VLA obs | 11,865 ± 53[d] | −50 |
| 2MASXJ02024375-0633453 | 60 minutes | 16,935 ± 45[c] | 40 |
| 2MASXJ02171125-0421194 | 60 minutes | 16,827 ± 45[c] | 0 |
| 2MASXJ02240162-0234269 | 60 minutes | 17,401 ± 45[c] | 0 |
| 2MASXJ02242997-0436135 | 120 minutes | 20,732 ± 45[c] | −45 |
| 2MASXJ02285454-0223466 | 80 minutes | 16,160 ± 45[c] | 55 |
| 2MASXJ02294973-0228590 | 60 minutes | 16,002 ± 45[c] | −110 |
| 2MASXJ02305122+0047190 | 130 minutes | 21,183 ± 3[e] | 35 |
| LEDA 1044960 | 30 minutes | 13,885 ± 4[e] | −40 |
| LEDA 1144613 | 120 minutes | 19,623 ± 3[e] | 0 |
| LEDA 1208506 (see Note) | 140 minutes | 23,534 ± < 323[f] | … |
| UGC 1876 | 60 minutes | 12,376 ± 45[b] | −30 |
| UGC 1382 | Literature | 5591 ± 2[g] | 0 |
| 2MASXJ02015377+0131087 | Literature | 12,390 ± 2[h] | 5 |
| UGC 1697 | Literature | 11,230 ± 45[b] | 37.5 |
| Mrk 592 | Literature | 7730 ± 35[i] | 30 |

**Note.** For LEDA 1208506, there was no detection of H I emission, so there is H I redshift offset.
**References.**
[a] J. P. Huchra et al. (2012).
[b] D. H. Jones et al. (2009).
[c] M. Sako et al. (2018).
[d] A. I. Zabludoff et al. (1993).
[e] F. D. Albareti et al. (2017).
[f] M. Yang et al. (2018).
[g] L. M. Z. Hagen et al. (2016).
[h] Our work (see the text).
[i] K. Abazajian et al. (2004).

(This table is available in its entirety in machine-readable form in the online article.)

**Table 2**
Relevant Details Derived from the HI Spectrum of the Observed gLSB Galaxies

| Galaxy | $M_{\rm H\,I}$ ($10^{10}\ M_{\odot}$) | $w_{50}$ (km $s^{-1}$) | incl (deg) | $v_{\rm rot}$ (km $s^{-1}$) | $R_{\rm H\,I}$ (kpc) | $M_{\rm dyn}$ ($10^{11}\ M_{\odot}$) |
|---|---|---|---|---|---|---|
| MCG-01-06-068 | 2.485 ± 0.046 | 388.3 ± 4.6 | 34.2 | 154.9 ± 5.0 | 51.48 ± 0.49 | 2.85 ± 0.13 |
| 2MASXJ02274954-0526053 | 1.134 ± 0.039 | 194.3 ± 4.1 | 45.3 | 75.7 ± 4.9 | 34.5 ± 0.6 | 0.456 ± 0.042 |
| UGC 2061 | 1.848 ± 0.062 | 533.9 ± 6.9 | 75.5 | 125.7 ± 5.3 | 44.26 ± 0.76 | 1.61 ± 0.10 |
| WISEA J020019.56-012210.9 | 0.279 ± 0.027 | 397 ± 90 | 36.3 | 188 ± 19 | 17.3 ± 0.4 | 1.40 ± 0.21 |
| 2MASXJ02024375-0633453 | 2.52 ± 0.07 | 228 ± 11 | 29.8 | 109.3 ± 5.8 | 109.3 ± 5.8 | 1.43 ± 0.11 |
| 2MASXJ02171125-0421194 | 0.442 ± 0.068 | 237 ± 49 | 45.3 | 77 ± 12 | 21.3 ± 1.7 | 0.292 ± 0.070 |
| 2MASXJ02240162-0234269 | 2.70 ± 0.20 | 230 ± 98 | 36.3 | 114 ± 55 | 53.7 ± 2.0 | 1.6 ± 1.1 |
| 2MASXJ02242997-0436135 | 3.03 ± 0.11 | 379 ± 14 | 37.2 | 131.9 ± 5.2 | 56.9 ± 1.1 | 2.29 ± 0.14 |
| 2MASXJ02285454-0223466 | 0.774 ± 0.040 | 95.7 ± 4.9 | 16.4 | 86.4 ± 5.5 | 28.40 ± 0.75 | 0.489 ± 0.046 |
| 2MASXJ02294973-0228590 | 0.643 ± 0.079 | 306 ± 120 | 57.4 | 89 ± 15 | 25.8 ± 1.6 | 0.47 ± 0.12 |
| 2MASXJ02305122+0047190 | 0.86 ± 0.088 | 199 ± 76 | 11.6 | 246 ± 35 | 30.0 ± 1.6 | 4.21 ± 0.88 |
| LEDA 1044960 | 2.27 ± 0.14 | 402 ± 37 | 57.4 | 116.7 ± 8.3 | 49.1 ± 1.5 | 1.54 ± 0.16 |
| LEDA 1144613 | 2.620 ± 0.077 | 357.6 ± 6.2 | 71.5 | 87.4 ± 4.3 | 52.89 ± 0.79 | 0.933 ± 0.067 |
| LEDA 1208506 | 0.075* | … | … | … | … | … |
| UGC 1876 | 1.540 ± 0.042 | 415.4 ± 2.6 | 58.8 | 111.8 ± 4.7 | 40.33 ± 0.56 | 1.165 ± 0.072 |
| UGC 1382 | 1.584 ± 0.026 | 435.2 ± 7.9 | 40.1 | 155.0 ± 4.8 | 40.92 ± 0.34 | 2.27 ± 0.10 |
| 2MASXJ02015377+0131087 | 2.06 ± 0.12 | 342.6 ± 34 | 40.1 | 138 ± 13 | 47.26 ± 1.4 | 1.93 ± 0.22 |
| UGC 1697 | 2.15 ± 0.10 | 771.8 ± 3.7 | 66.2 | 178 ± 10 | 47.9 ± 1.1 | 3.51 ± 0.29 |
| Mrk 592 | 1.005 ± 0.048 | 197 ± 10 | 60.9 | 59.1 ± 4.0 | 32.44 ± 0.79 | 0.262 ± 0.026 |

**Note.**
(This table is available in its entirety in machine-readable form in the online article.)

* The measurement here for LEDA 1208506 is a $2\sigma$ upper limit.

**Table 3**
Shape Parameters from the HI Spectra of the Observed gLSB Galaxies as well as the Enviroment of the Galaxies

| Galaxy | $A_F$ | $C_v$ | K | Environment |
|---|---|---|---|---|
| MCG-01-06-068 | $1.015 \pm 0.038$ | $2.745 \pm 0.12$ | $-0.06482 \pm 0.0099$ | Isolated within 500 kpc |
| 2MASXJ02274954-0526053 | $1.45 \pm 0.10$ | $2.82 \pm 0.29$ | $-0.05681 \pm 0.018$ | Isolated |
| UGC 2061 | $1.116 \pm 0.075$ | $3.832 \pm 0.4$ | $-0.002489 \pm 0.019$ | Companion at 300 kpc |
| WISEA J020019.56-012210.9 | $1.033 \pm 0.091$ | $4.38 \pm 0.57$ | $0.0210 \pm 0.028$ | Companion at 150 kpc |
| 2MASXJ02024375-0633453 | $1.349 \pm 0.078$ | $2.64 \pm 0.15$ | $-0.0383 \pm 0.016$ | Isolated |
| 2MASXJ02171125-0421194 | $1.03 \pm 0.31$ | $4.6 \pm 2.5$ | $0.0105 \pm 0.090$ | Isolated |
| 2MASXJ02240162-0234269 | $1.42 \pm 0.22$ | $4.8 \pm 2.4$ | $0.0609 \pm 0.047$ | Isolated within 500 kpc |
| 2MASXJ02242997-0436135 | $1.076 \pm 0.078$ | $3.02 \pm 0.21$ | $-0.0522 \pm 0.019$ | Isolated |
| 2MASXJ02285454-0223466 | $1.21 \pm 0.12$ | $3.41 \pm 0.18$ | $-0.01296 \pm 0.027$ | Isolated |
| 2MASXJ02294973-0228590 | $1.17 \pm 0.29$ | $2.54 \pm 0.79$ | $-0.0600 \pm 0.068$ | Isolated within 500 kpc |
| 2MASXJ02305122+0047190 | $1.25 \pm 0.26$ | $3.30 \pm 0.81$ | $-0.0109 \pm 0.059$ | Isolated within 500 kpc |
| LEDA 1044960 | $1.08 \pm 0.13$ | $3.07 \pm 0.47$ | $-0.039 \pm 0.033$ | Isolated |
| LEDA 1144613 | $1.456 \pm 0.089$ | $3.05 \pm 0.24$ | $-0.036 \pm 0.016$ | Isolated within 500 kpc |
| LEDA 1208506 | … | … | … | Isolated |
| UGC 1876 | $1.85 \pm 0.11$ | $2.52 \pm 0.12$ | $-0.0739 \pm 0.014$ | Isolated |
| UGC 1382 | $1.381 \pm 0.047$ | $2.649 \pm 0.083$ | $-0.07161 \pm 0.0086$ | Central galaxy in a group |
| 2MASXJ02015377+0131087 | $1.395 \pm 0.18$ | $2.532 \pm 0.42$ | $-0.06124 \pm 0.032$ | Isolated |
| UGC 1697 | $1.375 \pm 0.13$ | $3.684 \pm 0.49$ | $0.04286 \pm 0.029$ | Isolated within 500 kpc |
| Mrk 592 | $1.134 \pm 0.11$ | $3.415 \pm 0.26$ | $-0.03604 \pm 0.023$ | Isolated |

(This table is available in its entirety in machine-readable form in the online article.)

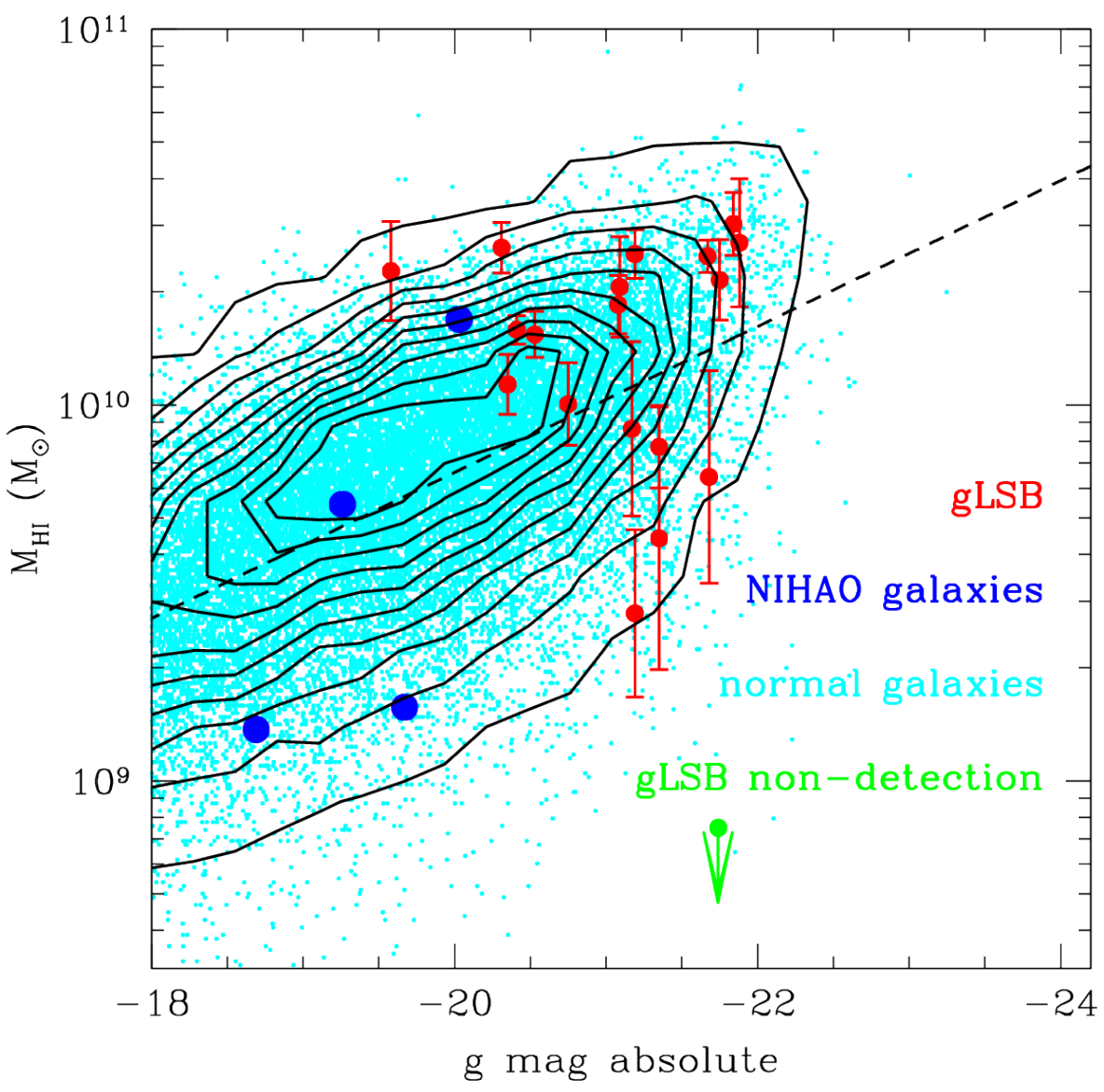

**Figure 3.** The correlation between absolute $g$ magnitude and H I mass derived by A. Durbala et al. (2020) using SDSS photometry and ALFALFA H I masses. The cyan points are their data. The black line is the linear least-squares fit to their data that we made and is shown as Equation (A1) in the text. This is what has been used to estimate the H I mass of galaxies where necessary. The red points are the gLSB galaxies in this work. The green upper limit at the $g$-band absolute magnitude of $-21.74^m$ is the value for LEDA 1208506 and is the $2\sigma$ upper limit for this galaxy. The blue points are the four simulated galaxies from NIHAO used as comparisons to the gLSB galaxies.

above the line fit for average galaxies. Those few galaxies that lie below this line are brighter than the average.

### 3.2. Scale Length and H I Mass

To determine the defining characteristic of gLSB galaxies, we compared the scale length of a sample of normal galaxies with their H I masses with these properties from our sample of gLSB galaxies. For the regular galaxies, we used the scale lengths from measurements made by K. Fathi et al. (2010) from SDSS imaging. While K. Fathi et al. (2010) measured the scale length in each of the five filter bands of SDSS, they only published the $r$-band results. However, they do publish a conversion function between the bands, enabling us to obtain an estimate of the $g$-band scale length. The SDSS imaging used by K. Fathi et al. (2010) is not that deep, so there is the possibility that they are underestimating the scale lengths for some galaxies. The H I data comes from A. Durbala et al. (2020), who matched ALFALFA galaxies with their SDSS counterparts. These two datasets were matched to obtain a dataset with scale lengths and H I masses. Unfortunately, due to the limited overlap in sky positions and redshifts, there are only 2089 galaxies in the combined dataset. These are the small blue points in Figure 4. Four times the scale length has been plotted here, as this is the length that A. S. Saburova et al. (2023) used to select gLSB galaxies. The normal galaxies form a clear locus of points, and they do extend up to the higher H I masses.

The big red points in this figure are the gLSB galaxies of our sample. For UGC 2061, the H I mass for the entire spectrum is used, ignoring the fact that it has a companion. Taking this into account would shift the point for this galaxy slightly lower in H I mass.

The large black points are gLSB galaxies from the literature. The extreme black point is Malin 1, which has a scale length significantly longer than anything else in the sample. The large magenta points are literature values for gLSB galaxies with scale lengths measured in the $R$ band rather than the $g$ band. While there is no simple way to convert these values to the $g$ band, scale lengths tend to decrease to redder bands, so the actual value for these points would be slightly higher (K. Fathi et al. 2010).

The small red points come from a low surface brightness galaxy sample from W. Du et al. (2019). These galaxies are not gLSB galaxies, but just low surface brightness galaxies. These galaxies were selected to have a central surface brightness $\mu_0(B)$ fainter than 22.5 mag arcsec$^{-2}$ in SDSS. They also had to have their H I mass measured in the ALFALFA survey. This sample is heavily affected by selection effects, both on the low- and high-scale length ends.

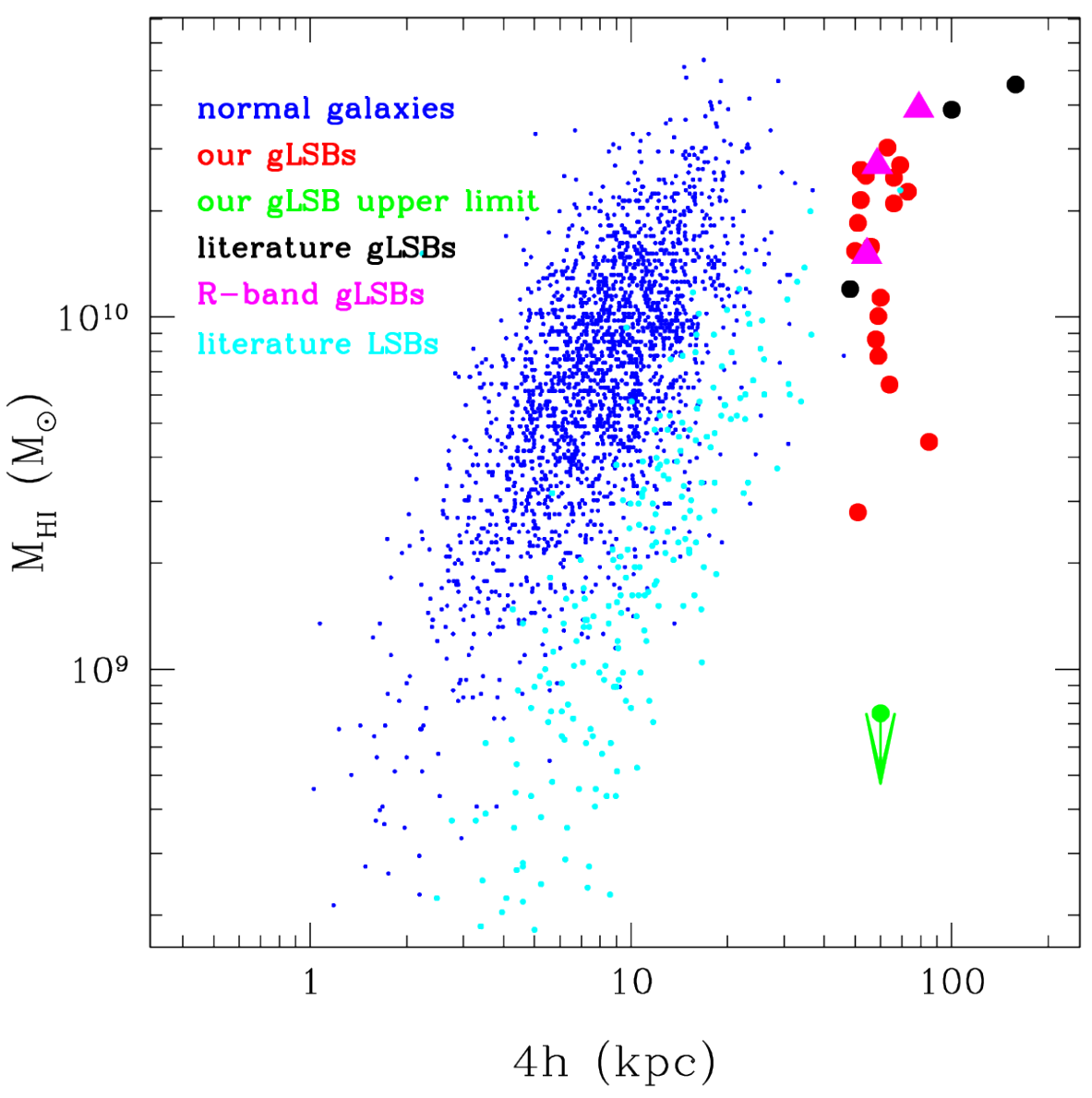

**Figure 4.** This figure shows the value of 4 times the *g*-band scale length, the length used by A. S. Saburova et al. (2023) to select a gLSB galaxy vs. the H I mass of the galaxy. The blue points are regular galaxies. The large red points are the gLSB galaxies in this work. The lower upper limit is for LEDA 1208506. The large black points are literature gLSB galaxies. The magenta triangle points are literature values with *R*-band scale lengths. The small red points are literature low surface brightness galaxies (not gLSB galaxies). See the text for more details.

What is clear from Figure 4 is that the gLSB galaxies are unusual only in their optical size; there are plenty of normal galaxies that have H I masses in a range similar to the values for the gLSB galaxies. Any formation mechanism for gLSB galaxies needs to explain their large optical size along with their high but not unusual H I mass.

### 3.3. *Shape of the H I Spectrum*

In the top left of Figure 5 is shown the H I mass of galaxies versus their symmetry parameter $A_F$, whose definition is described in Section 2. The data from N. Yu et al. (2022) show the range for normal galaxies drawn from the ALFALFA survey. The normal galaxies have a strong preference to be symmetric in their line profile, with only 17% of them having what N. Yu et al. (2022) define as asymmetric profiles. The dashed line in the top right of Figure 5 is the value beyond which N. Yu et al. (2022) define a galaxy as asymmetric. D. Espada et al. (2011) found a similar result in isolated galaxies from the Analysis of the Interstellar Medium of Isolated Galaxies project. This is in stark contrast to our gLSB galaxy sample, where nine out of the 18 galaxies with H I detections are asymmetric by this definition, 50% of the sample. There is only a one-dimensional Kolmogorov–Smirnov probability of 0.0165 that our gLSB galaxies are drawn from the same sample as the ALFALFA galaxies. A galaxy is unlikely to gain an asymmetric profile through steady accretion of gas. Steady accretion would happen slowly, and any asymmetry would work itself out over time. A major merger of two galaxies, where the H I gas in the disk come from both galaxies, could give this result. The large size of the disk of a gLSB galaxy will lead to longer dynamical times for the disk to become relaxed. This will mean that asymmetries will remain for longer than for a smaller galaxy.

N. Yu et al. (2022) provide inclinations for a subsample of the ALFALFA galaxies based on SDSS imaging. There does not appear to be any significant relationship between inclination and the symmetry parameter except at the extremes of the inclination, where there are fewer asymmetric galaxies. The distribution of the inclination of our gLSB galaxies is fairly similar to that of the ALFALFA galaxies expect for an overdensity near 40°, where there are six of our galaxies. This is well within the region where inclination and symmetry parameters have no relationship.

It is interesting to note that the gLSB galaxies with low H I masses are consistent with being symmetric. This could be due to how they formed or could be a result of evolution. If it is evolution, these galaxies may have formed with a higher H I mass and an asymmetric line profile. Over time, the H I mass decreased through such mechanisms as star formation, outflows, and ionization. At the same time, the asymmetry in the disk could have slowly been smoothed out, though whether this occurs needs to be tested in simulations. We currently do not have enough information to tell which path, formation, or evolution gave rise to these galaxies.

In the top right of Figure 5 is shown the H I mass of galaxies versus their profile shape parameter $C_v$, whose definition is described in Section 2. The data from N. Yu et al. (2022) show the range for normal galaxies from ALFALFA. The distribution they find for galaxies in their sample is 36.0% are double-horned profiles, 28.2% are flat-topped profiles, and 35.9% are single-peaked profiles. Our gLSB galaxy sample has 62.5% double-horned profiles, 25.0% flat-topped profiles, and 12.5% single-peaked profiles. There is a Kolmogorov–Smirnov probability of 0.24 that our gLSB galaxies are drawn from the same sample as the ALFALFA galaxies. This may indicate a preference for gLSB galaxies to have their H I in a rotating disk that gives rise to a double-horned profile.

In the bottom left of Figure 5 is shown the H I mass versus the profile shape parameter K, whose definition is described in Section 2. The data from N. Yu et al. (2022) show the range for normal galaxies from ALFALFA. The distribution they find for galaxies in their sample is 34.3% are double-horned profiles, 27.6% are flat-topped profiles, and 38.1% are single-peaked profiles. For our gLSB galaxies, the distribution is 61% double-horned profiles, 22% flat-topped profiles, and 17% single- peaked profiles. There is now a Kolmogorov–Smirnov probability of 0.025 that our gLSB galaxies are drawn from the same sample as the ALFALFA galaxies.

In the bottom right of Figure 5 is shown the distribution of the $C_v$ parameter versus the K parameter. These parameters measure the same thing in a different way. This figure highlights that they do not always give the same answer as to whether a galaxy has a double-horned profile or a Gaussian profile. However, there is a clear distribution for normal galaxies as shown by the extent of the N. Yu et al. (2022) data from ALFALFA. Our gLSB galaxies lie within this distribution.

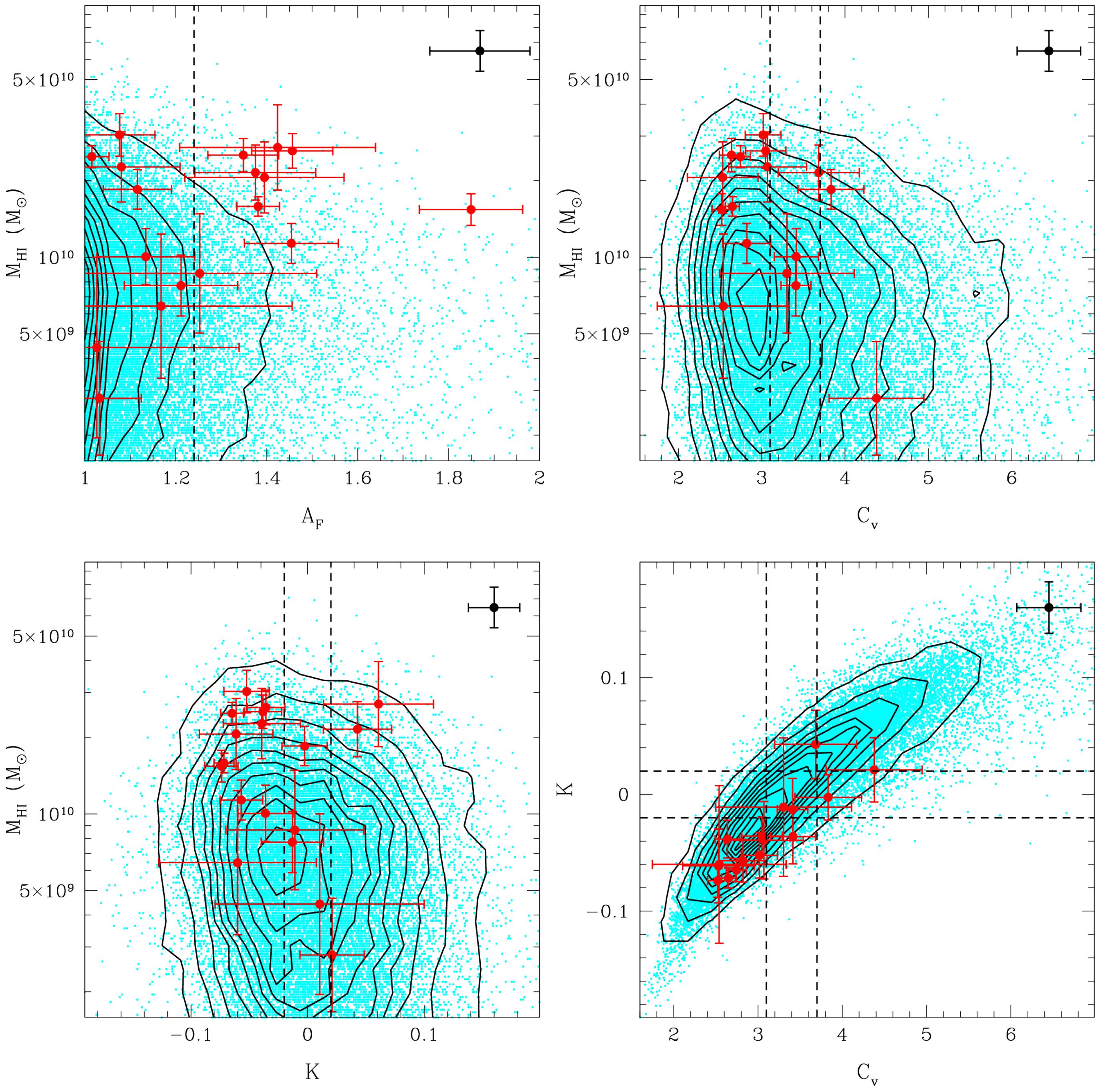

**Figure 5.** Top left: the H I mass of galaxies vs. their symmetry parameter $A_F$, whose definition is described in Section 2. The dashed line at $A_F = 1.24$ is the value beyond which N. Yu et al. (2022) define a galaxy as asymmetric. Top right: the H I mass of galaxies vs. their profile shape parameter $C_v$, whose definition is described in Section 2. There are two dashed lines at $C_v = 3.1$ and 3.7. N. Yu et al. (2022) define a spectrum below 3.1 as a double-horned profile. Between 3.1 and 3.7, the spectrum has a flat top. Above 3.7, the spectrum has a Gaussian profile. Bottom left: the H I mass of galaxies vs. their profile shape parameter K, whose definition is described in Section 2. The lines at $K = -0.02$ and 0.02 are marked on the figure. N. Yu et al. (2022) define their sample galaxies with $K < -0.02$ as a double-horned profile, $K > 0.02$ as a single-peaked profile, and between these profiles as a flat-topped profile. Bottom right: the distribution of $C_v$ parameter vs. the K parameter. The dashed lines are the selection limits as defined previously. The gLSB galaxies congregate mostly in the double-horned profile region. The cyan points are ALFALFA data from N. Yu et al. (2022) with 29,958 points ranging across all H I masses in that survey, and the black error in the top right corner is the median error bar of this sample. The red points are the 18 gLSB galaxies with H I detections from our sample. We note that for $C_v$, we exclude two galaxies with errors of more than 2.

## 4. Comparison with Simulation

gLSB galaxies are rare galaxies, so even large-volume galaxy simulations often have few examples of them. To better understand the evolution of gLSB galaxies, we use NIHAO (L. Wang et al. 2015; M. Blank et al. 2019; A. V. Macciò et al. 2020) to look for gLSB candidates. NIHAO is a set of ∼130 zoom-in cosmological simulations in a flat ΛCDM cosmology (Planck Collaboration et al. 2014) using the GASOLINE2.0

(J. W. Wadsley et al. 2017) tree smoothed particle hydrodynamics code to produce galaxies with stellar masses ranging from $10^6$ to $10^{12}\,M_\odot$ at $z = 0$. All simulations have similar resolution, containing $\sim10^6$ dark matter particles with a softening length of $\epsilon_{\rm dark} = 465$ pc and mass resolution of $2.1 \times 10^5\,M_\odot$. Each simulation also contains $\sim10^6$ gas particles, with a softening length of $\epsilon_{\rm gas} = 199$ pc and a typical particle mass of $3.1 \times 10^4\,M_\odot$. NIHAO galaxies form stars according to Kennicutt–Schmidt law (R. C. J. Kennicutt 1998) with suitable gas temperature and density thresholds, $T_{\rm SF} < 1.5 \times 10^4\,K$ and $n_{\rm SF} > 10.3\,{\rm cm}^{-3}$, and follow the Chabrier initial mass function (IMF; N. Arora et al. 2022). Massive stars with $8\,M_\odot < M_{\rm star} < 40\,M_\odot$ ionize the interstellar medium (ISM) before their supernova (SN) explosions (G. Stinson et al. 2006; L. Wang et al. 2015). This "early stellar feedback" (ESF) mode is set to inject 13% of the total stellar flux of $2 \times 10^{50}\,{\rm erg}\,M_\odot^{-1}$ into the ISM through blast-wave SN feedback that injects both energy and metals into the ISM. The stellar feedback does not have any variability with halo mass and/or redshift. NIHAO also includes subgrid models for turbulent mixing of metals and energy (B. Keller et al. 2014), UV heating, ionization, and metal cooling (S. Shen et al. 2010), and supermassive black hole growth and feedback (M. Blank et al. 2019; A. V. Macciò et al. 2020).

NIHAO simulations have proven successful at matching various observational aspects of galaxy formation and evolution such as the local galaxy velocity function (A. V. Macciò et al. 2016), high-redshift clumps in galaxy disks (T. Buck et al. 2017), the baryonic Tully–Fisher relation (A. A. Dutton et al. 2007; N. Arora et al. 2023), properties of low surface brightness galaxies (A. Di Cintio et al. 2019), the presence (or lack thereof) of diversity in galaxy rotation curves (I. M. Santos-Santos et al. 2018; M. Frosst et al. 2022), the star formation main sequence (M. Blank et al. 2021; M. Blank et al. 2022), and various structural scaling relations (N. Arora et al. 2023). In this study, we use the H I properties of these simulations. The H I fraction for each gas particle is calculated using the radiative transfer prescription from A. Rahmati et al. (2013), and H I masses are calculated using all gas particles enclosed within $4R_{\rm eff}$. Using the pynbody package (A. Pontzen et al. 2013), we calculate the face-on $g$-band surface brightness profiles for the stellar distribution of the simulated galaxies.

### 4.1. Sample Selection

To find galaxies comparable to the gLSB galaxies in the NIHAO simulation, we considered the 86 simulated NIHAO galaxies at $z = 0$ with $M_{200} > 10^{10}\,M_\odot$. We compared their simulated $g$-band surface brightness profiles with the observed light profiles for the 37 gLSB galaxies from A. S. Saburova et al. (2023; the majority of gLSB galaxies from A. S. Saburova et al. 2023 do not have H I mass measurements). We perform a least-squares fit between the groups of surface brightness profiles with a variable offset included as a parameter. The goal was to find NIHAO galaxies with light profiles of similar shape to the gLSB galaxies, even if they were not the same in terms of absolute brightness. Thus, an offset between the surface brightness was introduced that best matched the NIHAO galaxy profiles to the A. S. Saburova et al. (2023) profiles. Galaxies with a logarithmic least-squares fit less than 3.5 mag arcsec$^{-1}$ were taken as good matches, as galaxies with this matching criterion were good fits to the data. Through this method, four galaxies were determined to be similar in shape to the A. S. Saburova et al. (2023) light profiles and are shown in Figure 6. The four galaxies are labeled g1.12e12, g1.92e12, g1.77e12, and g2.79e12. g1.12e12 matched best with a gLSB profile with an offset shifting it down in magnitudes of 3.3, g1.77e12 an offset of 2.7, g1.92e12 an offset of 3.3, and g2.79e12 an offset of 2.9. g1.77e12 and g2.79e12 without an offset were within the lower range of surface brightnesses for the A. S. Saburova et al. (2023) sample (they were better fitted with the offset). g1.12e12 and g1.92e12 required the offset to reach the level of the A. S. Saburova et al. (2023) surface brightnesses (only their overall shape was similar). With these four simulated galaxies that have surface brightness profiles similar to those of gLSB galaxies (from the high surface brightness center to the low surface brightness disk), we can examine their simulated H I gas properties. All four galaxies have optical structure well beyond a radius of 50 kpc. The relevant properties of the simulated galaxies are listed in Table 4.

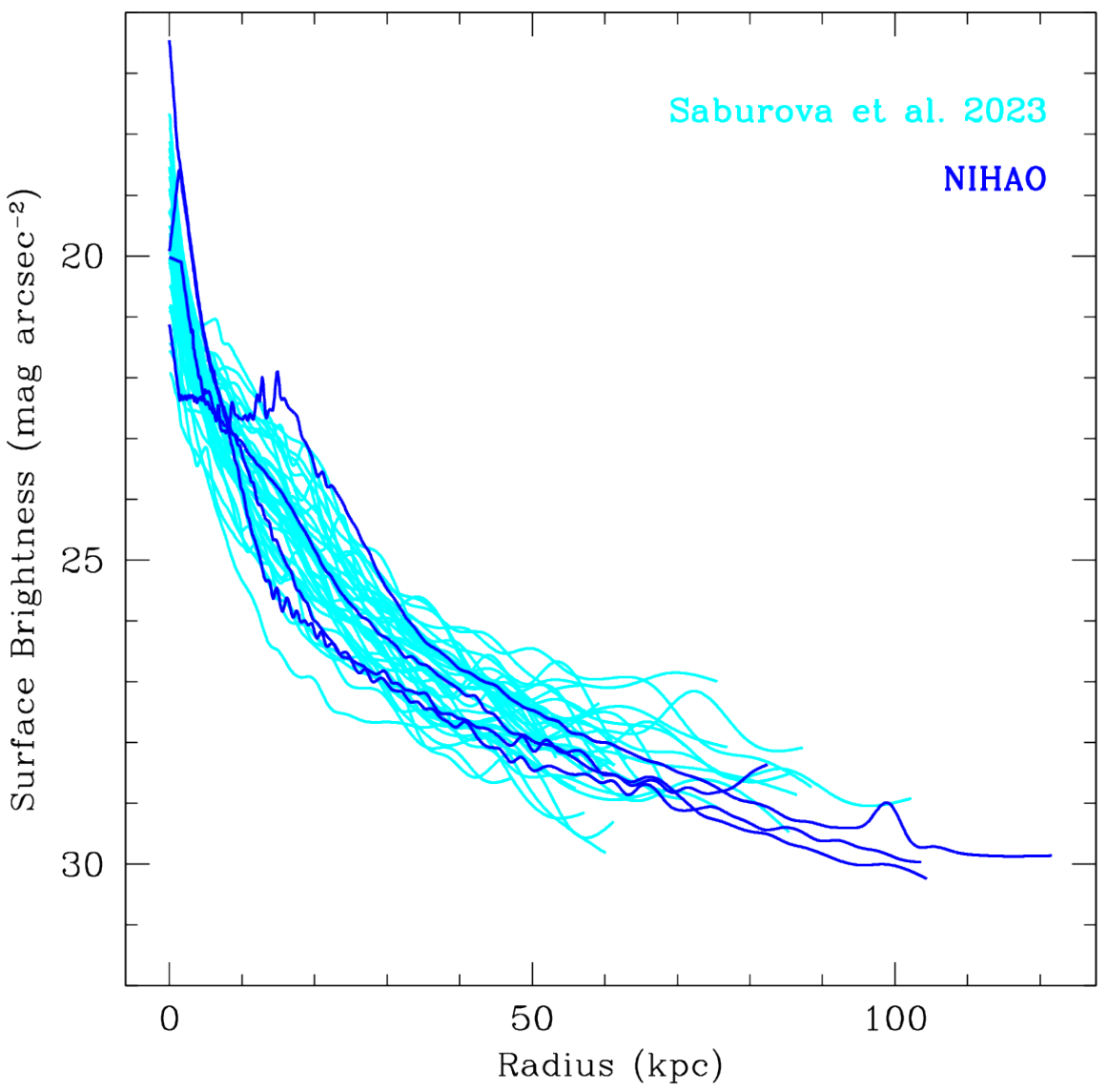

**Figure 6.** This figure displays all 37 $g$-band light profiles of gLSB galaxies from A. S. Saburova et al. (2023) in the cyan color against the four NIHAO galaxy $g$-band light profiles in the blue that best match the shape of the gLSB galaxy's profile after applying a normalization factor so that the shapes of the profiles match.

At $z = 0$, NIHAO galaxy g1.12e12 has a total H I mass of $1.37 \times 10^9\,M_\odot$ and a total mass of $9.6 \times 10^{11}\,M_\odot$. This is significantly lower in H I mass than the usual gLSB galaxies for which we have measurements, but the total mass is also larger than that of the dynamical masses found for the gLSB galaxies (it should be noted that the dynamical masses are dependent on a H I radius estimation that may be an underestimate for the observed gLSB galaxies). g1.12e12 had a gas-poor merger with an object 8.7 Gyr ago ($z = 1.3$) that was $1.1 \times 10^{11}\,M_\odot$ in total mass, which was 17% the size of the galaxy at that time. g1.12e12 had a more significant gas-rich merger 5.3 Gyr ago ($z = 0.53$) with an object with $2.5 \times 10^{11}\,M_\odot$ in total mass, which was 33% the size of the galaxy at that time. The left panel of Figure 7 shows the simulated optical image of g1.12e12 by combining $g$-, $r$-, and $i$-band images taken from data in Z. Jin et al. (2024). The galaxy has a clear bulge surrounded by an extended disk. In

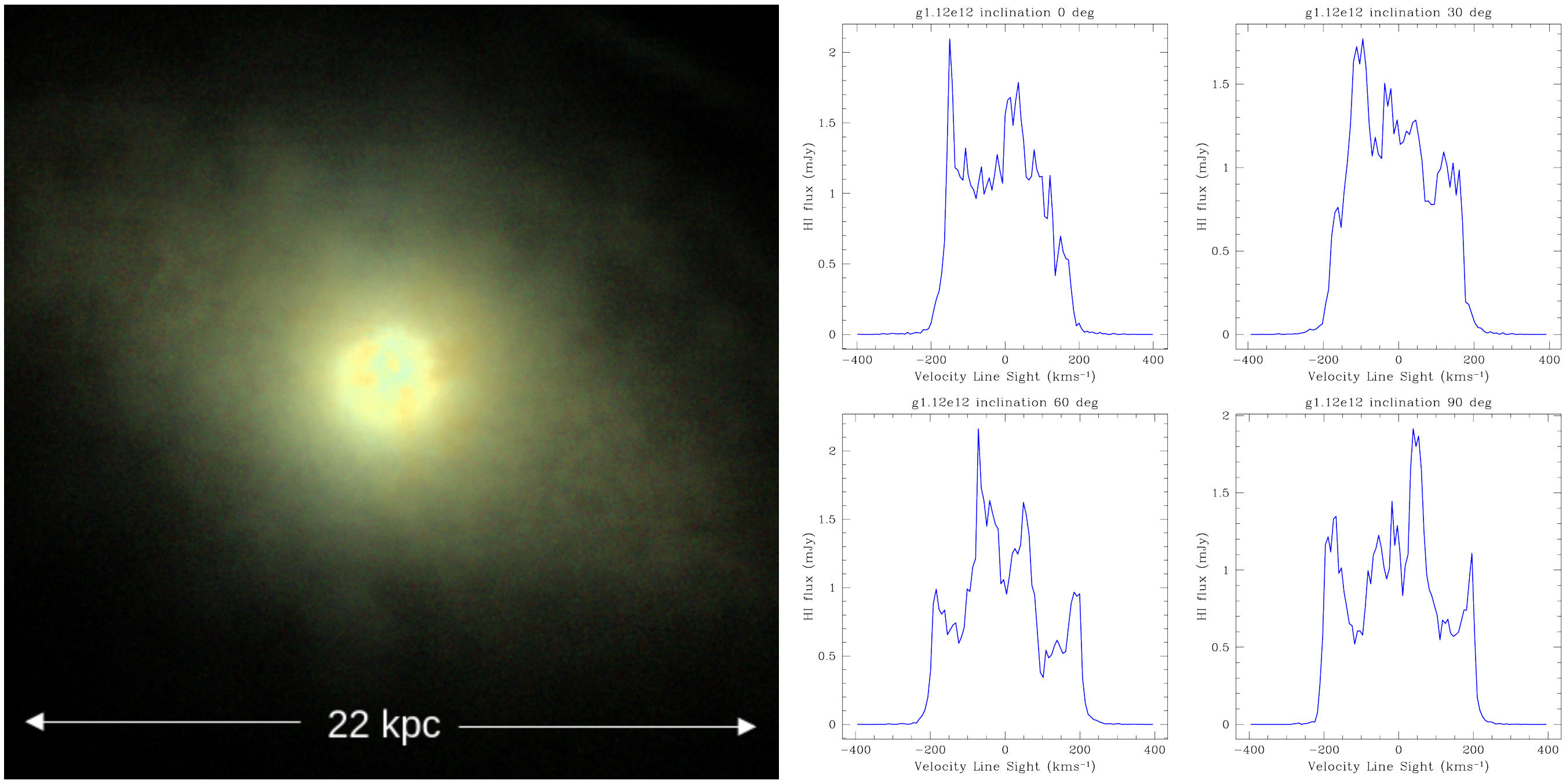

**Figure 7.** Left: the simulated optical image of NIHAO galaxy g1.12e12 based on the $g$, $r$, and $i$ bands, and is face-on. Right: the simulated H I spectrum of NIHAO galaxy g1.12e12 at a variety of inclinations and at a distance of 100 Mpc.

**Table 4**
Properties of the Simulated Galaxies That Are Similar to Those of gLSB Galaxies

| Galaxy Name | Absolute $g$ Mag | H I Mass at $z = 0$ | Total Mass at $z = 0$ | Maximum H I Mass | Redshift of Maximum H I Mass | Number of Major Mergers |
|---|---|---|---|---|---|---|
| g1.12e12 | −18.69 | $1.37 \times 10^9\,M_\odot$ | $9.60 \times 10^{11}\,M_\odot$ | $4.25 \times 10^9\,M_\odot$ | 1.66 | 2 |
| g1.77e12 | −19.67 | $1.57 \times 10^9\,M_\odot$ | $1.94 \times 10^{12}\,M_\odot$ | $1.18 \times 10^{10}\,M_\odot$ | 0.62 | 1 |
| g1.92e12 | −19.26 | $5.44 \times 10^9\,M_\odot$ | $1.90 \times 10^{12}\,M_\odot$ | $6.87 \times 10^9\,M_\odot$ | 2.50 | 0 |
| g2.79e12 | −20.03 | $1.69 \times 10^{10}\,M_\odot$ | $3.07 \times 10^{12}\,M_\odot$ | $1.69 \times 10^{10}\,M_\odot$ | 0.00 | 1 |

(This table is available in its entirety in machine-readable form in the online article.)

Figure 7, the right panel shows the simulated H I spectrum for g1.12e12 as observed at 100 Mpc at $z = 0$ at a variety of inclinations. 100 Mpc was chosen as the simulated distance to approximately match the observational H I spectra in this work. The H I spectrum shows a significant substructure and cannot simply be classified as a double-horned profile. What is interesting is that the H I velocity width of the galaxy does not change significantly with inclination, though the shape within this width does. This suggests that the H I gas does not simply reside in a rotating disk.

At $z = 0$, NIHAO galaxy g1.77e12 has a total H I mass of $1.57 \times 10^9\,M_\odot$ and a total mass of $1.94 \times 10^{12}\,M_\odot$. Again, this is less H I mass than most of the observational gLSB galaxies and has a higher total mass than the observational gLSB galaxies. g1.77e12 had a gas-rich major merger 6.7 Gyr ago ($z = 0.76$) with an object with a total mass of $1.35 \times 10^{12}\,M_\odot$, which was 159% of its mass at the time. g1.77e12 had a maximum H I mass of $1.18 \times 10^{10}\,M_\odot$ at lookback time 6.0 Gyr ago ($z = 0.62$). The left panel of Figure 8 shows the simulated optical image of g1.77e12 by combining $g$-, $r$-, and $i$-band images. The galaxy, while round, does not appear to have a central bulge and is quite different from the standard gLSB galaxy morphology (see the optical images of gLSB galaxies in the Appendix). There is a region of higher optical light in the outskirts of a galaxy. This may be related to the extra component seen in the H I spectrum for this galaxy. In Figure 8, the right panel shows the simulated H I spectrum for g1.77e12 as observed at 100 Mpc at $z = 0$ for a variety of inclinations. The galaxy is very unusual in how it appears in H I flux as one varies the inclination. At 90° inclination (edge-on), it is a simple Gaussian shape. Unexpectedly, with decreasing inclination, a second velocity component is seen and is strongest at 0° inclination (face-on). What is likely happening is that there is a component in the H I velocity left over from the major merger that is moving at a different velocity from the main H I gas in this galaxy.

At $z = 0$, NIHAO galaxy g1.92e12 has a total H I mass of $5.44 \times 10^9\,M_\odot$ and a total mass of $1.90 \times 10^{12}\,M_\odot$. Again, the H I mass is lower than most of the observed gLSB galaxies, and the total mass is higher than that of our observed gLSB galaxies. This NIHAO galaxy had no significant mergers in its past. The left panel of Figure 9 shows the simulated optical image of g1.12e12 by combining $g$-, $r$-, and $i$-band images. The galaxy has a redder bulge surrounded by a bluer disk-like structure. In Figure 9, the right panel shows the simulated H I spectrum for g1.12e12 as observed at 100 Mpc at $z = 0$.

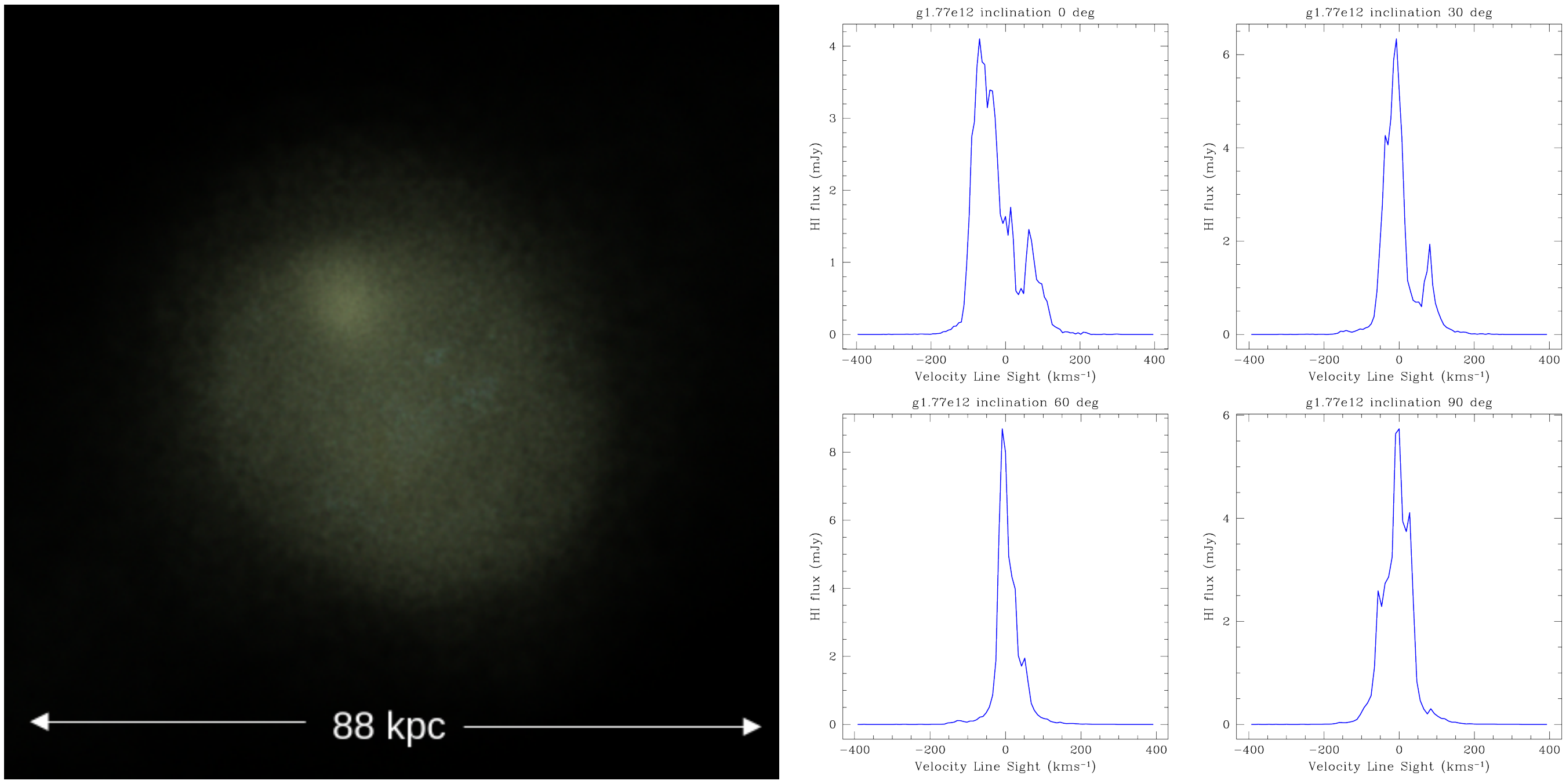

**Figure 8.** Left: the simulated optical image of NIHAO galaxy g1.77e12 based on the *g*, *r*, and *i* bands, and is face-on. Right: the simulated H I spectrum of NIHAO galaxy g1.77e12 at a variety of inclinations and at a distance of 100 Mpc.

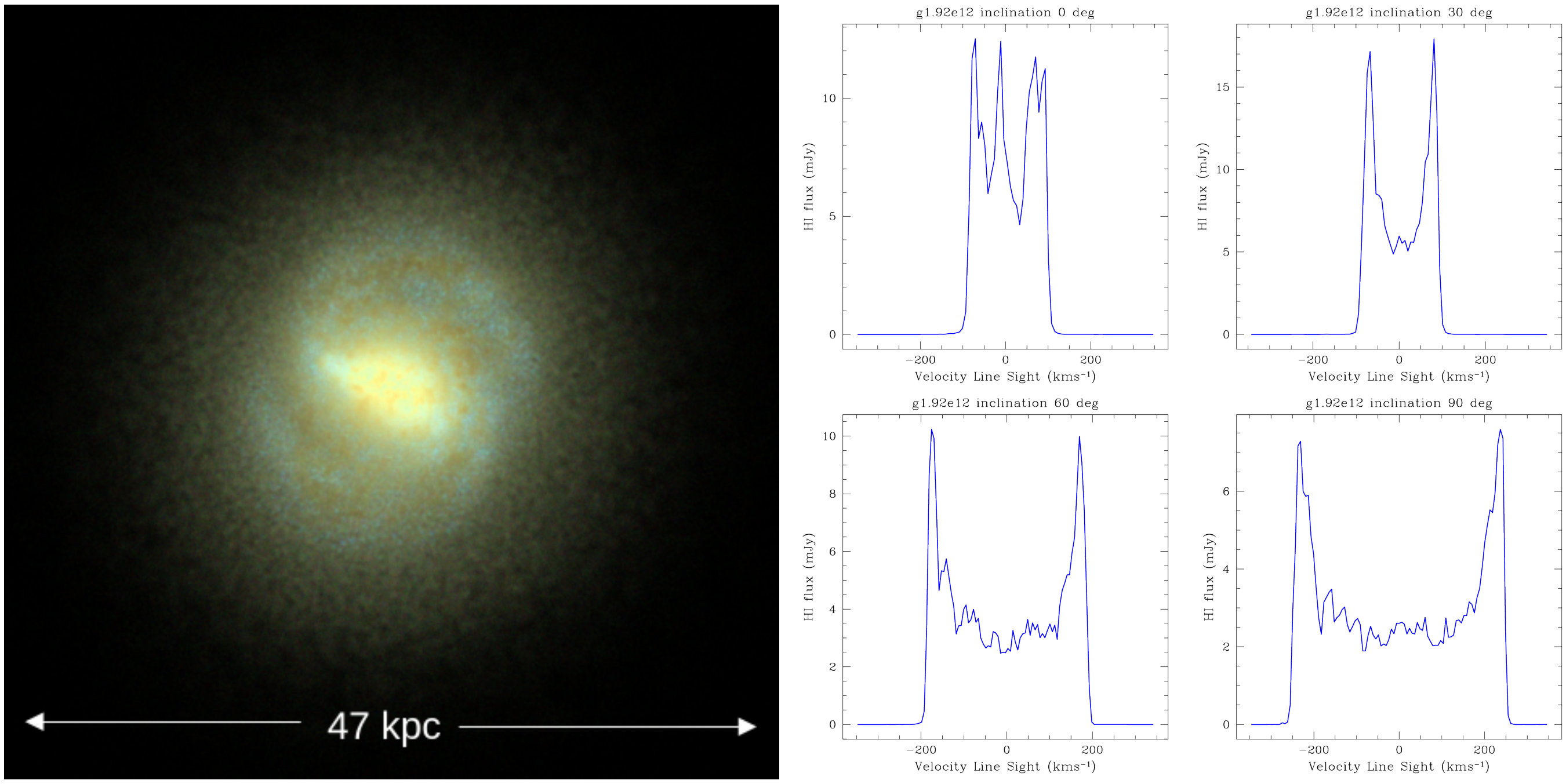

**Figure 9.** Left: the simulated optical image of NIHAO galaxy g1.92e12 based on the *g*, *r*, and *i* bands, and is face-on. Right: the simulated H I spectrum of NIHAO galaxy g1.92e12 at a variety of inclinations and at a distance of 100 Mpc.

The shape of this H I spectrum is usually a double-horned profile that decreases in velocity width as you decrease the inclination to 0° (face-on). This indicates that the majority of the H I gas for this system is in a rotating disk.

NIHAO galaxy g2.79e12 has a total H I mass of $1.69 \times 10^{10} M_\odot$ and a total mass of $3.07 \times 10^{12} M_\odot$. This H I mass is comparable to the H I mass found for the observed gLSB galaxies, but the total mass is still significantly higher. g2.79e12 had a major merger 7.7 Gyr ago ($z = 0.97$) with an object $1.20 \times 10^{12} M_\odot$ which was 60% of its mass at the time. This was a gas-rich merger. In Figure 10, the left panel shows the simulated optical image of g2.79e12 by combining *g*-, *r*-, and *i*-band images. The galaxy shows a clear bulge with a disk with blue, spiral arms surrounding it. In Figure 10, the right panel is the simulated H I spectrum for g2.79e12 as observed at 100 Mpc at $z = 0$ for a variety of inclinations. As one starts at inclination 90° (edge-on) the shape of the H I spectrum is an asymmetric profile which changes to a Gaussian shape as you approach 0° (face-on). This suggests that the H I gas is in a disk that is asymmetric in shape.

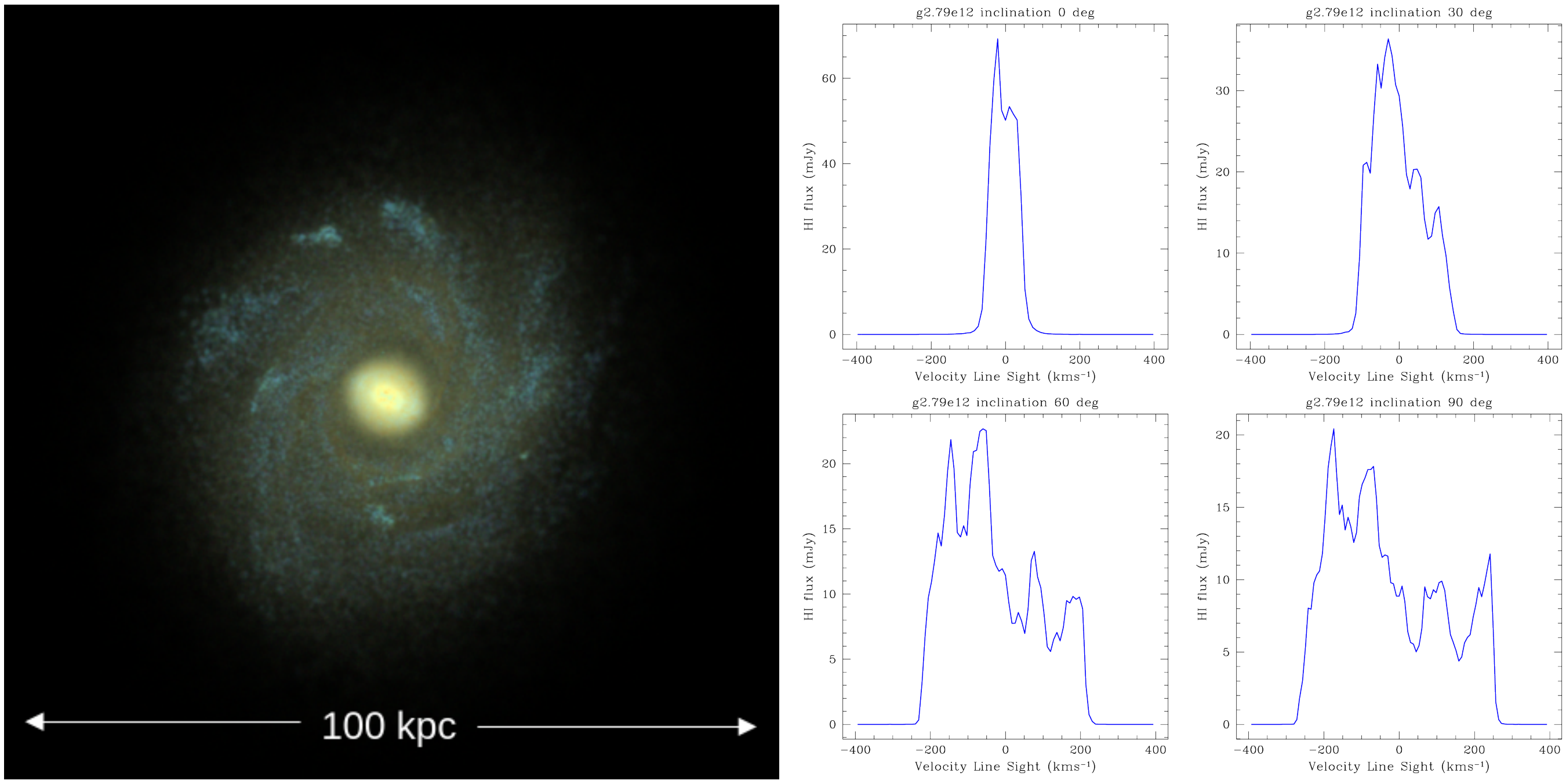

**Figure 10.** Left: the simulated optical image of NIHAO galaxy g2.79e12 based on the $g$, $r$, and $i$ bands, and is face-on. Right: the simulated H I spectrum of NIHAO galaxy g2.79e12 at a variety of inclinations and at a distance of 100 Mpc.

Figure 11 shows the $g$-band radius containing 90% (red solid line) and 95% (blue solid line) of the flux as a function of time for the four NIHAO galaxies. The dashed regions are where the mergers occur, and the radius has little meaning there. The significant difference between the 90% and 95% values shows that there is a faint optical disk surrounding the NIHAO galaxies. Usually, after a merger, there is an increase in radius. The exception is the second merger of g1.12e12, where the radius is unchanged. The galaxy without mergers, g1.92e12, reaches a large optical radius early (around lookback time 10 Gyr) and then its 90% radius is fairly stable while its 95% varies greatly. The growth of radius after mergers suggests that they play a key role in forming the large optical disks of gLSB galaxies.

### 4.2. *H I Spectra Shape Properties*

The top left panel in Figure 12 displays the symmetry parameter $A_F$, whose definition is described in Section 2, as a function of inclination for the four NIHAO galaxies. With observed H I spectra, one cannot as easily vary the inclination as one can with simulated data. Whether a galaxy is asymmetric partially depends on inclination. At 0° inclination (face-on), three of the galaxies have their H I profiles being symmetric. Only g1.77e12 is asymmetric, and this is likely due to a velocity component left over from the recent major merger. However, being face-on is the least likely inclination from a random distribution of inclinations. At 90° inclination (edge-on), three galaxies are symmetric. However, now it is g1.77e12 that is symmetric, and g2.79e12 is asymmetric. At inclinations between these extremes, there is a lot of variety. g1.92e12 with its double-horned profile is symmetric at all inclinations. g1.12e12 occasionally increases above the $A_F = 1.24$ line that N. Yu et al. (2022) defines a galaxy as asymmetric, but only by a small amount. g1.77e12 with its extra velocity component, which is small compared to the main component, is almost always severely asymmetric. Galaxy g2.79e12 is asymmetric for inclinations 20° and above. Based on this, the four NIHAO galaxies, which are similar to the observed gLSB galaxies, are more likely to be asymmetric than the 17% figure that N. Yu et al. (2022) found for galaxies from the ALFALFA survey. This matches the trend seen for the observed gLSB galaxy sample. The only galaxy without any major mergers, g1.92e12, is symmetric, while the others are asymmetric at least for some values of their inclination.

The top right panel in Figure 12 displays the shape parameter $C_v$ (see Equation 10) versus inclination for the four NIHAO galaxies. Galaxy g1.92e12 is typical for a galaxy with a rotating disk of H I gas. At 0° inclination (face-on), it is a flat-topped profile, and then moves into a double-horned profile with increasing inclination. Galaxy 1.77e12 at 0° inclination starts off with a double-horned profile and moves into a Gaussian profile at higher inclinations. This is due to the secondary velocity component in its H I spectrum, likely the remains of the major merger it underwent. g1.1e12 always has a Gaussian profile. g2.79e12 starts out as a Gaussian profile at low inclination and then moves into a double-horned profile at high inclination, but its journey is very chaotic. This may be due to the asymmetric nature of its H I velocity profile. Compared to our observed gLSB galaxies, the NIHAO galaxies do not show the same level of preference for double-horned profiles, though one could say that the simulated galaxies are roughly evenly split between double-horned profiles and other shapes. It is possible that some of the values for $C_v$ are not seen in nature, as will be discussed below.

The bottom left panel in Figure 12 displays the shape parameter K, whose definition is described in Section 2, versus inclination for the four NIHAO galaxies. The K values for g1.92e12 place it always in the double-horned profile region without the flat values at low inclination seen with $C_v$. g1.12e12 has values in the flat-topped region for low inclinations, which are not present in the $C_v$ values. The K values for g1.77e12 behave chaotically, swinging between the possible line profile types without the clear trend seen with $C_v$. g2.79e12 swings between flat-topped and double-horned profiles, also chaotically. The reason for the chaotic behavior with inclination seen here is likely due to the way the H I gas is clumped in the simulation, rather than being smoothed out as seen for observed galaxies. Furthermore, some of the K values are not seen in observed galaxies, as will be detailed next.

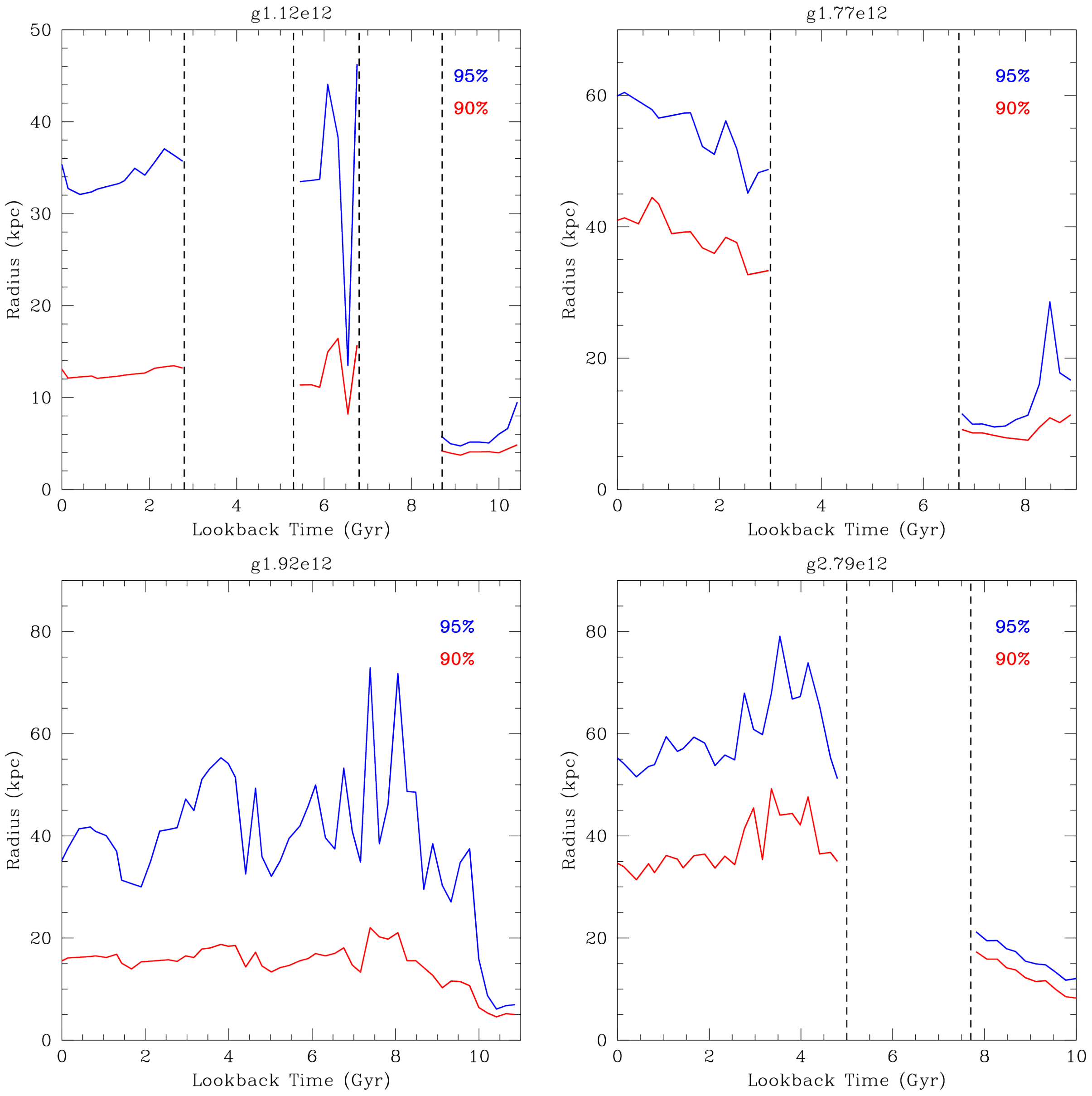

**Figure 11.** The $g$-band radius containing 90% and 95% of the flux as a function of time for the four NIHAO galaxies. The dashed regions are where the mergers occur, and the radius has little meaning there.

Included in these three figures are points for values of the observed gLSB galaxies whose optical surface brightness profile most closely matches the individual NIHAO galaxy. These points are at the inclination of the galaxy at its value of the specific parameter. g1.12e12 matches with Mrk 592, MCG-01-06-068 matches g1.77e12, UGC 1382 matches g1.92e12, and UGC 1697 matches g2.79e12. What is most noticeable about the distribution of the points in all three figures is how poorly the points usually match the NIHAO galaxy they are associated with. This indicates that while the NIHAO galaxies may have similar properties to the gLSB galaxies as a whole, when trying to match their H I properties to a specific galaxy, they do poorly. What should be noted is that while the matched gLSB to the NIHAO galaxy is the closest to its optical surface brightness profile, the other gLSB profiles are not that dissimilar in shape.

The bottom right panel in Figure 12 displays the shape parameters $C_v$ versus K for the NIHAO galaxies, with different points being for the different inclinations. The cyan points are the values from N. Yu et al. (2022) for ALFALFA galaxies. While most of the values lie in the region of physical values described by N. Yu et al. (2022), there are values that lie outside this region. Clearly, the simulation does not describe reality for these values. There are points at extremely low K and $C_v$ for galaxy g2.79e12 in the double-horn region. There are also values for g1.77e12 and g2.79e12 that lie in a region of high $C_v$ values but low K values that are also not physical. With g1.77e12, the second velocity component seen in its

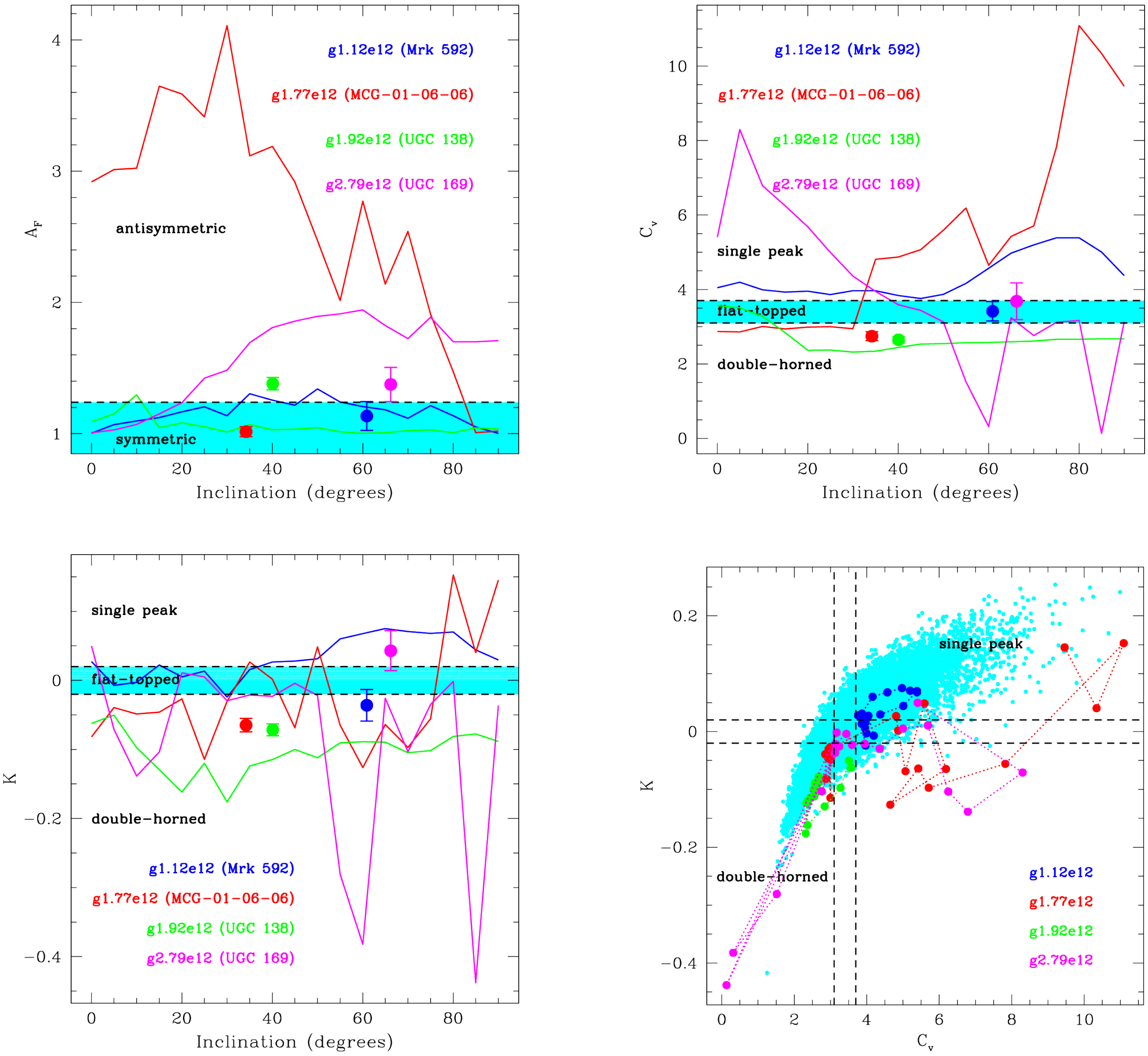

**Figure 12.** Top right: $A_F$, as defined in Section 2, vs. inclination for the four NIHAO galaxies. The dashed line is at $A_F = 1.24$. Galaxies with $A_F$ higher than this value are defined by N. Yu et al. (2022) as having asymmetric H I line profiles. The points are the values for observed gLSB galaxies whose surface brightness profile most closely matches the individual NIHAO galaxy. They are given at the gLSB galaxy's inclination and $A_F$ value. Top left: this figure displays the shape parameter $C_v$ (see Section 2) vs. inclination for the four NIHAO galaxies. The two dashed black lines at $C_v = 3.1$ and 3.7, taken from N. Yu et al. (2022), define a spectrum below 3.1 as a double-horned profile; between 3.1 and 3.7, the spectrum has a flat top, and above 3.7, the spectrum has a Gaussian profile. The points are the values for observed gLSB galaxies whose surface brightness profile most closely matches the individual NIHAO galaxy. They are given at the gLSB galaxy's inclination and $C_v$ value. Bottom left: this figure displays the shape parameter K (see Section 2) vs. inclination for the four NIHAO galaxies. The black dashed lines at $K = -0.02$ and $K = 0.02$ are marked on the figure. N. Yu et al. (2022) defines for their sample galaxies with $K < -0.02$ as double-horned profiles, $K > 0.02$ as single-peaked profiles, and between these profiles as flat-topped profiles. The points are the values for observed gLSB galaxies whose surface brightness profile most closely matches the individual NIHAO galaxy. They are given at the gLSB galaxy's inclination and K value. Bottom right: this figure displays the shape parameters $C_v$ vs. K for the NIHAO galaxies, with different points being values at different inclinations. Dotted lines connect the points showing the direction of the inclination change. The cyan points are the N. Yu et al. (2022) values for ALFALFA galaxies. The dashed lines mark the boundaries between the double-horned, flat-topped, and Gaussian profiles as described previously. Figure 5 shows this same plot for the observed gLSB galaxies in this paper. All the observed gLSB galaxies lie in the region covered by the ALFALFA galaxies. The points are the observed gLSB galaxies that most closely match the NIHAO profiles. g1.12e12 matches with Mrk 592, MCG-01-06-068 matches g1.77e12, UGC 1382 matches g1.92e12 and UGC 1697 matches g2.79e12.

H I spectrum is likely causing problems with the calculation of these values. g2.79e12 has a very chaotic K value with inclination due to structures appearing and disappearing as one steps across inclination. This is likely the cause of the nonphysical values seen here. The main cause of these problems is likely the way the simulated H I gas is clumped rather than smoothly distributed throughout the galaxy, as one expects for an actual galaxy.

## 5. Conclusion

We have examined the H I properties of 19 gLSB galaxies selected from the optical survey of A. S. Saburova et al. (2023). The optical selection condition for being a gLSB galaxy was that the galaxy had a $g$-band 27.7 mag arcsec$^{-2}$ isophotal radius or had four disk scale lengths of $4h \geqslant 50$ kpc. Selecting the sample of gLSB galaxies in the optical rather than by H I mass allows us to test whether the H I properties of gLSB galaxies are a defining property of this galaxy class. For 13 of 19 our galaxies, the H I mass of the gLSB galaxy is indeed high, $>10^{10}$ $M_{\odot}$. Five of the gLSB galaxies have lower H I mass, but are still of a reasonable size, being at least $2.8 \times 10^{9} M_{\odot}$. For one of the galaxies, only an upper limit on the H I mass can be determined. Based on our sample of optical selected gLSB galaxies, it is clear that not all gLSB galaxies have high H I mass, though the vast majority do (13 out of 19 have H I masses above $10^{10} M_{\odot}$). What is not clear is whether the gLSB galaxies start with a high H I mass and then evolve to a lower mass, or whether the galaxies formed as a gLSB galaxy with the lower H I mass. The fact that the galaxy with only an upper limit has evidence of young, hot stars in its optical imaging suggests that the process may be one of evolution rather than formation.

From the comparison of the optical scale length with the H I mass of the galaxies, it is clear that the defining property of gLSB galaxies is not their high H I mass, as many other, more normal galaxies have as much gas. It is their optical size that defines the uniqueness of the sample.

Information on the properties of the H I gas within the gLSB galaxies can be determined from the shape of the H I spectrum. The sample of gLSB galaxies shows a preference to being asymmetric in their H I spectrum beyond what is seen for normal galaxies. This could indicate that a major merger is the likely formation mechanism for these galaxies. Ongoing interactions are unlikely, as most of the galaxies are fairly isolated. The sample of gLSB galaxies shows a preference for having a double-horned profile somewhat more than for more than normal galaxies. This indicates that most of the H I gas in these systems is in a rotating disk.

Four galaxies with similar surface brightness profiles to gLSB galaxies, though not perfect analogs, have been selected from the NIHAO simulation. These galaxies usually have lower H I mass and higher total mass than the gLSB galaxies examined here. Most of these galaxies have significant galaxy mergers in their past. The simulated radius of the galaxies usually grows in size after a merger (for three out of four major mergers for these galaxies, this is the case). The NIHAO galaxies show similarly asymmetric H I spectra as the gLSB galaxies examined here. However, it is not as clear that the NIHAO galaxies have substantial amounts of their H I gas in a rotating disk like the gLSB galaxies examined here. Based on the asymmetry of the H I spectra of the gLSB galaxies and the history of mergers increasing the radius of the NIHAO galaxies, the scenario of major mergers forming gLSB galaxies seems to be favored.

## Acknowledgments

We wish to acknowledge the help of Andrea Maccio and the NIHAO team with this work.

The contributions of P.L., I.K., and J.D.G. are supported by Tamkeen through a grant to NYUAD to support basic faculty research.

I.G. and D.G. acknowledge support from the Russian Science Foundation (RSCF, grant No. 23-12-00146) for their work on the analysis and fitting of long-slit spectral data for 2MASXJ02015377+0131087 and the search for available H I data on gLSB galaxies.

The National Radio Astronomy Observatory and Green Bank Observatory are facilities of the US National Science Foundation operated under cooperative agreement by Associated Universities, Inc.

The Legacy Surveys consist of three individual and complementary projects: the Dark Energy Camera Legacy Survey (DECaLS; Proposal ID 2014B-0404; PIs: David Schlegel and Arjun Dey), the Beijing-Arizona Sky Survey (BASS; NOAO Prop. ID 2015A-0801; PIs: Zhou Xu and Xiaohui Fan), and the Mayall $z$-band Legacy Survey (MzLS; Prop. ID 016A-0453; PI: Arjun Dey). DECaLS, BASS, and MzLS together include data obtained, respectively, at the Blanco telescope, Cerro Tololo Inter-American Observatory, NSF's NOIRLab; the Bok telescope, Steward Observatory, University of Arizona; and the Mayall telescope, Kitt Peak National Observatory, NOIRLab. Pipeline processing and analyses of the data were supported by NOIRLab and the Lawrence Berkeley National Laboratory (LBNL). The Legacy Surveys project is honored to be permitted to conduct astronomical research on Iolkam Du'ag (Kitt Peak), a mountain with particular significance to the Tohono O'odham Nation.

NOIRLab is operated by the Association of Universities for Research in Astronomy (AURA) under a cooperative agreement with the National Science Foundation. LBNL is managed by the Regents of the University of California under contract to the US Department of Energy.

This project used data obtained with the Dark Energy Camera (DECam), which was constructed by the Dark Energy Survey (DES) collaboration. Funding for the DES Projects has been provided by the US Department of Energy, the US National Science Foundation, the Ministry of Science and Education of Spain, the Science and Technology Facilities Council of the United Kingdom, the Higher Education Funding Council for England, the National Center for Supercomputing Applications at the University of Illinois at Urbana-Champaign, the Kavli Institute of Cosmological Physics at the University of Chicago, Center for Cosmology and Astro-Particle Physics at the Ohio State University, the Mitchell Institute for Fundamental Physics and Astronomy at Texas A&M University, Financiadora de Estudos e Projetos, Fundacao Carlos Chagas Filho de Amparo, Financiadora de Estudos e Projetos, Fundacao Carlos Chagas Filho de Amparo a Pesquisa do Estado do Rio de Janeiro, Conselho Nacional de Desenvolvimento Cientifico e Tecnologico and the Ministerio da Ciencia, Tecnologia e Inovacao, the Deutsche Forschungsgemeinschaft and the Collaborating Institutions in the Dark Energy Survey. The Collaborating Institutions are Argonne National Laboratory, the University of California at Santa Cruz, the University of Cambridge, Centro de Investigaciones Energeticas, Medioambientales y Tecnologicas-Madrid, the University of Chicago, University College London, the DES-Brazil Consortium, the University of Edinburgh, the Eidgenossische Technische Hochschule (ETH) Zurich, Fermi

National Accelerator Laboratory, the University of Illinois at Urbana-Champaign, the Institut de Ciencies de l'Espai (IEEC/CSIC), the Institut de Fisica d'Altes Energies, Lawrence Berkeley National Laboratory, the Ludwig Maximilians Universitat Munchen and the associated Excellence Cluster Universe, the University of Michigan, NSF's NOIRLab, the University of Nottingham, the Ohio State University, the University of Pennsylvania, the University of Portsmouth, SLAC National Accelerator Laboratory, Stanford University, the University of Sussex, and Texas A&M University.

BASS is a key project of the Telescope Access Program (TAP), which has been funded by the National Astronomical Observatories of China, the Chinese Academy of Sciences (the Strategic Priority Research Program "The Emergence of Cosmological Structures" grant XDB09000000), and the Special Fund for Astronomy from the Ministry of Finance. The BASS is also supported by the External Cooperation Program of the Chinese Academy of Sciences (grant 114A11KYSB20160057), and the Chinese National Natural Science Foundation (grant Nos. 12120101003, 11433005).

The Legacy Survey team makes use of data products from the Near-Earth Object Wide-field Infrared Survey Explorer (NEOWISE), which is a project of the Jet Propulsion Laboratory/California Institute of Technology. NEOWISE is funded by the National Aeronautics and Space Administration.

The Legacy Surveys imaging of the DESI footprint is supported by the Director, Office of Science, Office of High Energy Physics of the US Department of Energy under contract No. DE-AC02-05CH1123, by the National Energy Research Scientific Computing Center, a DOE Office of Science User Facility under the same contract; and by the US National Science Foundation, Division of Astronomical Sciences under contract No. AST-0950945 to NOAO.

This research was undertaken thanks in part to funding from the Canada First Research Excellence Fund through the Arthur B. McDonald Canadian Astroparticle Physics Research Institute. The NIHAO simulations are based upon work supported by Tamkeen under the NYU Abu Dhabi Research Institute grant CASS. The authors gratefully acknowledge the Gauss Centre for Supercomputing e.V. (www.Gauss-center.eu) for enabling this project via computing time on the GCS Supercomputer SuperMUC at Leibniz Supercomputing Centre (www.lrz.de), along with the high-performance computing resources at New York University Abu Dhabi.

## Appendix A
## Individual Galaxies

In this appendix, the properties of the individual galaxies are discussed. The H I spectrum for each galaxy is displayed next to the optical image in Figures 13, 14, 16, 20, 21, 22, 23, 24, 25, 26, 27, 28, 29, 30, 32, 33, and 34. The ellipse around the galaxy in the optical image shows the extent of the four exponential disk scale lengths (4 hr). The central velocity of each H I spectrum is the optical redshift, which often is not exactly the same as the H I redshift. The error on the optical redshift is displayed as dotted lines on either side of the main optical redshift line. Green dotted lines show the H I redshift.

### A.1. Galaxy UGC 2061

The H I spectrum for galaxy UGC 2061, shown in Figure 14, is more unusual as it appears almost flat across its width if one smooths out the noise. It is possible to generate such a H I spectrum by having a double-horned profile along with a Gaussian, as has been shown by B. Peng et al. (2023). The values in the tables in Section 3 assume that all of the

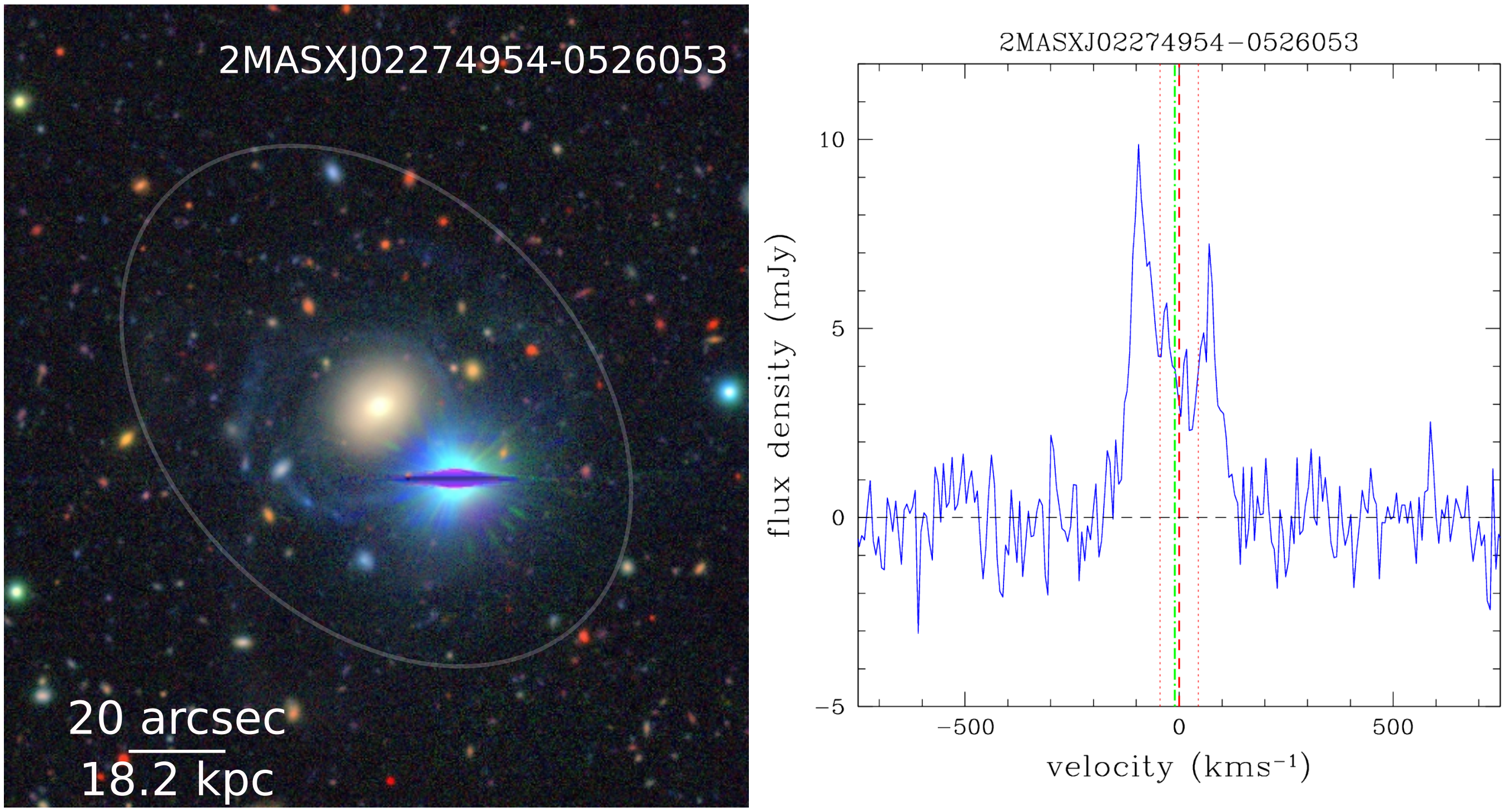

**Figure 13.** This figure displays the properties of galaxy 2MASXJ02274954-0526053. On the left is a color composite of the HSC images. On the right is the GBT H I spectrum centered at the optical redshift. The galaxy has H I mass of $1.13 \times 10^{10}\, M_\odot$ and a $g$-band absolute magnitude of $-20.35$. The H I spectrum is an asymmetric double-horned profile, indicating a rotating disk of H I gas that is asymmetric. This is also a pretty standard H I spectrum. Again, one should note that the H I mass of this gLSB galaxy is indeed high above $10^{10}\, M_\odot$.

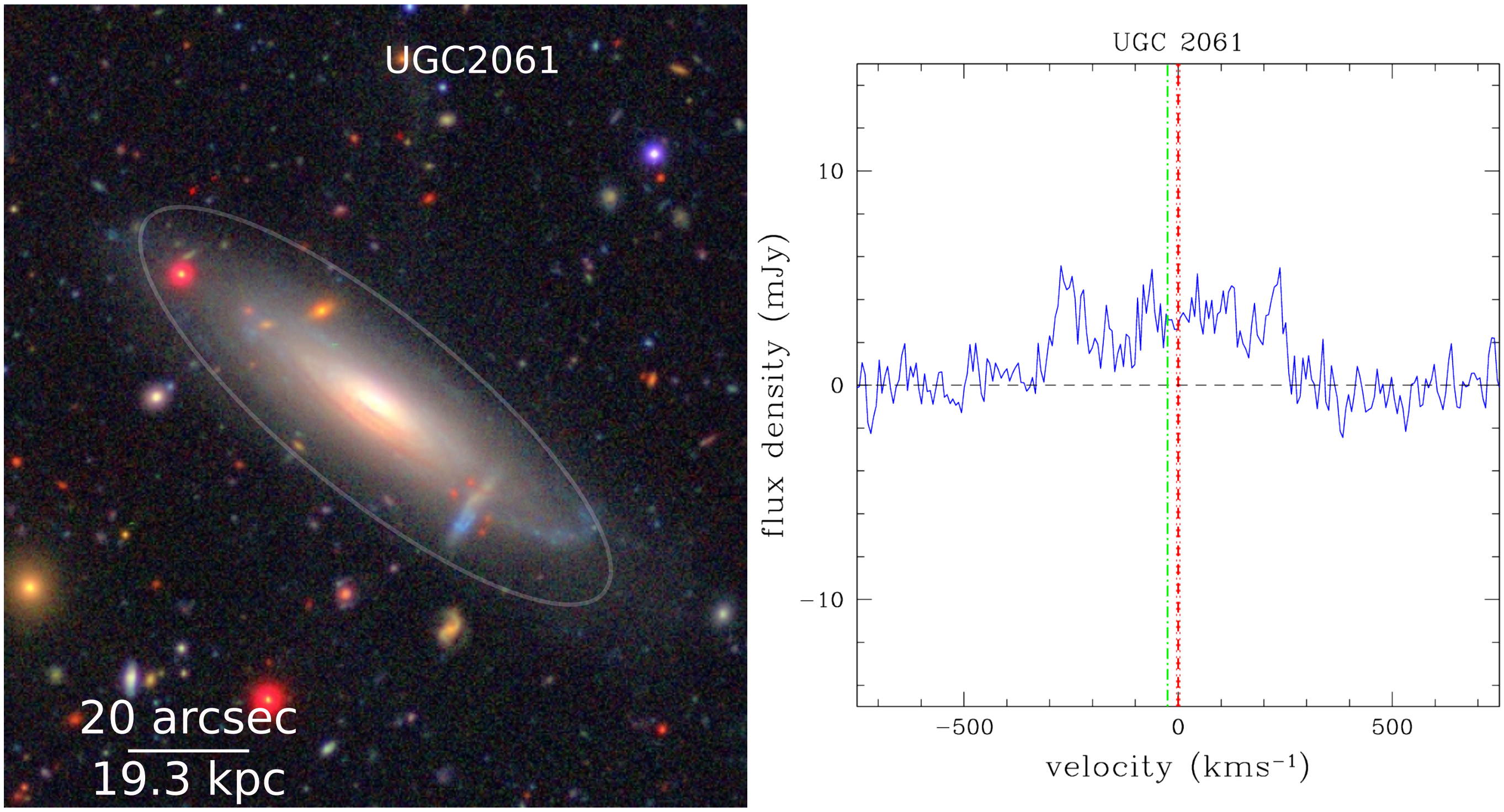

**Figure 14.** This figure displays the properties of galaxy UGC 2061. On the left is a color composite of the HSC images. On the right is the GBT H I spectrum centered at the optical redshift. The galaxy has H I mass of $1.85 \times 10^{10}\,M_\odot$ and $g$-band absolute magnitude of −21.08.

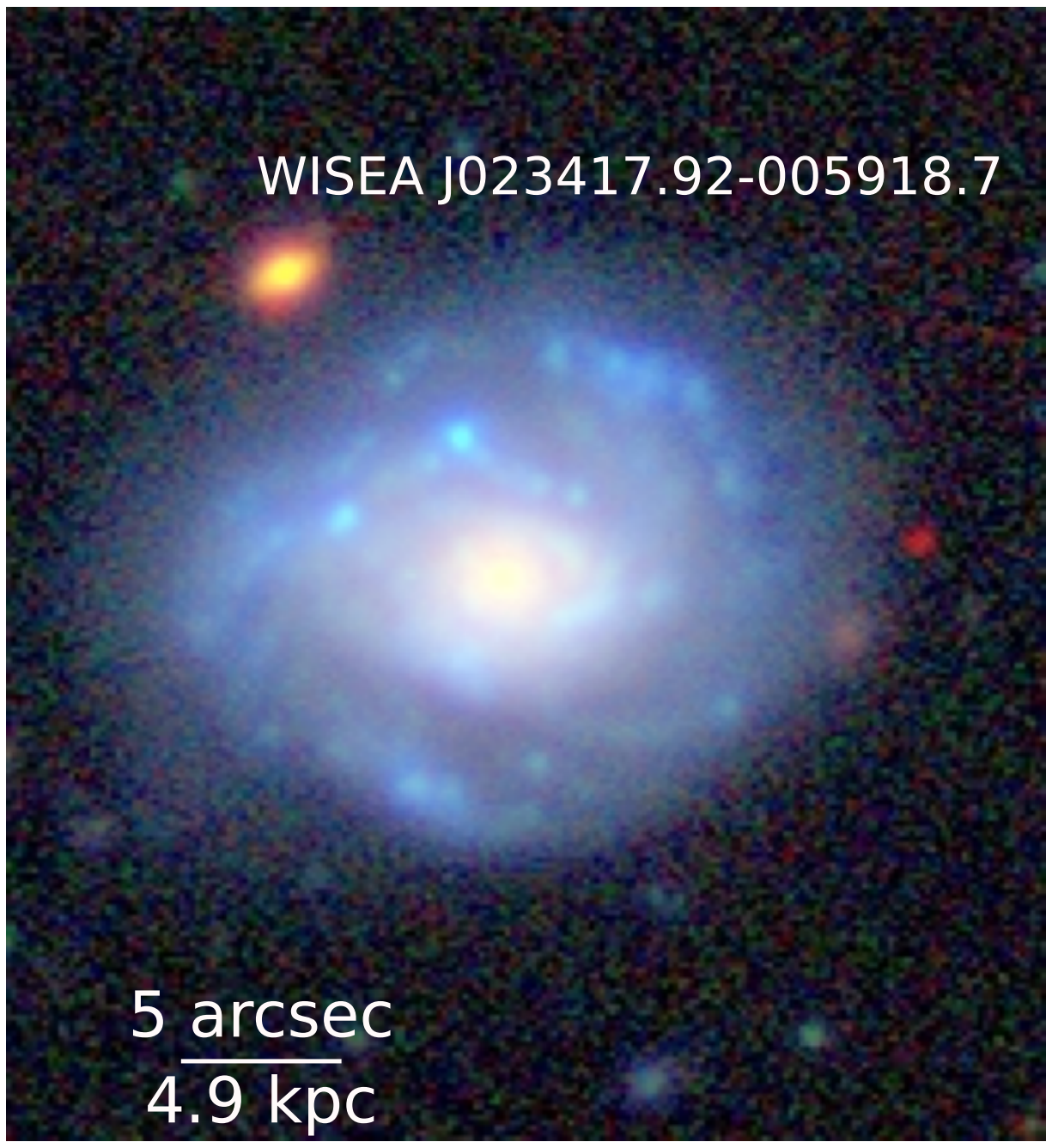

**Figure 15.** Color composite of the HSC image of WISEA 023417.92-005918.7, the companion galaxy that may be adding H I flux to the spectrum of UGC 2061.

H I signal is due to galaxy UGC 2061. However, there is an additional factor here. There is a small galaxy, WISEA J023417.92-005918.7, within the GBT beam, and it is at a redshift similar to that of UGC 2061, so it may be adding H I flux to the spectrum.

WISEA J023417.92-005918.7 is $4\overset{\prime}{.}97$ away from UGC 2061 on the sky (the GBT beam FWHM at this frequency is $9\overset{\prime}{.}31$). Its redshift is 14,773 km s$^{-1}$, which is only a 9 km s$^{-1}$ difference from that of UGC 2061. This makes it likely that this smaller galaxy is adding flux to the H I spectrum. Being at the same redshift, the projected distance between the galaxies is 318 kpc. Figure 15 shows an optical composite image of WISEA J023417.92-005918.7, the companion galaxy. This image is quite blue, indicating the presence of many massive stars, a signal of recent star formation that should mean the galaxy has significant reservoirs of H I gas.

Taking the apparent model magnitude for the two galaxies from the SDSS $g$ band (K. Abazajian et al. 2004), UGC 2061 is $15.73^m$ and WISEA J023417.92-005918.7 is $16.65^m$. This means that the companion is ∼0.5 times as bright in the optical as UGC 2061.

There is a correlation between absolute $g$ magnitude and H I mass derived by A. Durbala et al. (2020) using SDSS photometry and ALFALFA H I masses (see Figure 3). There is a reasonable amount of scatter in this relationship, and it does not apply for all galaxies (early-type galaxies often have no H I gas), but it can be used to estimate the H I masses of the two galaxies. The correlation is

$$M_{\rm HI} = -0.246 g_{\rm abs} + 4.91. \quad \text{(A1)}$$

UGC 2061 has a $g$-band absolute magnitude of $-20.98^m$, which means it should have H I mass of $\sim 1 \times 10^{10}\,M_\odot$. WISEA J023417.92-005918.7 has a $g$-band absolute magnitude of $-20.07^m$, which means it should have H I mass of $\sim 7 \times 10^9\,M_\odot$. Not being at the center of the GBT beam, this signal will be reduced by a factor of 0.45, meaning that the contribution to the H I spectrum would be that of a galaxy of $\sim 3 \times 10^9\,M_\odot$. The measured signal is $1.848 \times 10^{10}\,M_\odot$ so there is more H I gas than one would expect from using the simple relationship from A. Durbala et al. (2020). This is probably what one would expect for gLSB galaxies, as they tend to have higher H I masses than average. However, the optical image of the companion WISEA J023417.92-005918.7 is almost symmetric, indicating that it is nearly face-on. As such, its H I spectrum is likely fairly narrow velocity width compared to UGC 2061, which is highly inclined, giving rise to a wider velocity width. This may mean that some of the flux in the center of H I spectrum is due to the companion, and it is this that leads to the flatness of the spectrum. In any case, based on the estimates of the size of the companion, it is still reasonable to describe UGC 2061 as having a high H I mass probably well over $10^{10}$ $M_{\odot}$.

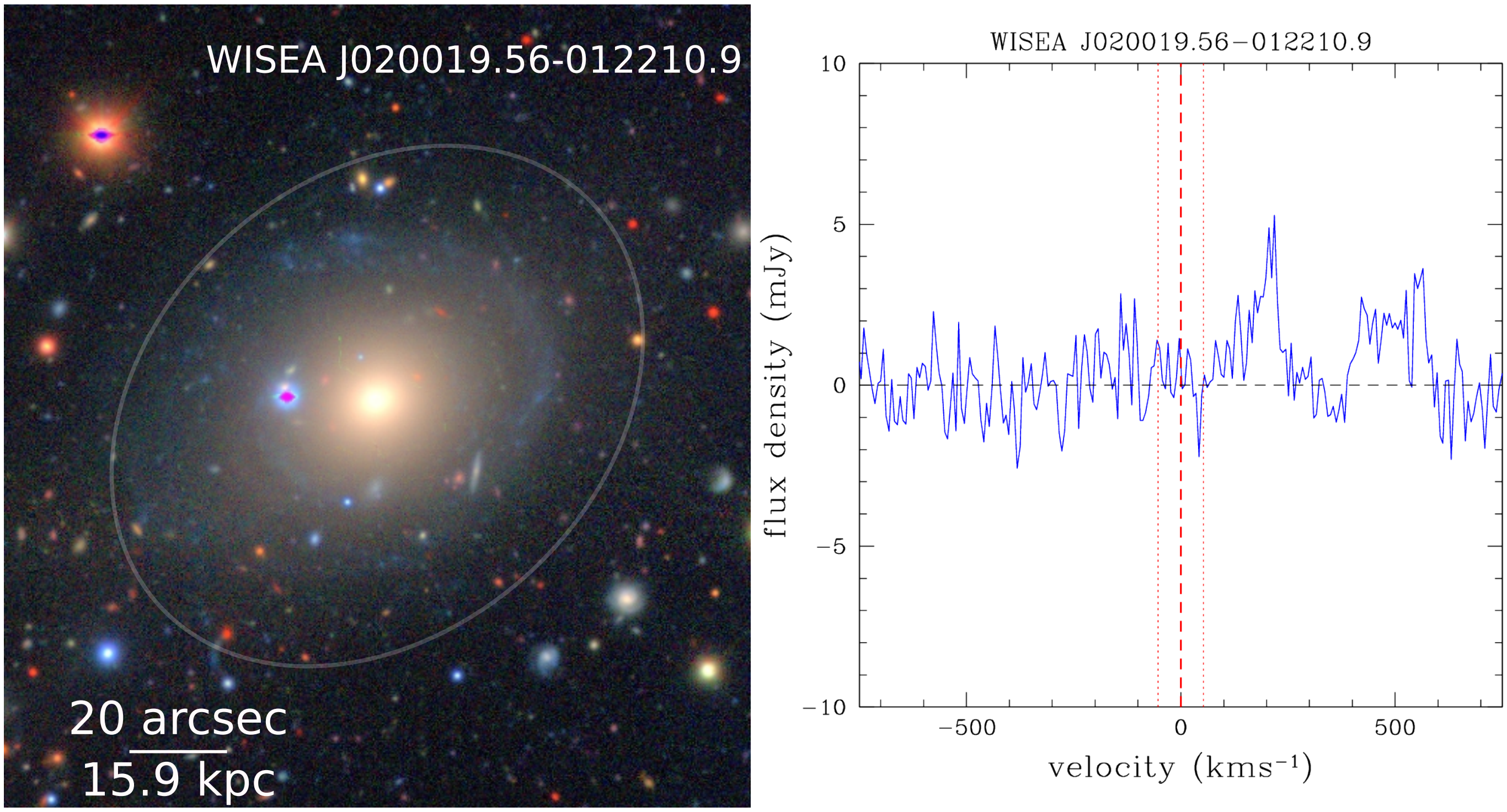

**Figure 16.** This figure displays the properties of galaxy WISEA J020019.56-012210.9. On the left is a color composite of the HSC images. On the right is the GBT H I spectrum centered at the optical redshift. This spectrum was observed with the GBT on-off for 60 minutes. The galaxy has a $g$-band absolute magnitude of −21.19.

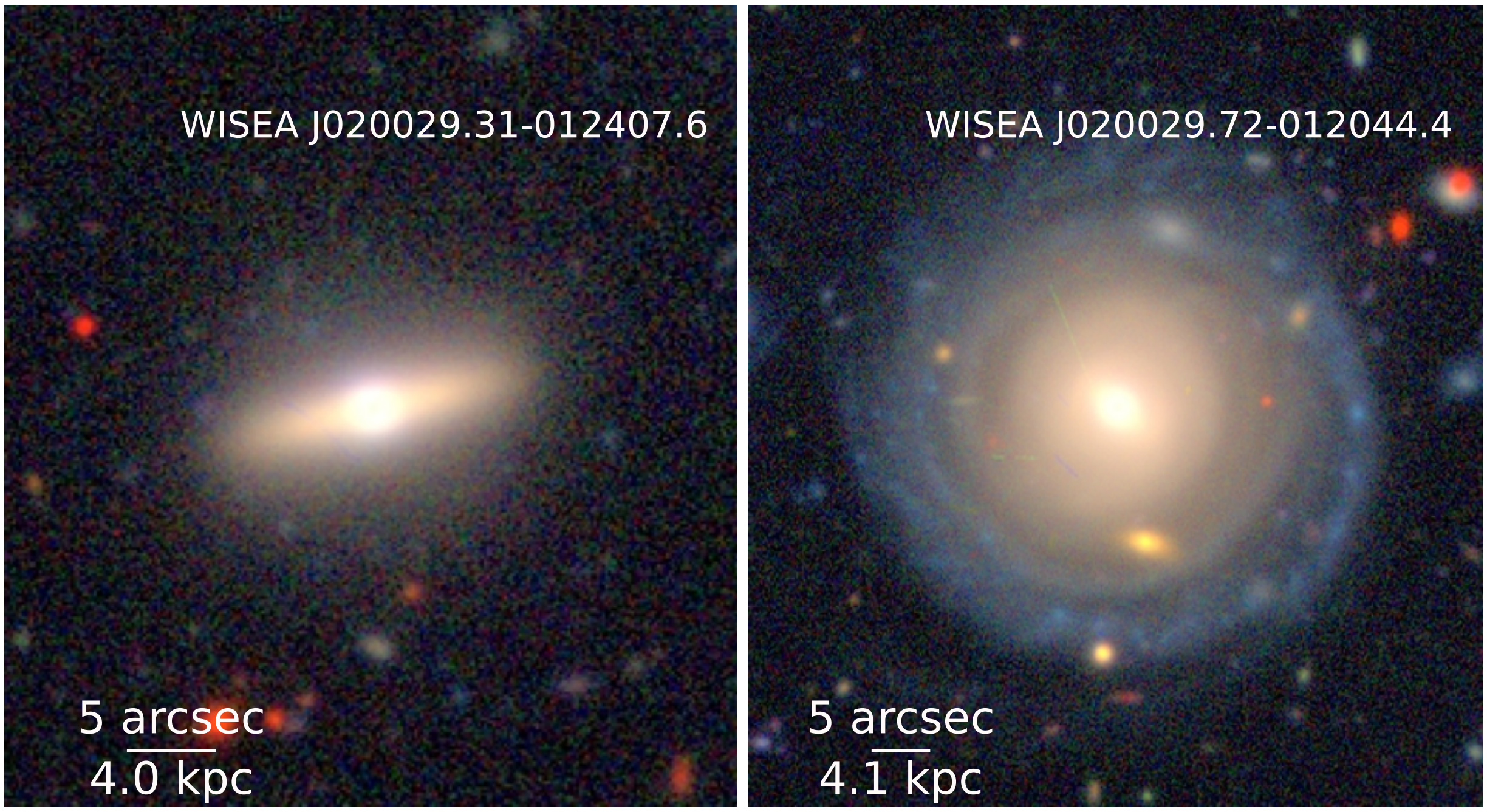

**Figure 17.** Optical images of the two companion galaxies to WISEA J020019.56-012210.9 from the HSC Subaru Strategic Program (SSP). They are combined $g$, $r$, and $z$ filter images. The galaxy on the left is WISEA J020029.31-012407.6, and the galaxy on the right is WISEA J020029.72-012044.4.

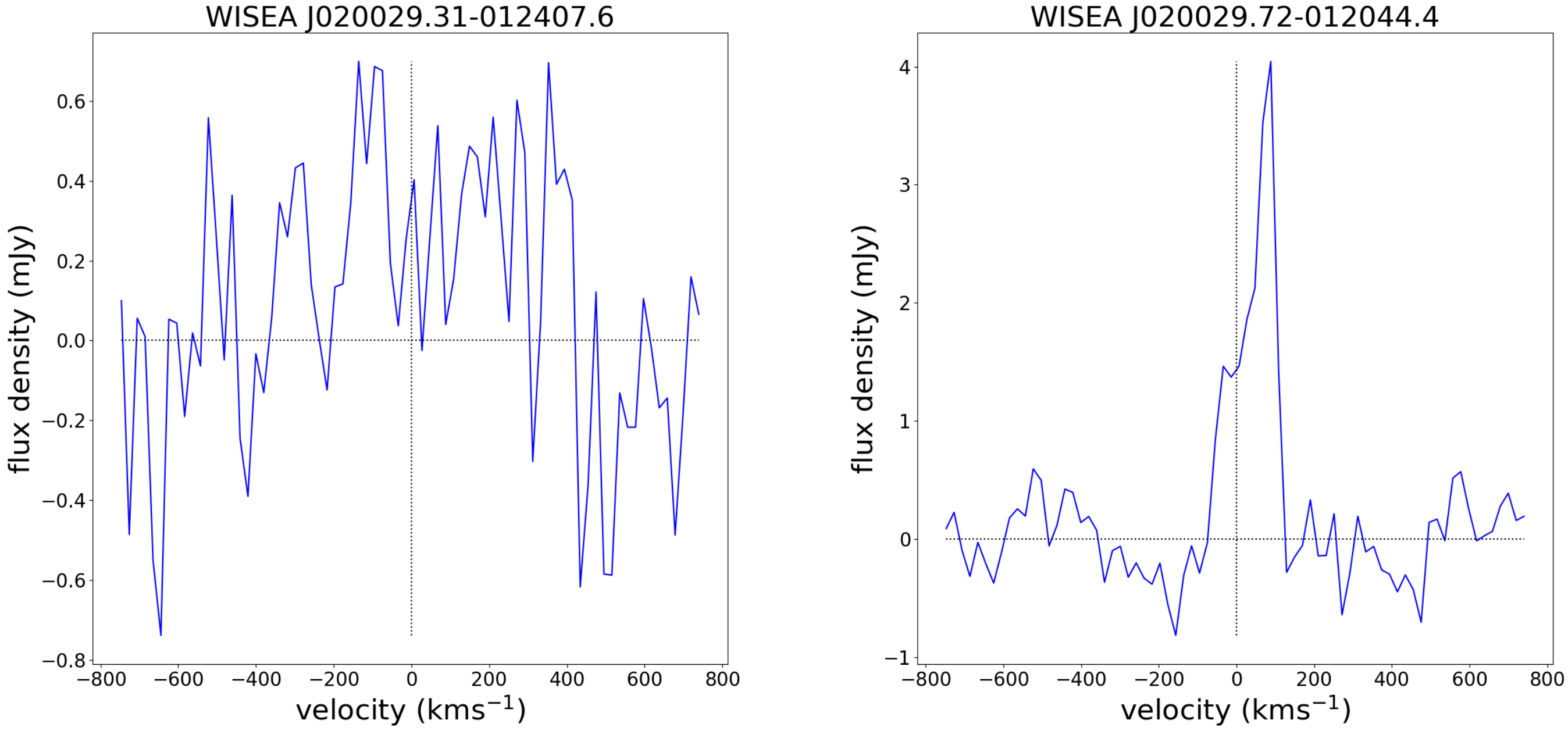

**Figure 18.** This figure shows the VLA H I spectra for the two companions.

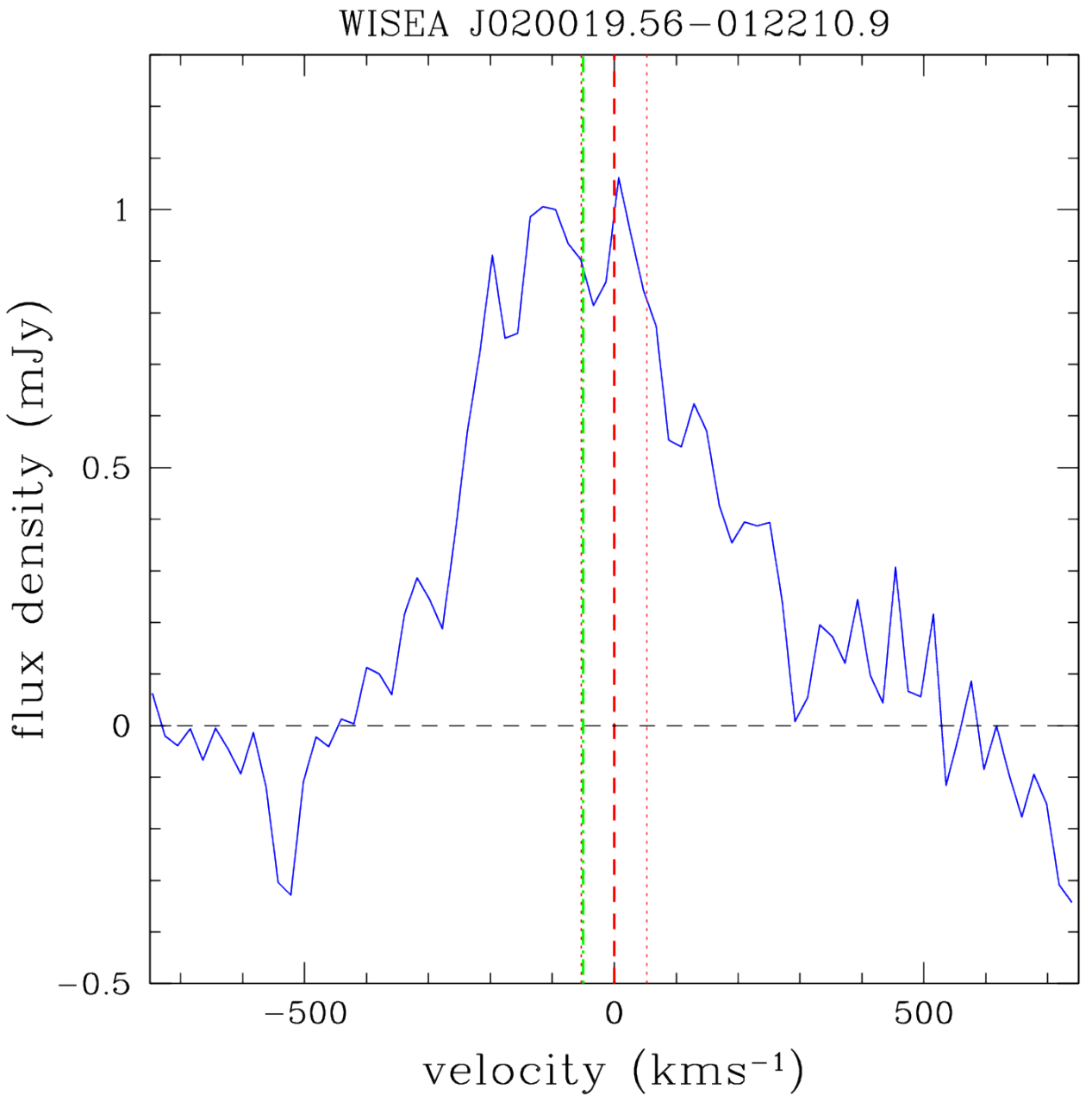

**Figure 19.** This figure shows the H I spectrum of galaxy WISEA J020019.56-012210.9 from the VLA observations. The H I spectrum is centered at the optical redshift. The velocity step size is 20 km s$^{-1}$. The galaxy has H I mass of $2.79 \times 10^9$ $M_\odot$.

### A.2. Galaxy WISEA J020019.56-012210.9

The GBT H I spectrum for galaxy WISEA J020019.56-012210.9 is shown in Figure 16, and it is unusual. The first thing one notices is that there is a significant offset between the H I signal and the optical redshift. The other thing one notices is that the H I spectrum goes to zero (actually below zero, but this could just be noise) between the two peaks at ∼200 and ∼500 km s$^{-1}$. The H I emission seen here may not be due to the gLSB galaxy, but to two companion galaxies within the GBT beam. The two companion galaxies lie at the redshifts of the two peaks in the H I spectrum. This, however, would mean that WISEA J020019.56-012210.9 is not contributing any significant H I flux to the spectrum, which would indicate it has little to no H I gas, making it a very anomalous gLSB galaxy.

The optical redshift for WISEA J020019.56-012210.9 is (11,865 ± 53) km s$^{-1}$ from A. I. Zabludoff et al. (1993). The GBT beam has an FWHM size of 9′.2 at this frequency. The companion galaxy WISEA J020029.31-012407.6 is 3′.93 away and has redshift (12,080 ± 3) km s$^{-1}$ from F. D. Albareti et al. (2017). This is 5 km s$^{-1}$ away from the redshift of the first peak in the H I spectrum. The companion galaxy WISEA J020029.72-012044.4 is 2′.94 away and has redshift (12,379 ± 45) km s$^{-1}$ from D. H. Jones et al. (2009). This is 29 km s$^{-1}$ away from the second peak in the H I spectrum. The companion galaxy WISEA J020029.31-012407.6 is a projected distance of 159 kpc from WISEA J020019.56-012210.9 and WISEA J020029.72-012044.4 is 150 kpc. The optical images of the two galaxies can be seen in Figure 17. Assuming that the peaks in the H I spectrum are due to these companions, then WISEA J020029.31-012407.6 has H I mass $(2.96 \pm 0.33) \times 10^9$ $M_\odot$ and WISEA J020029.72-012044.4 has H I mass $(3.20 \pm 0.34) \times 10^9$ $M_\odot$.

To disentangle the H I gas in this system, interferometric observations of the system were taken with the VLA using the D-configuration over 2 days in 2025 March for a total time of 11 hr. The synthesis beam size of the observations was 67″.06 × 51″.68. This means that we cannot make useful maps of the galaxies, but do have enough resolution to separate the galaxies spatially as well as by redshift and thus make measurements of the total H I mass of the galaxies. The data cubes produced automatically by the National Radio Astronomy Observatory were of sufficient quality for the measurements we wanted to make.

Figure 18 shows the VLA H I spectra of the two companions. These galaxies were unresolved by the VLA beam. The H I spectrum of the companion WISEA J020029.31-012407.6 is rather noisy and at a much lower flux density than the signal seen in the GBT observations. Assuming a H I spectral width of 200 km s$^{-1}$, the galaxy would have H I mass of $(4.45 \pm 2.0) \times 10^8\, M_\odot$. The H I spectrum for the companion WISEA J020029.72-012044.4 is much more definite with a clear signal. This galaxy has H I mass of $(2.72 \pm 0.20) \times 10^9\, M_\odot$ not too far from the GBT measurement.

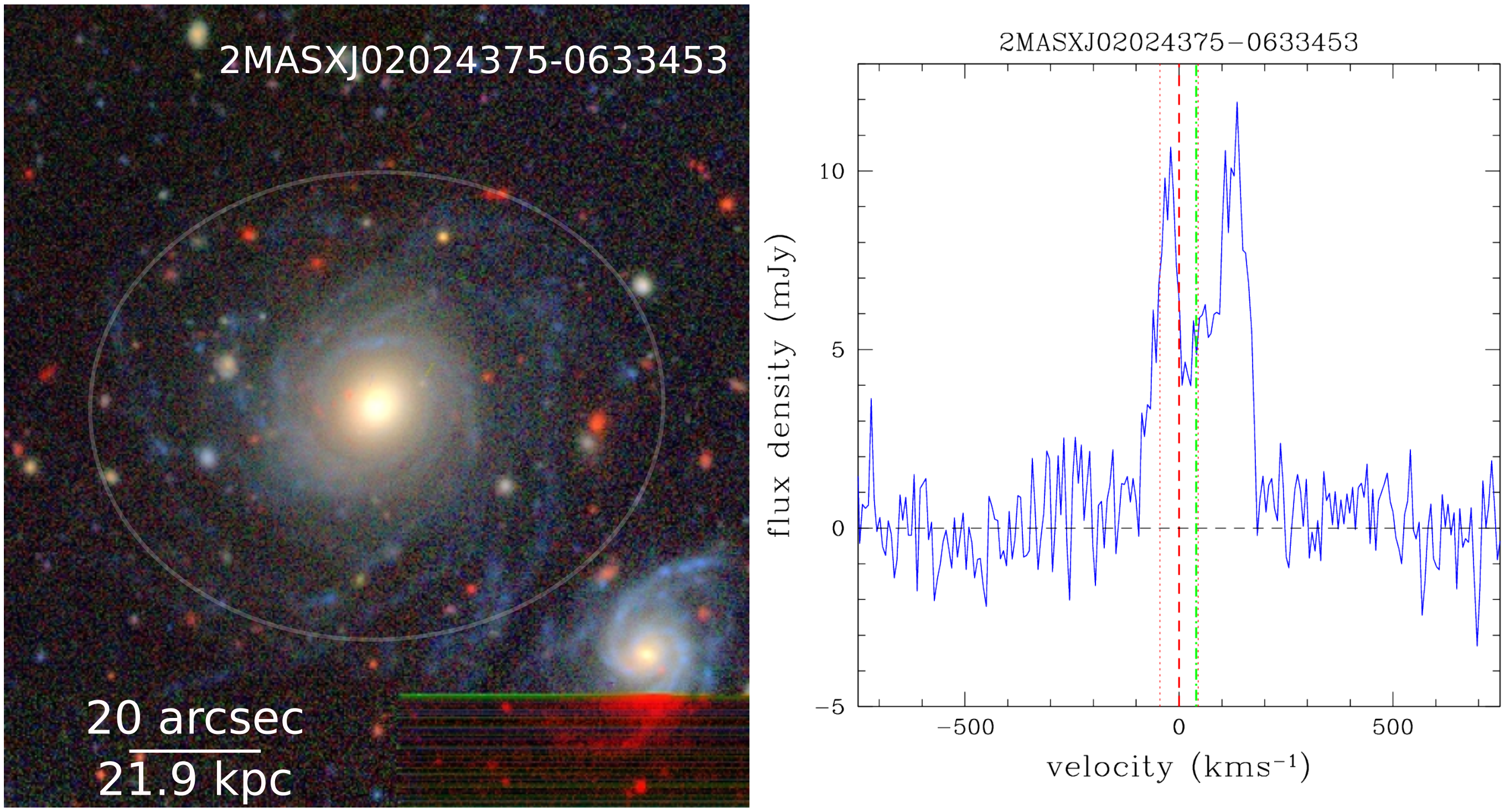

**Figure 20.** This figure displays the properties of galaxy 2MASXJ02024375-0633453. On the left is a color composite of the HSC images. The galaxy on the bottom right is at a significantly higher redshift than the gLSB galaxy. The gLSB galaxy is at redshift $cz = 16{,}935$ km s$^{-1}$. The companion WISEA J020240.80-063425.1 is at redshift $cz = 38{,}565$ km s$^{-1}$ (D. R. Law et al. 2016). On the right is the GBT H I spectrum centered at the optical redshift. The galaxy has H I mass of $2.52 \times 10^{10}\ M_{\odot}$ and a $g$-band absolute magnitude of −21.19. There is a slight asymmetry in the double-horned profile of the H I spectrum of 2MASXJ02024375-0633453, but otherwise this is a standard gLSB spectrum. Again, the H I mass is above $10^{10}\ M_{\odot}$.

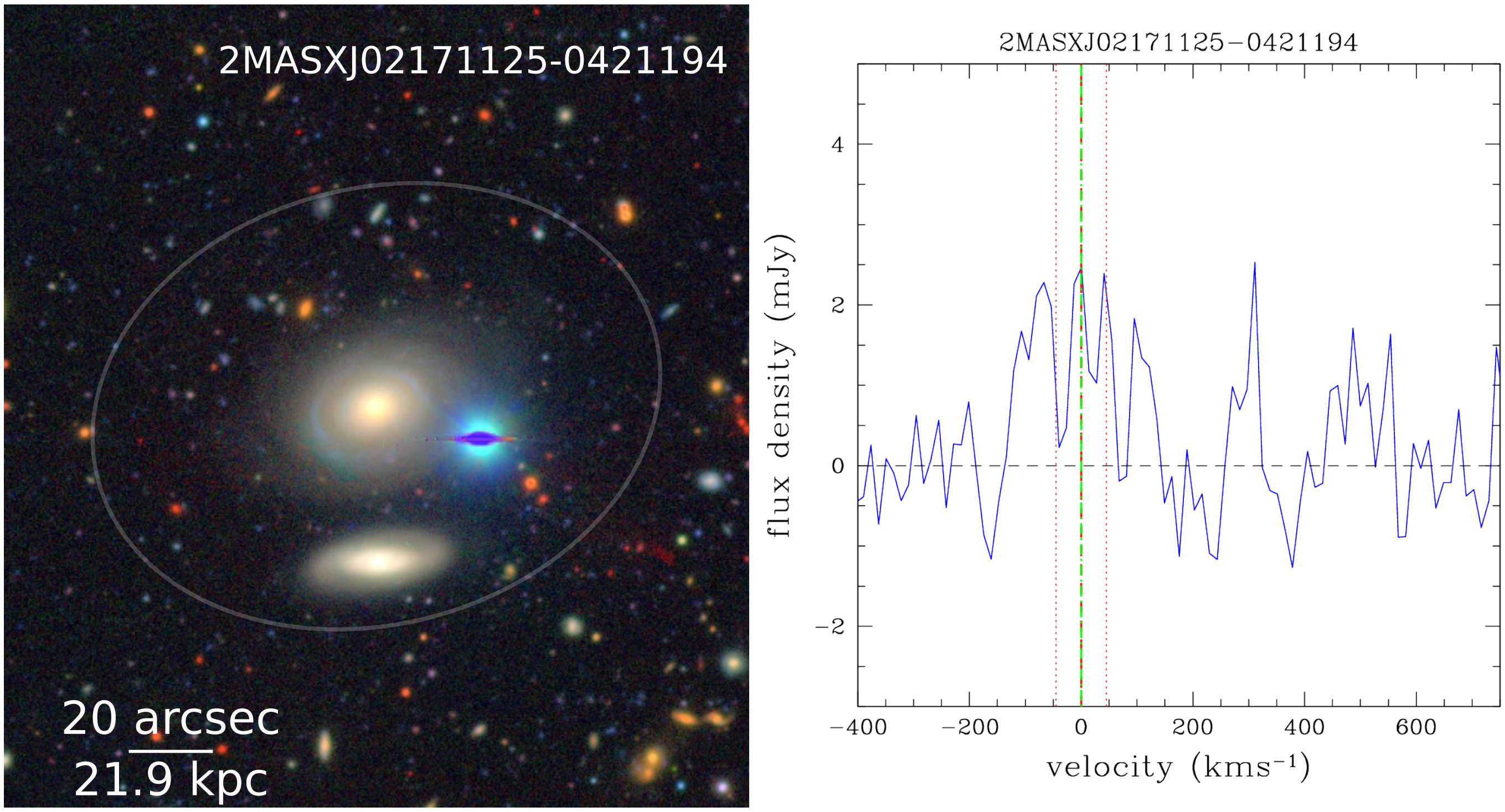

**Figure 21.** This figure displays the properties of galaxy 2MASXJ02171125-0421194. On the left is a color composite of the HSC images. On the right is the GBT H I spectrum centered at the optical redshift. The galaxy has H I mass of $2.52 \times 10^{9}\ M_{\odot}$ and a $g$-band absolute magnitude of −21.35. The H I emission signal from 2MASXJ02171125-0421194 is weaker than the other gLSB galaxies as the galaxy has H I mass less than $10^{10}\ M_{\odot}$. The H I spectrum has been Gaussian smoothed over 10 channels and then decimated to make the H I signal clearer (the values in tables in Section 3 were derived from the five-channel smoothed spectrum, like the other gLSB galaxies). Whether the galaxy started with a lower H I mass than the other gLSB galaxies or it evolved to have a lower H I mass is not clear.

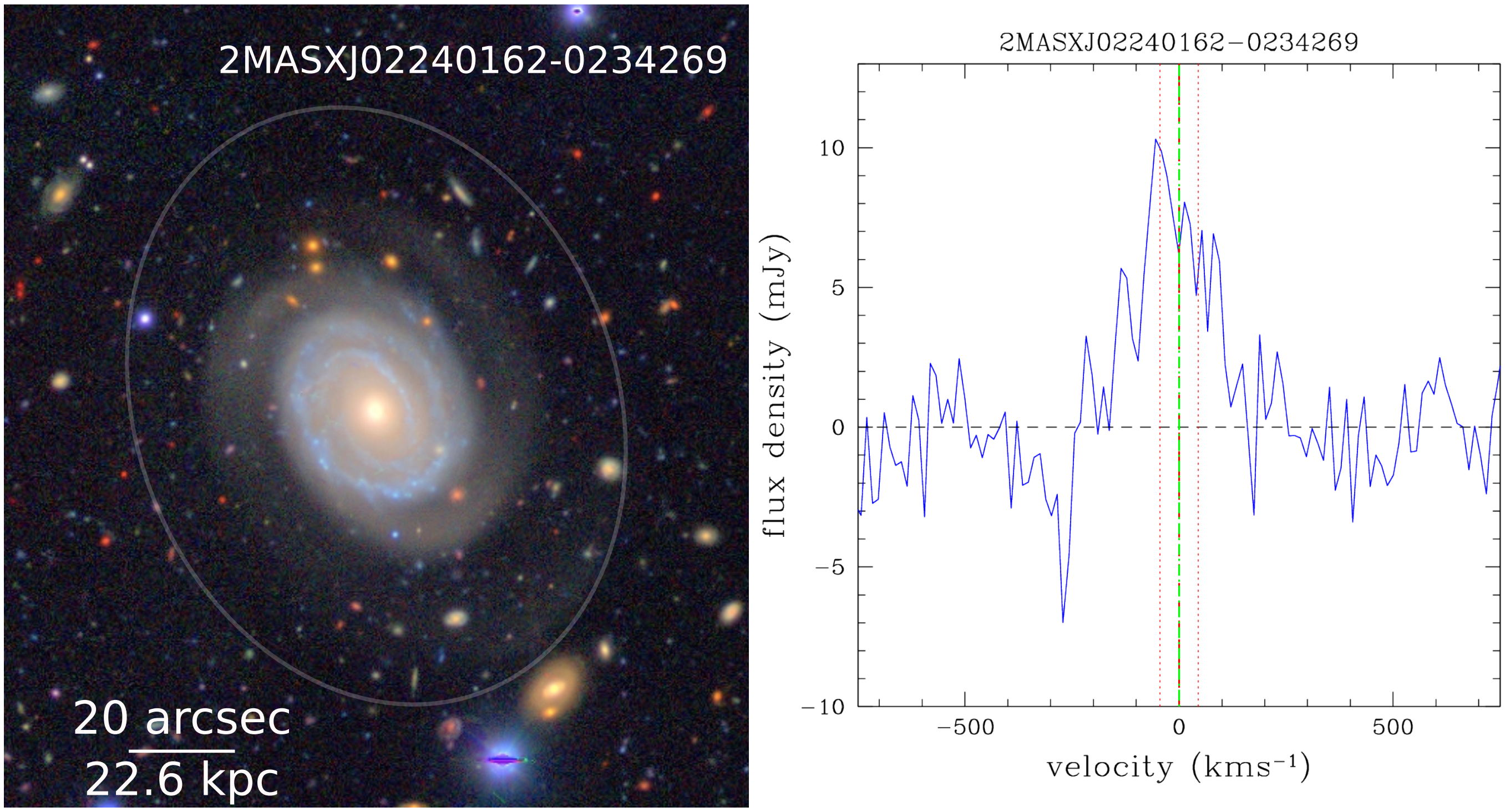

**Figure 22.** This figure displays the properties of galaxy 2MASXJ02240162-0234269. On the left is a color composite of the HSC images. On the right is the GBT H I spectrum centered at the optical redshift. The H I spectrum of 2MASXJ02240162-0234269 has been Gaussian smoothed over 10 channels and then decimated to highlight the emission features (again, the values in the tables in Section 3 are from the five-channel smoothing, like the other galaxies). The galaxy has H I mass of $2.70 \times 10^{10}\ M_{\odot}$ and a $g$-band absolute magnitude of −21.88. This, however, is a more normal gLSB with H I mass over $10^{10}\ M_{\odot}$ with a Gaussian spectral profile. The error in the $C_v$ value is unusually large, making it practically unusable. The central surface brightness for this galaxy is 23.31 mag arcsec$^{-2}$ in the $g$ band.

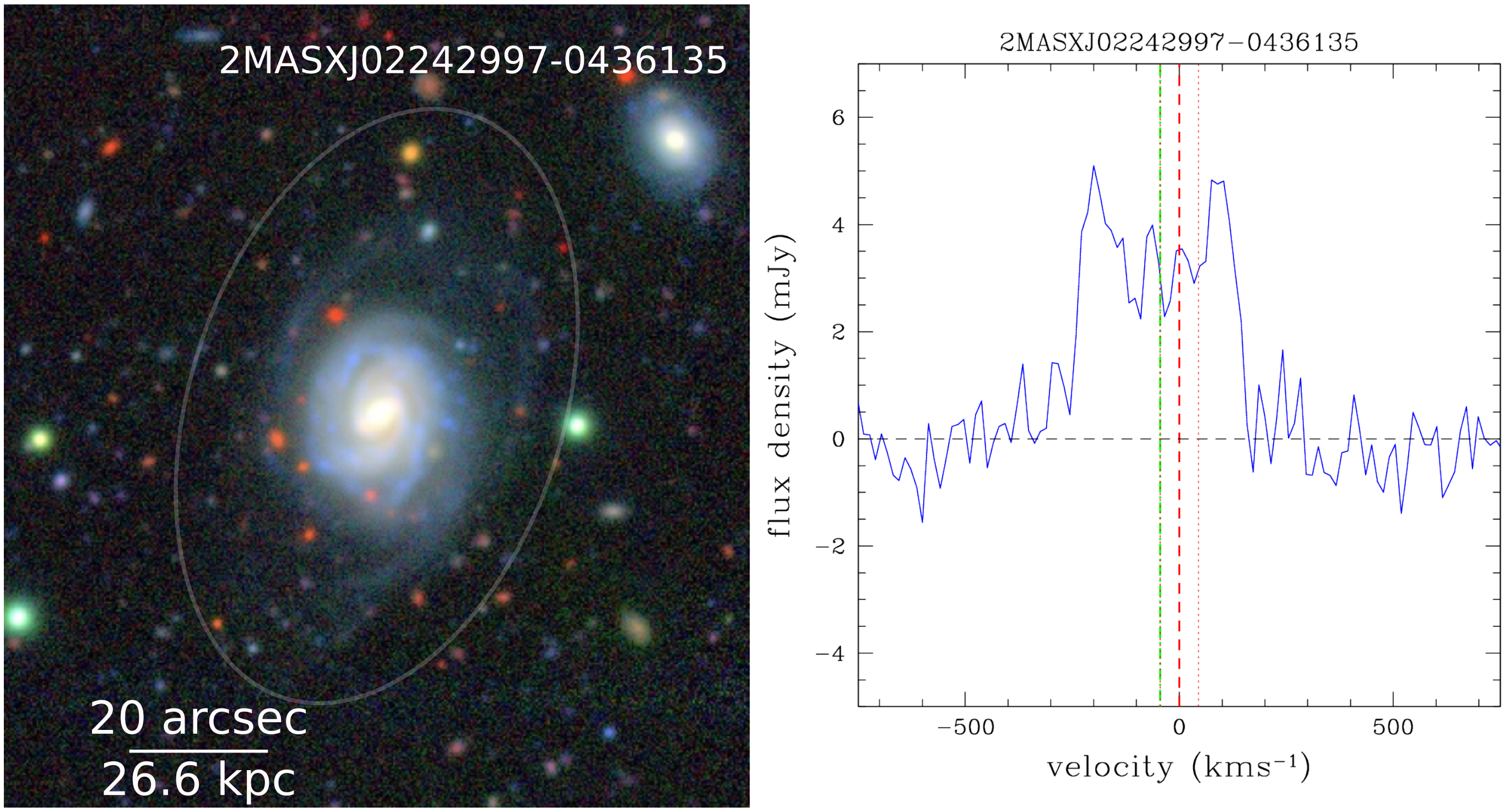

**Figure 23.** This figure displays the properties of galaxy 2MASXJ02242997-0436135. On the left is a color composite of the HSC images. On the right is the GBT H I spectrum centered at the optical redshift. The galaxy has H I mass of $3.03 \times 10^{10}\ M_{\odot}$ and a $g$-band absolute magnitude of −21.84. The galaxy has H I mass of $3.03 \times 10^{10}\ M_{\odot}$ and a $g$-band absolute magnitude of −21.84. 2MASXJ02242997-0436135 is a fairly standard gLSB, which looks like a noisy but still recognisable double-horned profile. Its H I mass is large, above $10^{10}\ M_{\odot}$.

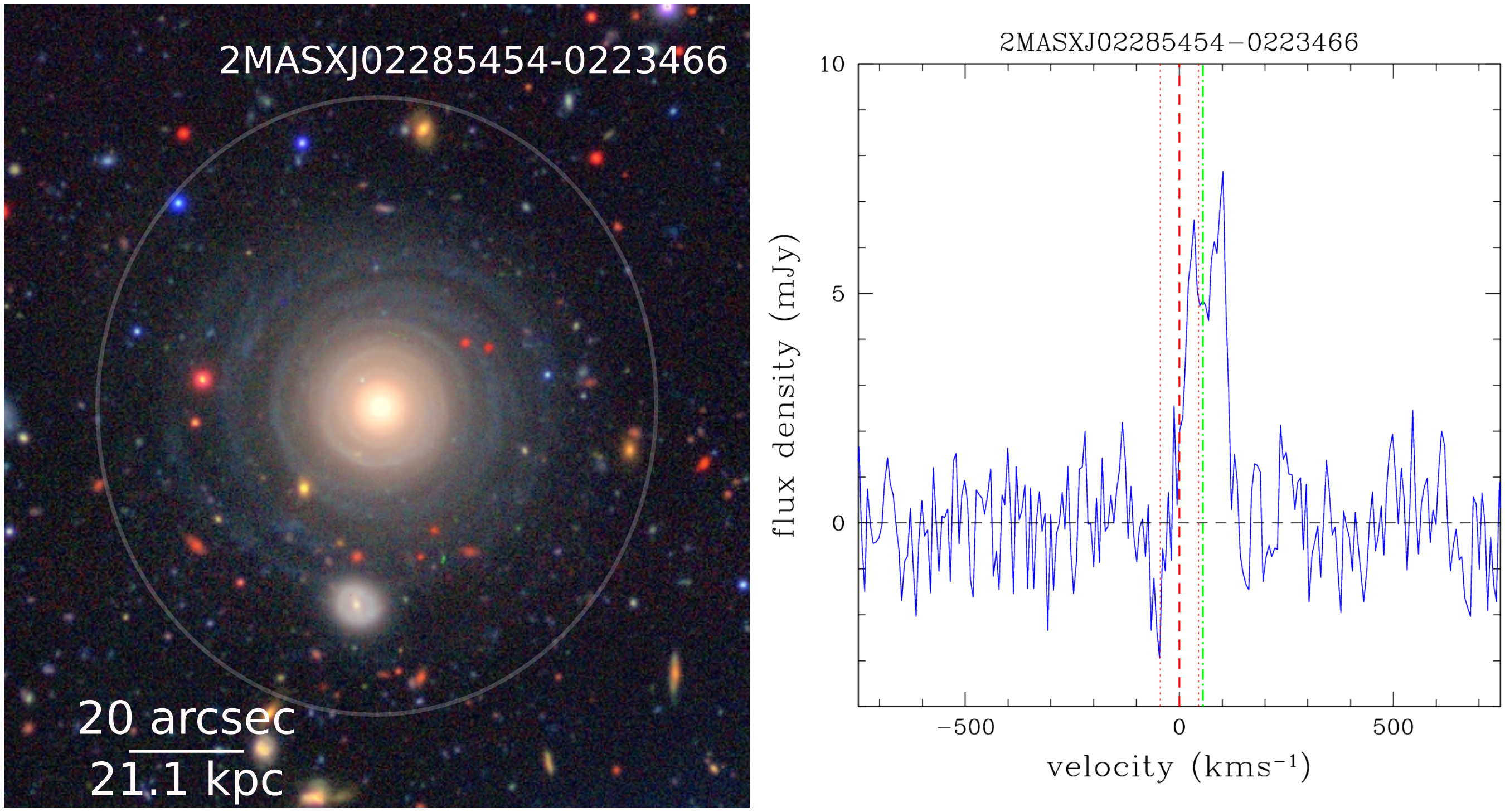

**Figure 24.** This figure displays the properties of galaxy 2MASXJ02285454-0223466. On the left is a color composite of the HSC images. On the right is the GBT H I spectrum centered at the optical redshift. The galaxy has H I mass of $7.74 \times 10^9\, M_\odot$ and $g$-band absolute magnitude of −21.35.

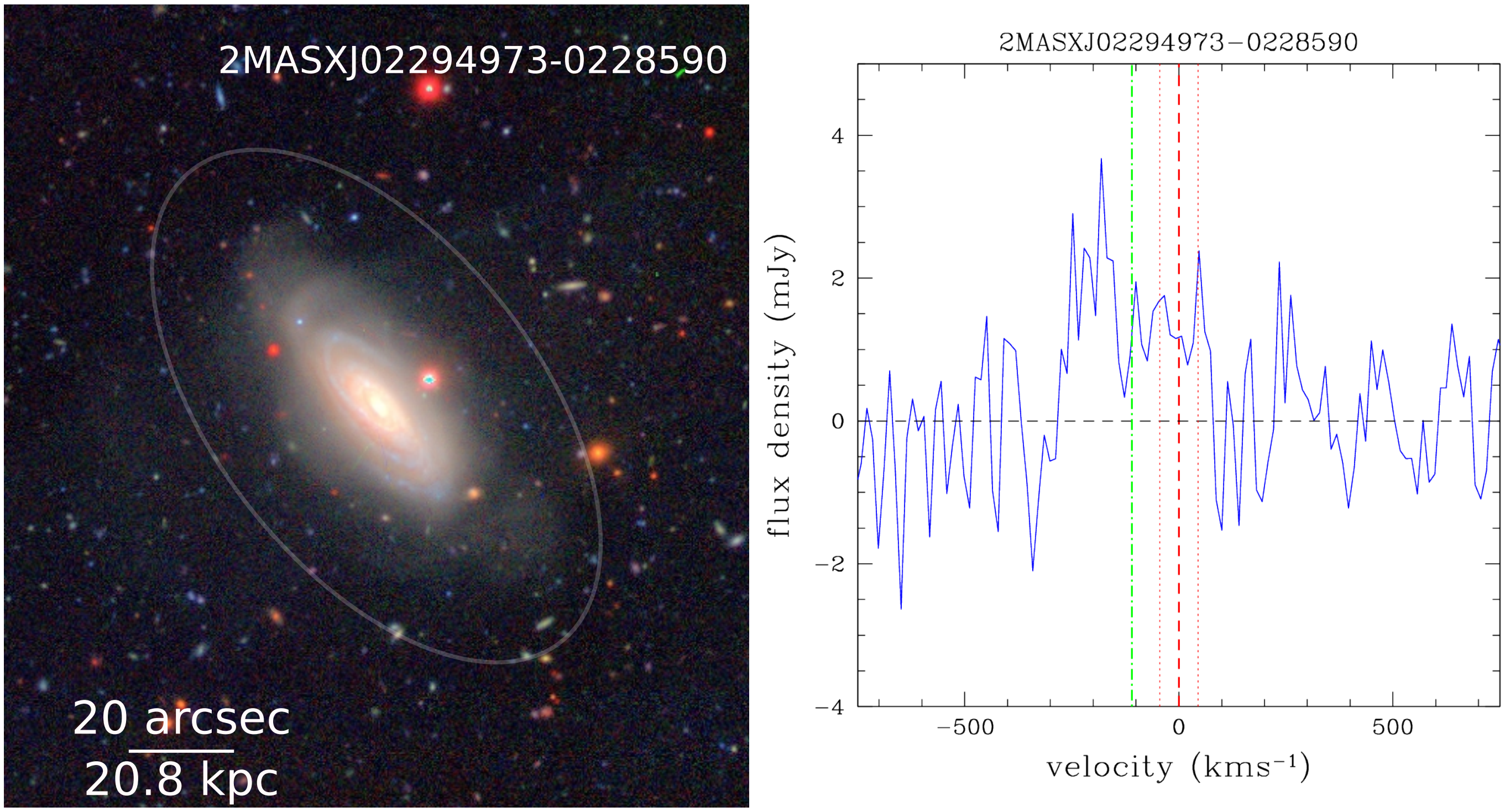

**Figure 25.** This figure displays the properties of galaxy 2MASXJ02294973-0228590. On the left is a color composite of the HSC images. On the right is the GBT H I spectrum centered at the optical redshift. The H I spectrum of 2MASXJ02294973-0228590 was Gaussian smoothed over 10 channels and then decimated to highlight the emission (the values in the Section 3 were derived for the five-channel smoothing). The galaxy has H I mass of $6.43 \times 10^9\, M_\odot$ and a $g$-band absolute magnitude of −21.68. The $w_{50}$ has a very high error for this galaxy, mainly due to the number of dips in the five-channel spectrum. This galaxy has a lower H I mass than the other gLSB galaxies, being below $10^{10}\, M_\odot$. Again, whether the galaxy started with this H I mass or evolved to it is uncertain.

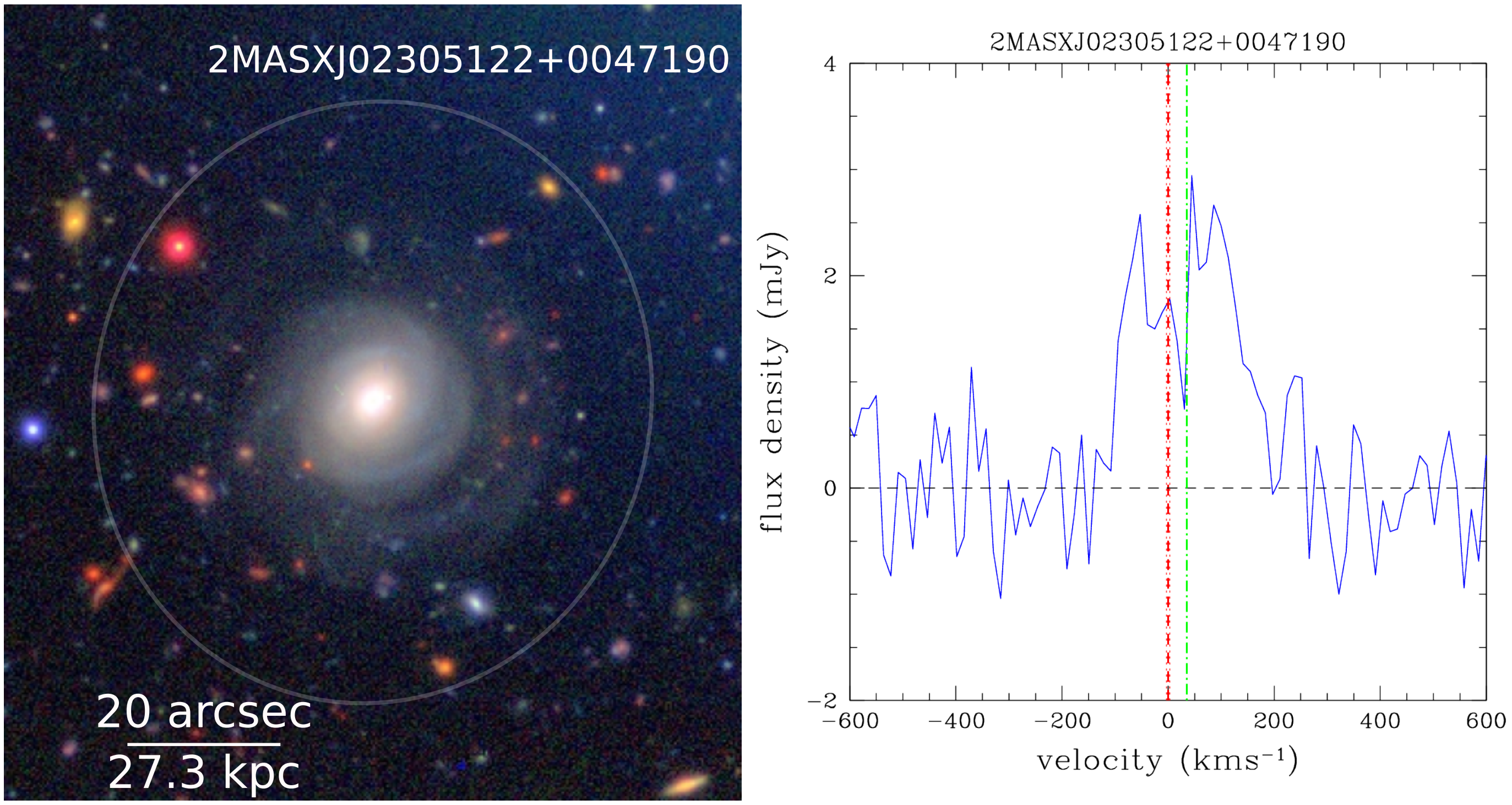

**Figure 26.** This figure displays the properties of galaxy 2MASXJ02305122+0047190. On the left is a color composite of the HSC images. On the right is the GBT H I spectrum centered at the optical redshift. The H I spectrum of 2MASXJ02305122+0047190 has been Gaussian smoothed over 10 channels and then decimated to highlight the H I emission (the values in the tables in Section 3 are from the five-channel smoothing as always). The galaxy has H I mass of $8.64 \times 10^9\ M_\odot$ and $g$ absolute magnitude of −21.17. This looks like a standard double-horned profile. Its H I mass is lower than we have come to expect, being less than $10^{10}\ M_\odot$. Again, whether this is where the galaxy started or where it evolved to is unknown.

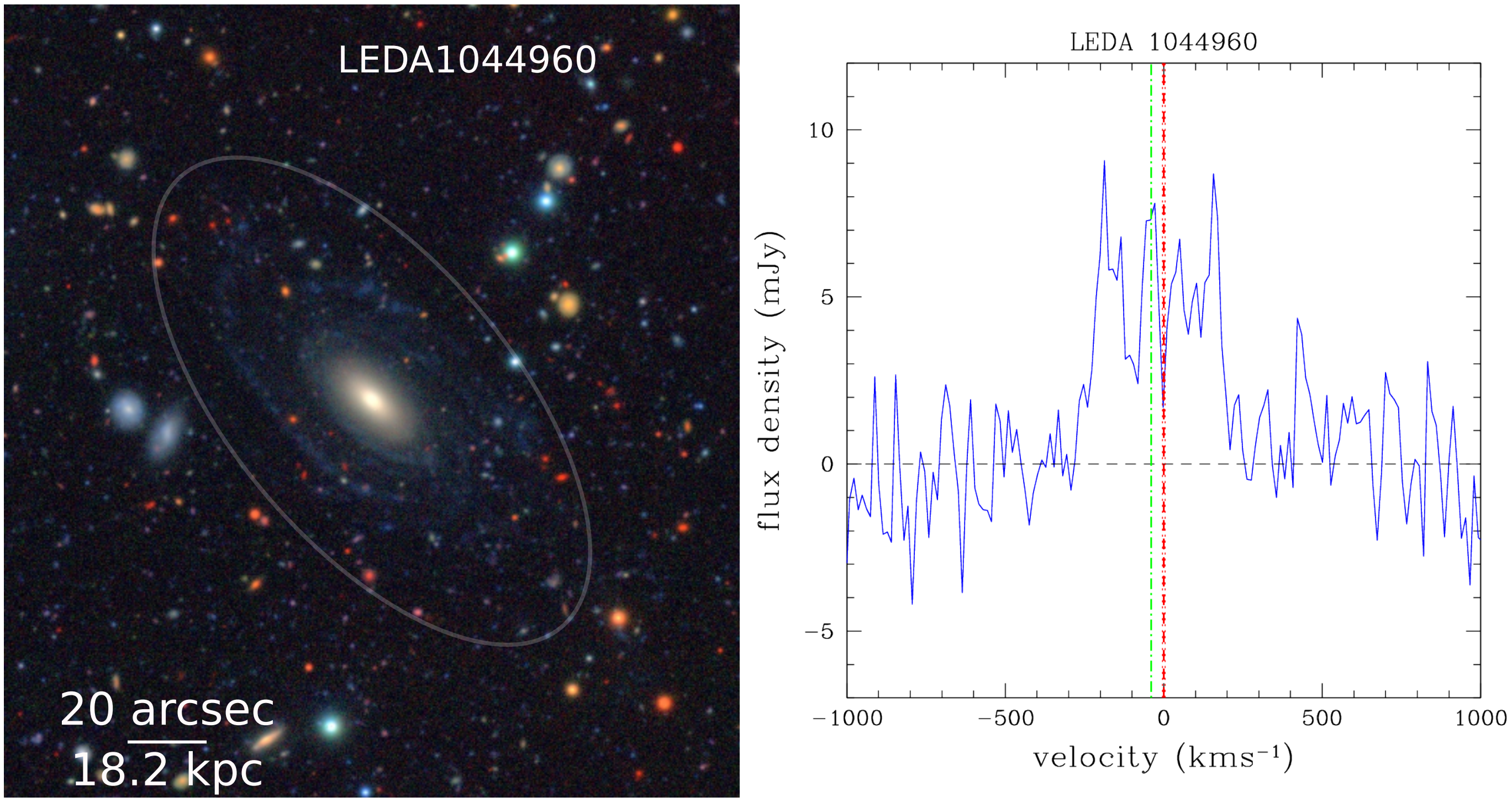

**Figure 27.** This figure displays the properties of galaxy LEDA 1044960. On the left is a color composite of the HSC images. On the right is the GBT H I spectrum centered at the optical redshift. Again, the H I spectrum of LEDA 1044960 has been Gaussian smoothed over 10 channels and then decimated to clarify the H I emission signal (the values in the tables in Section 3 come from the five-channel smoothing). The galaxy has H I mass of $2.27 \times 10^{10}\ M_\odot$ and a $g$-band absolute magnitude of −19.58. This is a fairly normal gLSB galaxy with H I mass above $10^{10}\ M_\odot$ and from the $C_V$ and K values, mostly close to a double-horned profile.

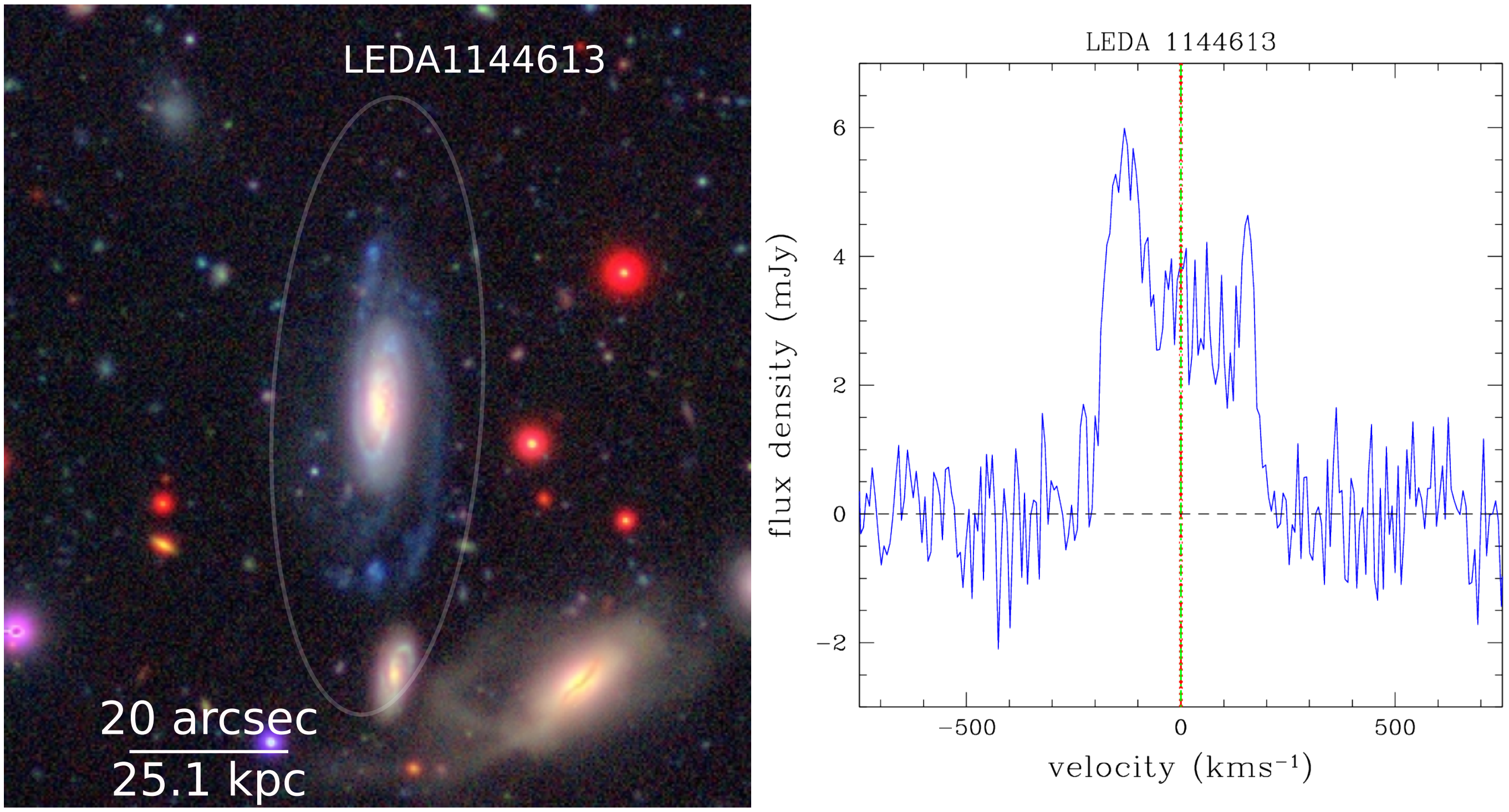

**Figure 28.** This figure displays the properties of galaxy LEDA 1144613. On the left is a color composite of the HSC images. There are two large companions in this image, but they are not near the gLSB galaxy. LEDA 1144613, the gLSB galaxy, is at $cz = 19{,}623$ km s$^{-1}$. WISEA J020727.60-002347.9, one of the companions, is at $cz = 24{,}943$ km s$^{-1}$ (Y. A. Li et al. 2023). The other companion is WISEA J020729.20-002346.9 at $cz = 49{,}119$ km s$^{-1}$ (D. R. Law et al. 2016). On the right is the GBT H I spectrum centered at the optical redshift. The galaxy has H I mass of $2.62 \times 10^{10}\,M_{\odot}$ and g absolute magnitude of −20.31. LEDA 1144613 is a fairly normal gLSB as it has H I mass over $10^{10}\,M_{\odot}$. Its H I spectrum is an asymmetric double-horned profile, indicating a possible asymmetry in the location of the gas within the galaxy.

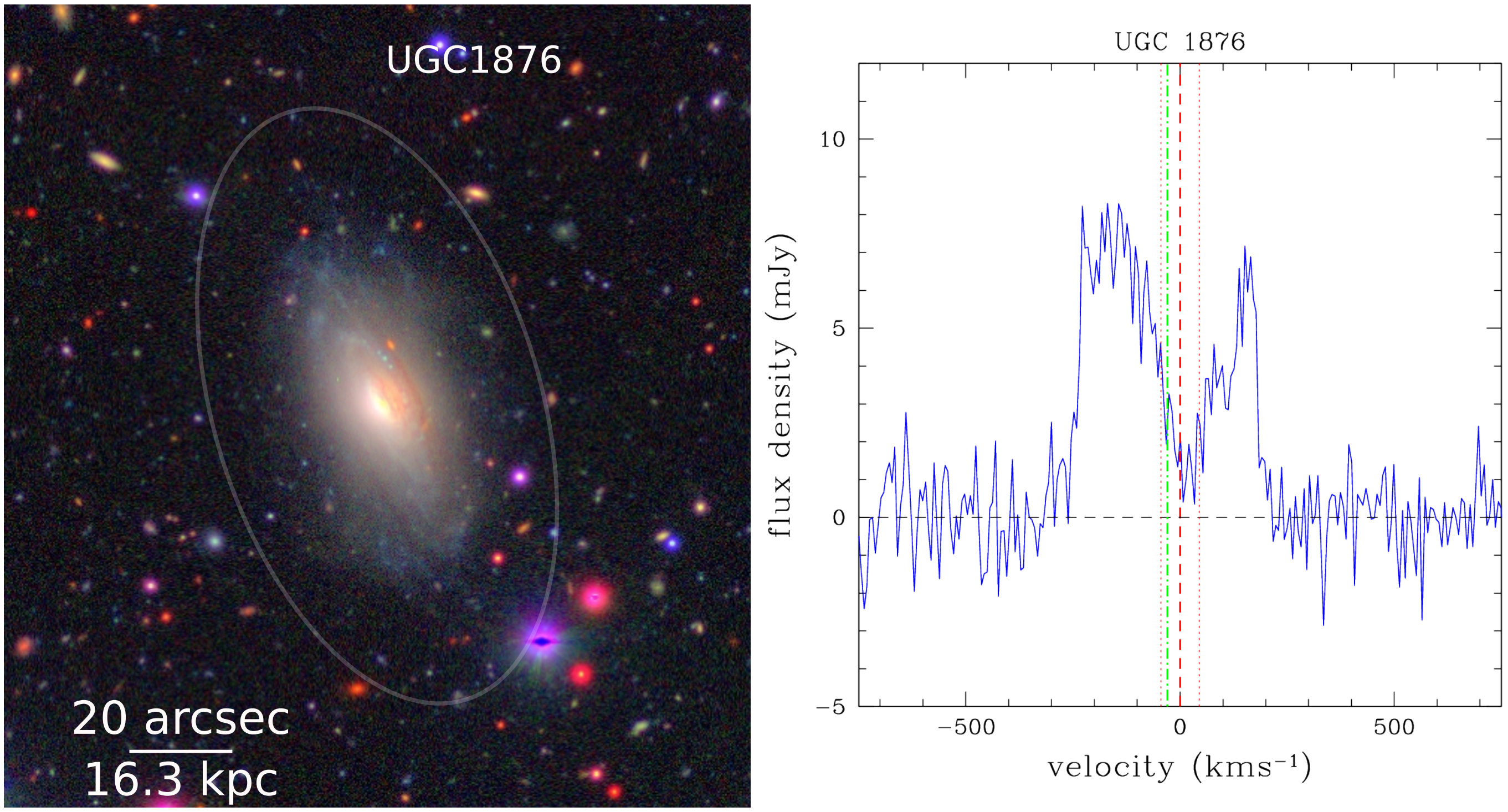

**Figure 29.** This figure displays the properties of galaxy UGC 1876. On the left is a color composite of the HSC images. On the right is the GBT H I spectrum centered at the optical redshift. The galaxy has H I mass of $1.54 \times 10^{10}\,M_{\odot}$ and a $g$-band absolute magnitude of −20.53. UGC 1876 is a fairly normal gLSB with H I mass greater than $10^{10}\,M_{\odot}$. Its H I spectrum shows an asymmetric double-horned profile, indicating a gas disk that likely has an asymmetry to it.

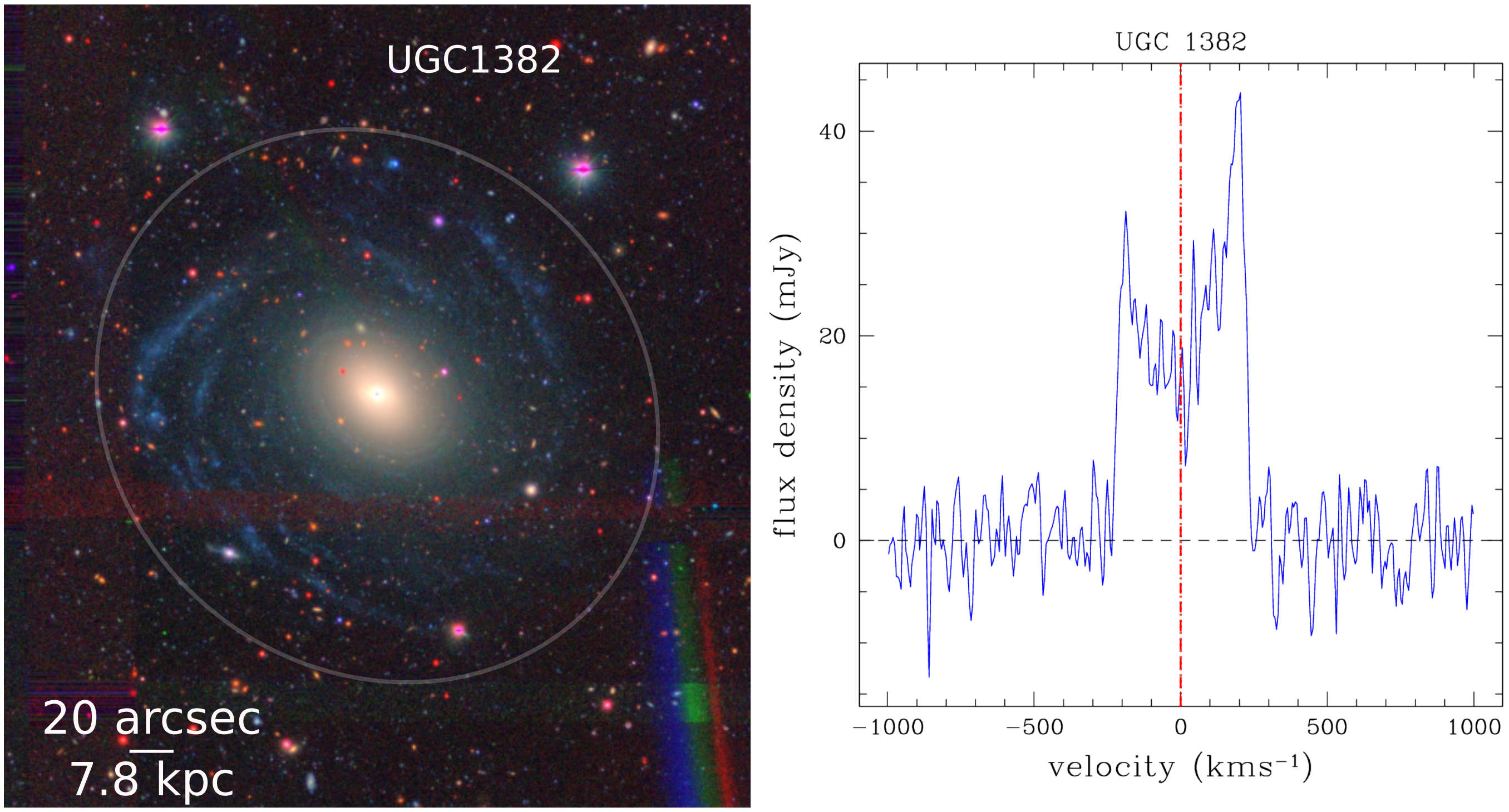

**Figure 30.** This figure displays the properties of galaxy UGC 1382. On the left is a color composite of the HSC images. On the right is the ALFALFA H I spectrum centered at the optical redshift. The galaxy has H I mass of 1.58 × $10^{10}$ $M_{\odot}$ and a $g$-band absolute magnitude of −20.41.

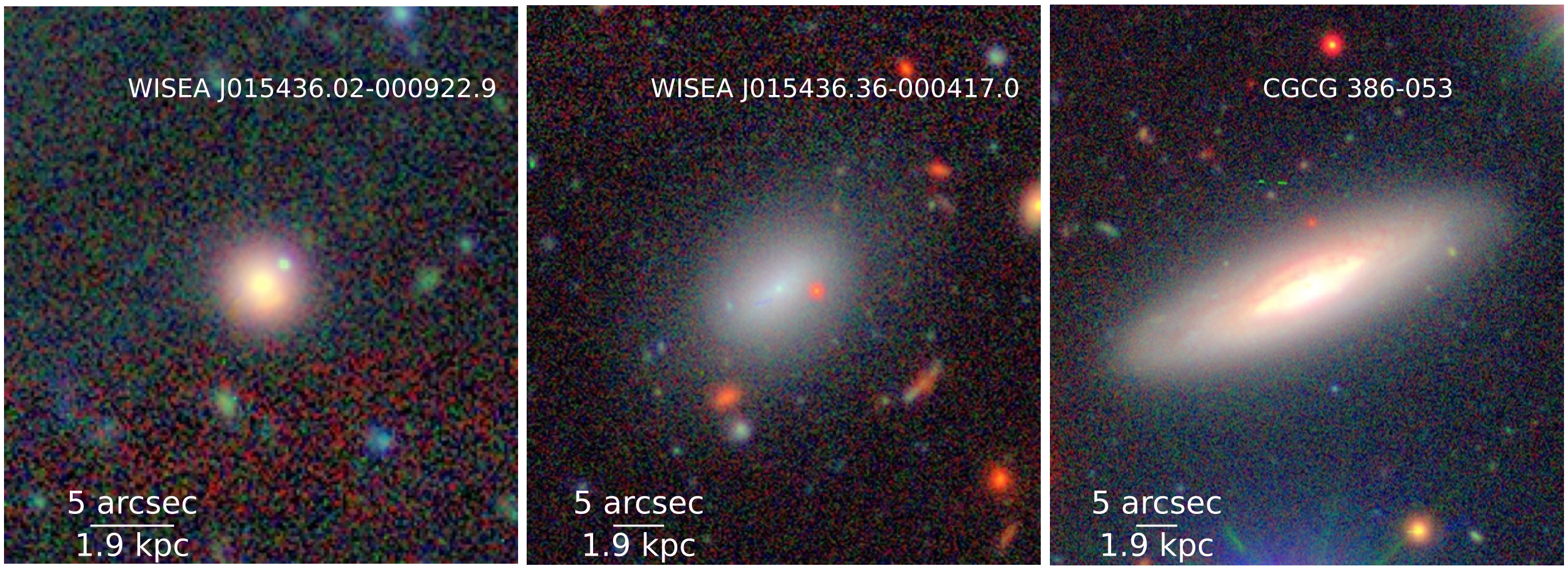

**Figure 31.** Optical images of the companion galaxies to UGC 1382 from the HSC SSP survey. They are combined $g$, $r$, and $z$ filter images.

For the measurement of the gLSB galaxy, which is much larger in optical size than the companions, the signal from a box 150″ across was integrated. This size was based on an estimate from the optical size of the galaxy and reviews of the VLA data cube. The H I spectrum from this observation can be seen in Figure 19. There is evidence of H I signal, but it is at a level that would be seen as noise in the GBT spectrum. Based on this spectrum, the gLSB galaxy WISEA J020019.56-012210.9 has H I mass of (0.279 ± 0.027) × $10^{10}$ $M_{\odot}$. This is considerably lower than what one would normally expect for a gLSB galaxy. The various quantities measured for the other gLSB galaxies have been computed based on this spectrum and put in the tables in Section 3.

The fact that WISEA J020019.56-012210.9 has a much lower H I mass than most of the other gLSB galaxies is at odds with its optical properties. The optical image of WISEA J020019.56-012210.9 shows clear evidence in the disk of blue light, a sign of recent star formation, which usually indicates the presence of H I gas. The galaxy may be in a transition state, having used up all the H I gas for star formation, but the high-mass stars, produced in recent star formation, have not yet died out.

### *A.3. Galaxy 2MASXJ02285454-0223466*

2MASXJ02285454-0223466 is almost face-on, so its H I spectrum is quite narrow at $w_{50}$ = (95.7 ± 4.9) km $s^{-1}$.

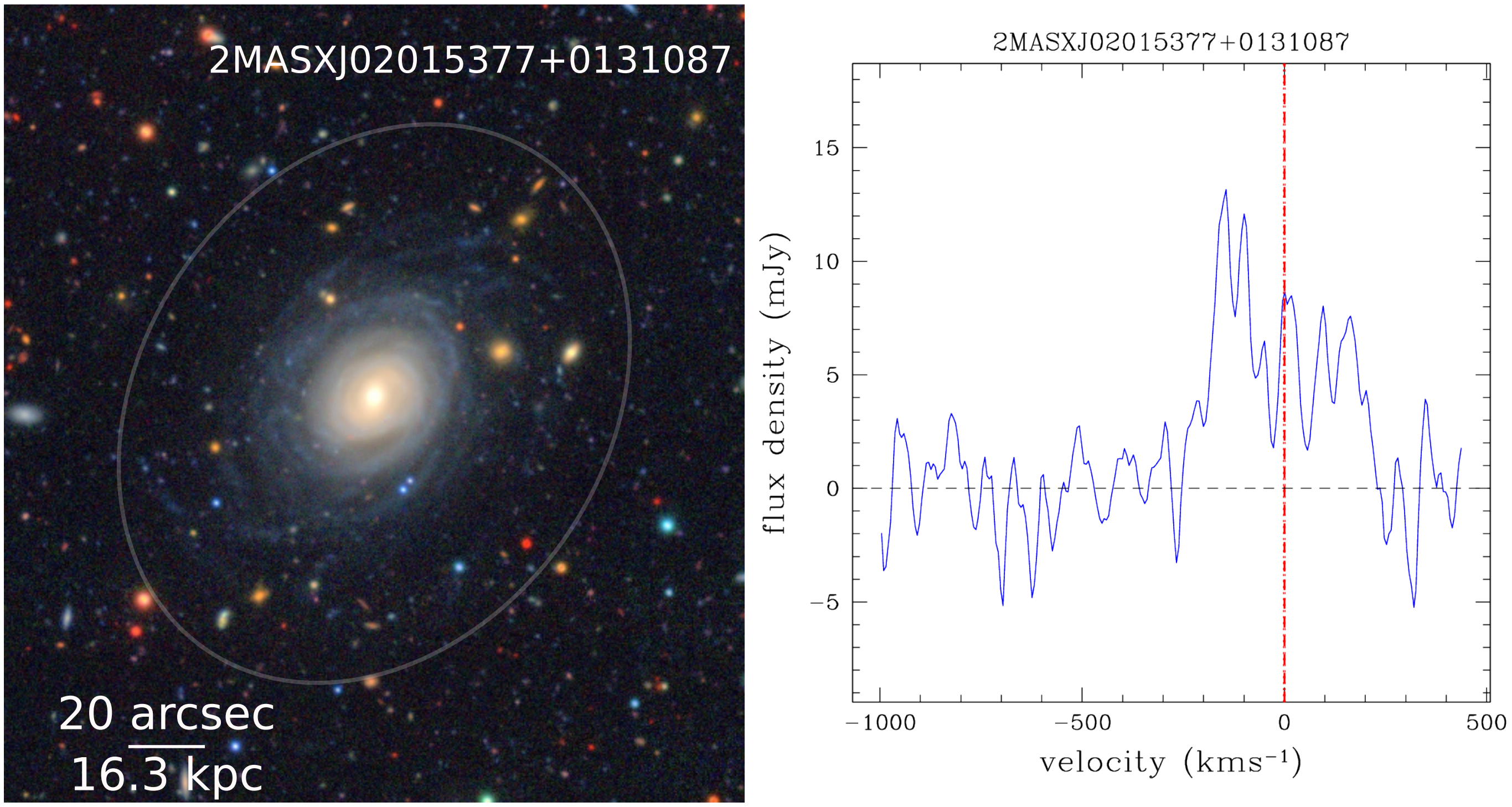

**Figure 32.** This figure displays the properties of galaxy 2MASXJ02015377+0131087. On the left is a color composite of the HSC images. On the right is the ALFALFA H I spectrum centered at the optical redshift. The galaxy has H I mass of $2.06 \times 10^{10}\, M_{\odot}$ and a $g$-band absolute magnitude of −20.09.

**Table 5**
Information on the gLSB Galaxy UGC 1382 and Its Companion Galaxies

| Galaxy | Velocity Difference (km s$^{-1}$) | $g$ mag abs | $M_{\rm H\,I}$ Estimated ($M_{\odot}$) | Distance From UGC 1382 (arcmin) | Distance From UGC 1382 (kpc) |
|---|---|---|---|---|---|
| UGC 1382 | 0 | −20.26 | $8 \times 10^9$ | 0 | 0 |
| WISEA J015436.02-000922.9 | −123 | −15.31 | $5 \times 10^8$ | 1.47 | 34.8 |
| WISEA J015436.36-000417.0 | 82 | −16.79 | $1 \times 10^9$ | 4.46 | 106 |
| CGCG 386-053 | −122 | −19.01 | $4 \times 10^9$ | 5.51 | 131 |

**Note.** The velocity difference is the difference between the optical redshift of each galaxy and UGC 1382. The $g$-band magnitudes are from SDSS K. Abazajian et al. (2004). The $M_{\rm H\,I}$ estimated is the H I mass estimated using the correlation from A. Durbala et al. (2020) between the $g$-band absolute magnitude and H I mass (see Figure 3). The Arecibo beam has an FWHM of $\sim 3\overset{\prime}{.}6$ at the relevant frequencies.

The optical redshift looks like it is offset from the H I emission in Figure 24 but when you compare the redshift offset of 55 km s$^{-1}$ to the optical redshift error of 45 km s$^{-1}$ this is not that significant. It is just that the H I spectrum is narrower than previous spectra. However, the H I mass of this galaxy is less than we have come to expect for gLSB galaxies being less than $10^{10}\, M_{\odot}$. There was a problem with the measurement of the K value that gave it a value beyond what was possible for a galaxy. This was probably due to the reduced number of points within the H I spectrum. To fix this, a spline fit was used to sample the spectrum at a higher frequency, and the K value from this fit became a more reasonable number.

### A.4. Galaxy 2MASXJ02305122+0047190

### A.5. Galaxy UGC 1382

The H I spectrum for galaxy UGC 1382 shown in Figure 30 is an asymmetric double-horned profile, indicating a rotating disk of H I gas that is asymmetric. This spectrum comes from the ALFALFA Survey (M. P. Haynes et al. 2018). No smoothing has been done on the spectrum. There are three companion galaxies close to this galaxy: they are described in Table 5, and their optical images are shown in Figure 31. ALFALFA was a drift scan survey with each of the seven beams having an FWHM of an $\sim 3\overset{\prime}{.}5$ beam (R. Giovanelli et al. 2005). Only the companion WISEA J015436.02-000922.9 would be within one Arecibo beam FWHM at the same time as UGC 1382, and this galaxy likely has a very low H I masses, less than an order of magnitude below the estimated value for UGC 1382. None of these companion galaxies was detected in ALFALFA. One can be fairly confident that there is minimal contamination from the companion galaxies in the H I spectrum of UGC 1382.

### A.6. Galaxy 2MASXJ02015377+0131087

The H I spectrum for galaxy 2MASXJ02015377+0131087 shown in Figure 32 is fairly noisy but could be an asymmetric double-horned profile. The displayed spectrum has been boxcar smoothed to size 5 to highlight the H I signal. This spectrum comes from the ALFALFA survey (M. P. Haynes et al. 2018). While there is significant noise in the spectrum, one can still measure the H I mass of the galaxy, which comes out to be higher than $10^{10}$ $M_\odot$ making it a fairly standard gLSB galaxy.

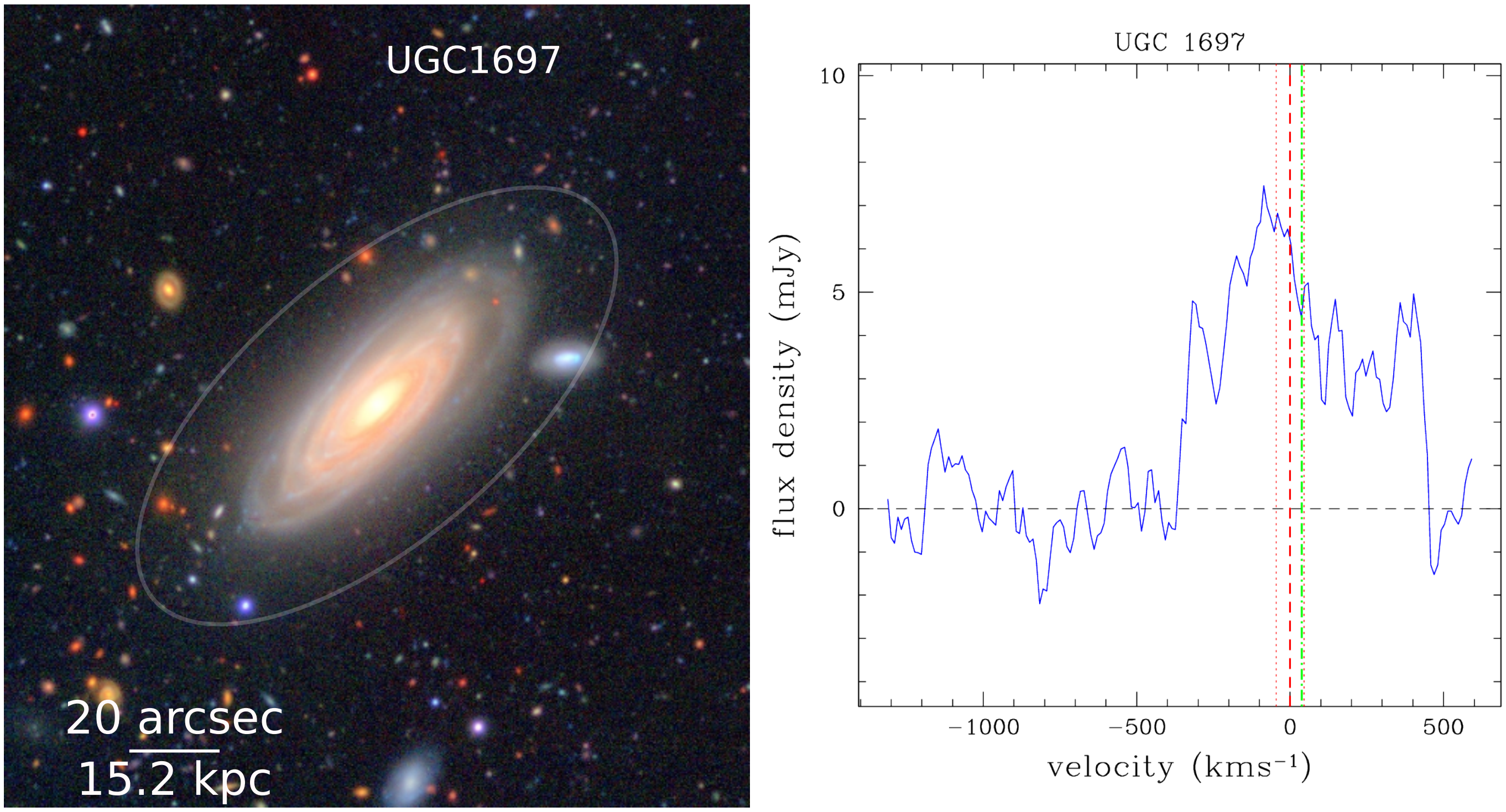

**Figure 33.** This figure displays the properties of galaxy UGC 1697. On the left is a color composite of the HSC images. On the right is the Nançay Telescope H I spectrum centered at the optical redshift. The galaxy has H I mass of 2.15 × $10^{10}$ $M_\odot$ and g absolute magnitude of −21.75. The H I spectrum for galaxy UGC 1697 is unusual in its large velocity width, $w_{50} = (771.8 \pm 3.7)$ km $s^{-1}$. The original spectrum was quite noisy, so the displayed spectrum has been boxcar smoothed with a size of 5 to highlight the H I signal. This spectrum comes from the Nançay Telescope observations (G. Theureau et al. 2005). The beam size of the Nançay Telescope is $3'.6$ (east–west) × $22'$ (north–south). There are no companions to explain the longer velocity width and its H I mass, while large, it is in line with other gLSB galaxies being greater than $10^{10}$ $M_\odot$.

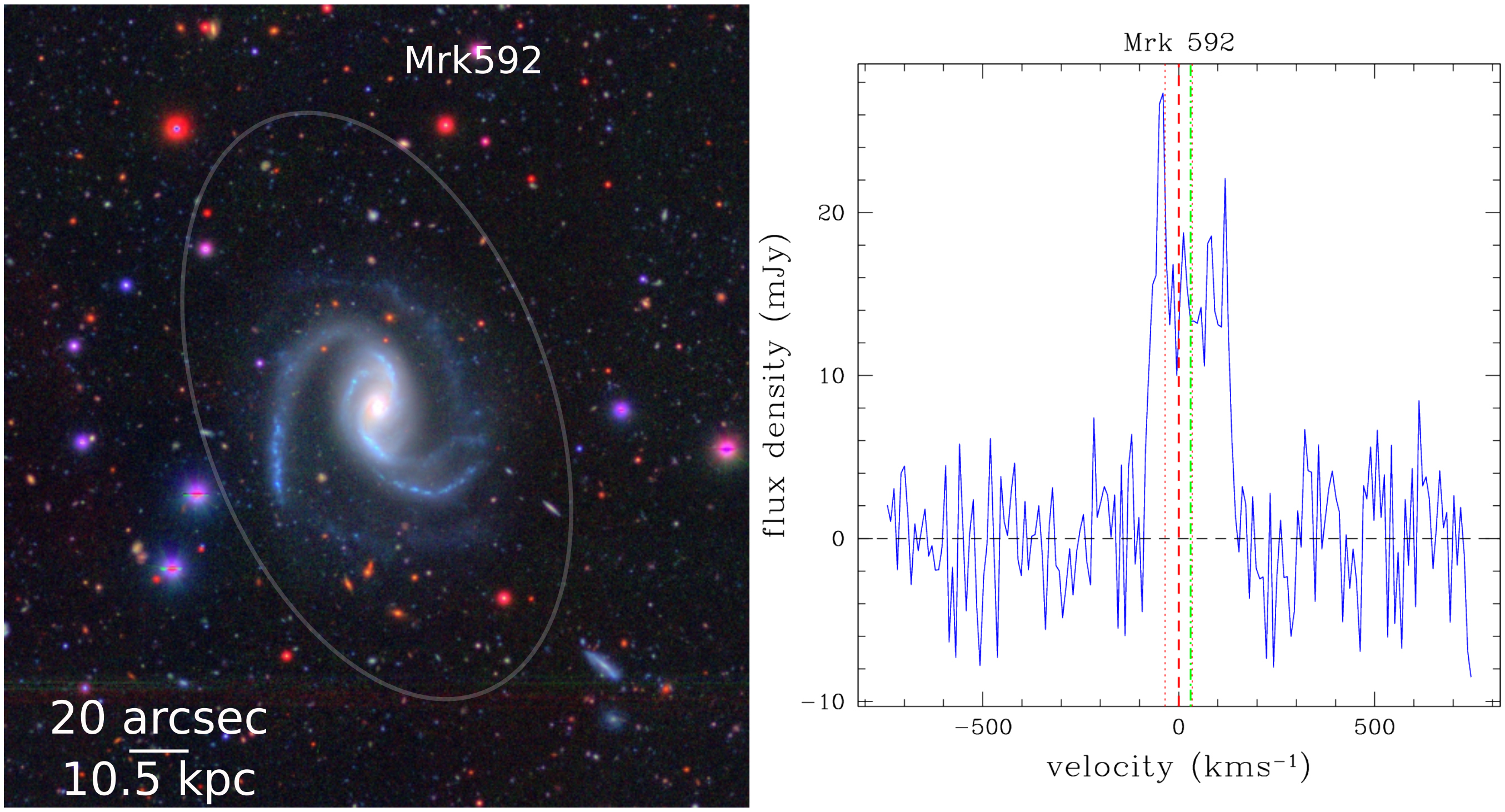

**Figure 34.** This figure displays the properties of galaxy Mrk 592. On the left is a color composite of the HSC images. On the right is the Arecibo H I spectrum centered at the optical redshift. The galaxy has H I mass of 1.01 × $10^{10}$ $M_\odot$ and *g*-band absolute magnitude of −20.75. The H I spectrum for galaxy Mrk 592 is between an asymmetric double-horned profile and a flat top curve. This spectrum comes from a compilation of H I spectrum made by C. M. Springob et al. (2005), with this spectrum observed with Arecibo. No smoothing has been done on the spectrum. This is a gLSB galaxy that has H I mass just above $10^{10}$ $M_\odot$ but is otherwise similar to the other gLSB galaxies.

We have observed the optical long-slit spectrum of this galaxy with multi-mode focal reducer SCORPIO-2 (V. L. Afanasiev & A. V. Moiseev 2011) mounted in the primary focus at the 6 m Big Telescope Alt-azimuthal of the Special Astrophysical Observatory of the Russian Academy of Science on 18.12.2020 with 1″ seeing. The scale along the slit is 0.″36 pixel$^{-1}$, the slit width was 1″. We utilized the grism VPHG1200@540, which covers the spectral range of 3600–7070 Å and has a dispersion of 0.87 Å pixel$^{-1}$. Using analysis of emission line spectrum (by `NBursts`; I. Chilingarian et al. 2007a; I. V. Chilingarian et al. 2007b) of ionized gas (for H$\beta$, [O III] and [S II] lines), we clarified the redshift (corrected to barycenter with `astropy.coordinates`) and rotational velocity of the galaxy: $v_{\rm redshift}$ = (12,390 ± 2) km s$^{-1}$ and $v_{\rm rot}$ = (161 ± 4) km s$^{-1}$.

### *A.7. Galaxy UGC 1697*

## Appendix B
## Additional H I Detection

In the H I spectrum of 2MASXJ02274954-0526053, there was an additional H I signal from a companion galaxy near the edge of the bandpass as seen in Figure 35. This H I galaxy has a redshift of 16,150 km s$^{-1}$ some 2311 km s$^{-1}$ distant from 2MASXJ02274954-0526053. This large velocity difference pretty much rules out that the galaxies are interacting in any way. As it has a Gaussian-like shape, it may not have a H I disk, or if it does, it is nearly face-on. The GBT beam has an FWHM of 9.′4 at this frequency, so the companion could come from quite a wide area. A search in the NASA Extragalactic Database found no galaxy anywhere in the GBT beam with the redshift of this object. A visual inspection of the Legacy images near this object also was unable to locate a likely galaxy. If we assume that the companion is at the center of the GBT beam, it would have H I mass of (1.01 ± 0.058) × $10^{10}$ $M_\odot$. It is likely higher than this, though. The companion has a $w_{50}$ of (253.0 ± 68) km s$^{-1}$.

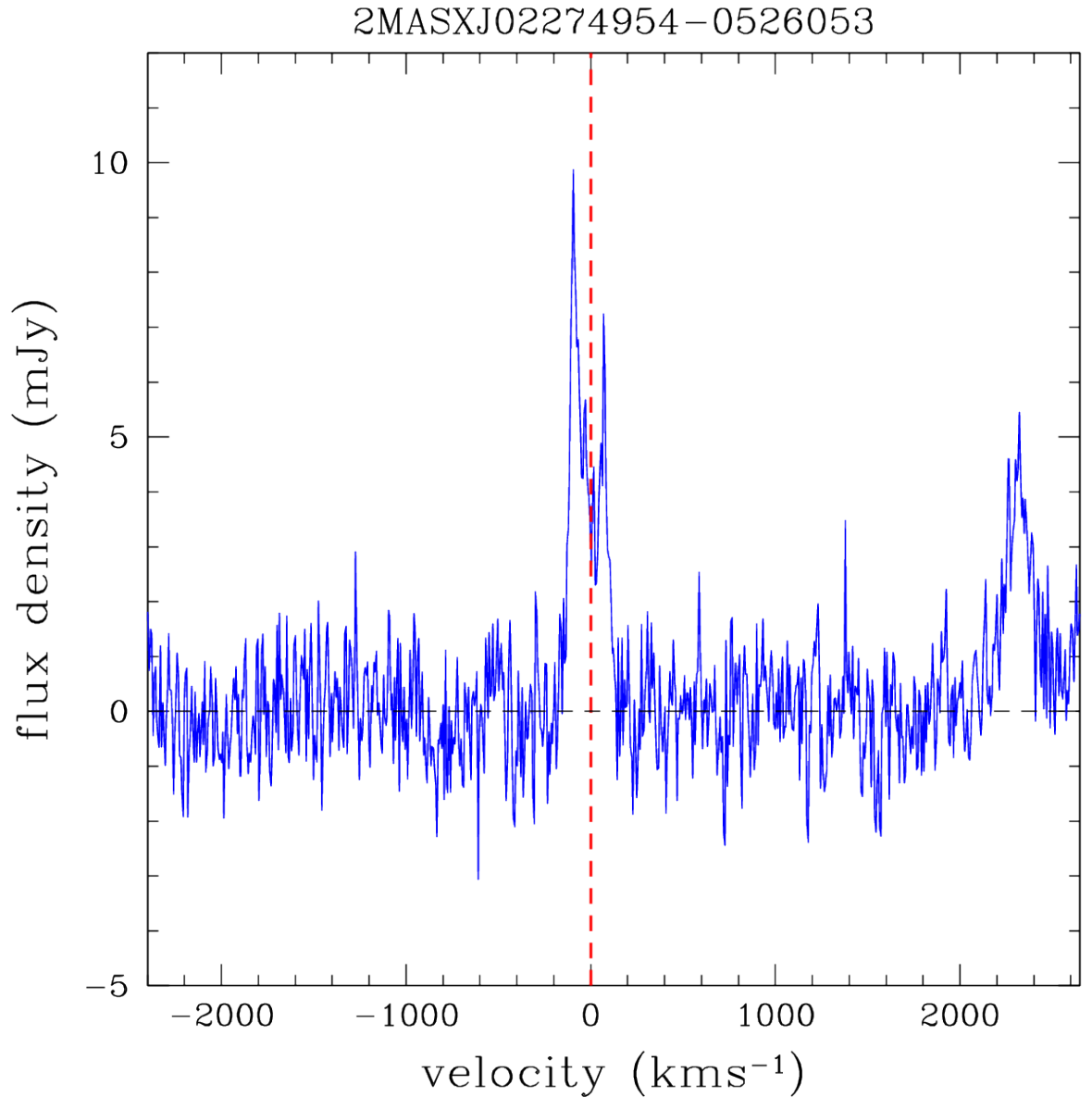

**Figure 35.** This is the H I spectrum for 2MASXJ02274954-0526053 extended to show the companion galaxy near the end of the bandpass.

## ORCID iDs

Philip Lah https://orcid.org/0000-0001-6841-6553
Nikhil Arora https://orcid.org/0000-0002-3929-9316
Ivan Yu. Katkov https://orcid.org/0000-0002-6425-6879
Joseph D. Gelfand https://orcid.org/0000-0003-4679-1058
Anna S. Saburova https://orcid.org/0000-0002-4342-9312
Igor V. Chilingarian https://orcid.org/0000-0002-7924-3253
Ivan Gerasimov https://orcid.org/0000-0001-7113-8152
Damir Gasymov https://orcid.org/0000-0002-1750-2096